\documentclass[pss]{wiley2sp}
\usepackage{amssymb}
\usepackage{amsmath}
\usepackage{booktabs}
\usepackage{caption}
\usepackage{subfloat}%
\input{tcilatex}

\begin{document}

\title{The Microscopic Response Method: theory of transport for systems with both topological and thermal disorder}
\author{ MingLiang Zhang, D. A. Drabold\textsuperscript{\Ast}}




\authorrunning{M.-L. Zhang and D. A. Drabold}

\mail{e-mail
  \textsf{zhangm@ohio.edu,drabold@ohio.edu}, Phone:
  +01-740-5931715, Fax: +01-740-5930433}

\institute{    Department of Physics and Astronomy, Ohio University, Athens, Ohio 45701\\
   }

\received{XXXX, revised XXXX, accepted XXXX}
\published{XXXX} 



\abstract{\abstcol{In this paper, we review and substantially develop the recently proposed "Microscopic Response Method", which has been devised to compute transport coefficients and especially associated temperature dependence in complex materials. The conductivity and Hall mobility of amorphous semiconductors and semiconducting polymers are systematically derived, and shown to be more practical than the Kubo formalism. The effect of a quantized lattice (phonons) on transport coefficients is fully included and then integrated out, providing the primary}{ temperature dependence for the transport coefficients.
For higher-order processes, using a diagrammatic expansion, one can consistently include all important contributions to a given order and directly write out the expressions of transport coefficients for various processes.}
}

%
%

\maketitle
\section{Introduction}
The career of O. F. Sankey exemplifies the ideals of scientific exploration, and even adventure. His many contributions to semiconductor, materials and biophysics are testimony to a man of ingenuity, energy and integrity. He has also been a rigorous but patient mentor to many contributors to this volume, including D.A.D. This paper on transport in complex materials is offered with affection and gratitude on the occasion of his sixtieth birthday.

The Kubo formula
\footnote{In this paper the Kubo formula refers to the original treatment of Kubo\cite{ku57}, not the subsequent approximate form of Greenwood\cite{Gre}, which is discussed elsewhere\cite{pre,jpcm}.}
 has been used to calculate the conductivity and Hall mobility of small polarons in molecular crystals\cite{lan62,fir,sch65,hol68,ald,lyo}.
The results obtained by different authors are inconsistent. The reasons are: (i) the imaginary time integral in the Kubo formula is complicated; and (ii) there is no systematic way to classify various transport processes induced by the external field and by the residual interactions.
\begin{figure}[ptb]
\begin{center}
\includegraphics[width=9cm]{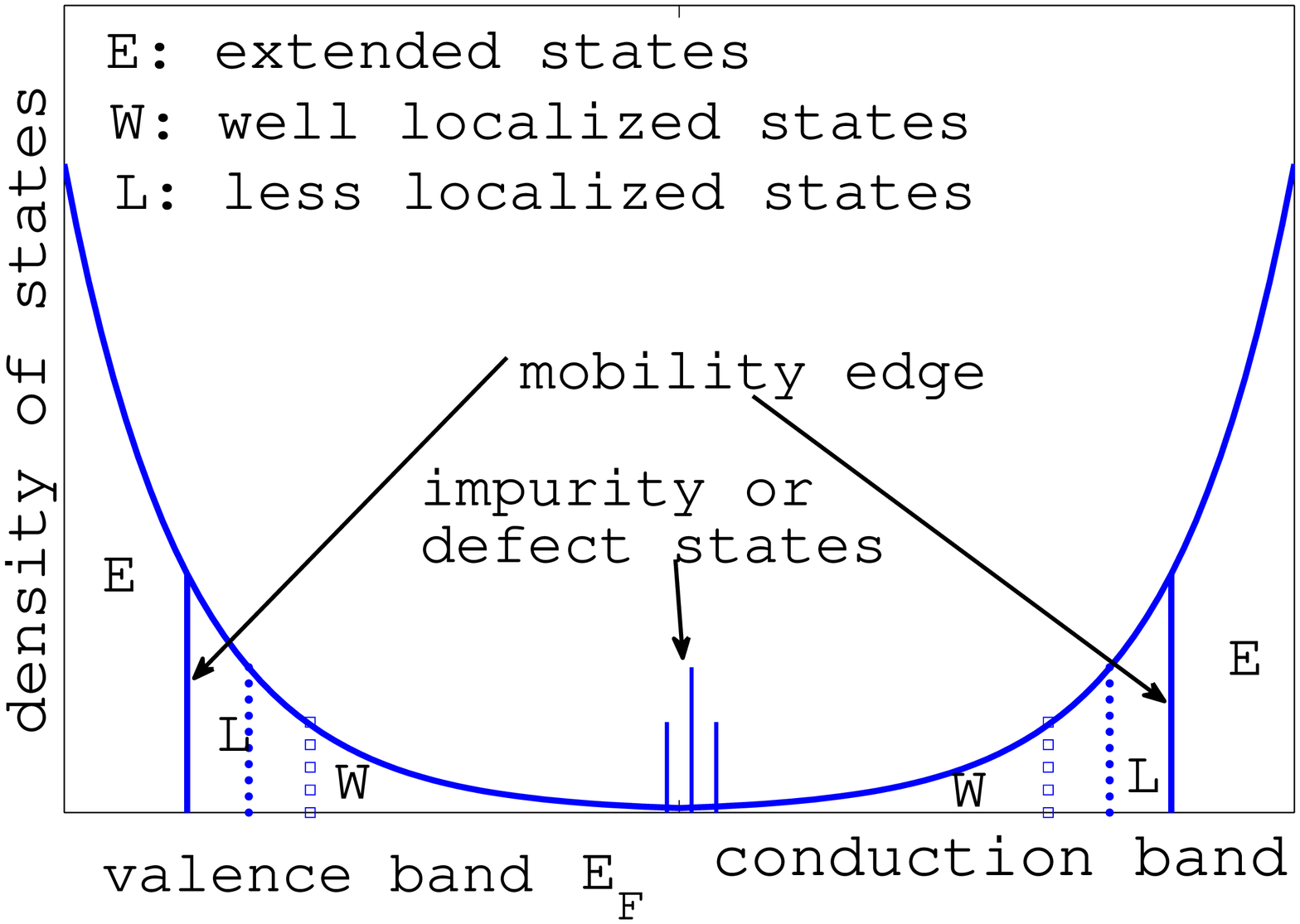}
\end{center}
\caption{Schematic energy spectrum of amorphous semiconductors or semiconducting polymers.
}%
\label{spectrum}%
\end{figure}
The small polaron model is inadequate for amorphous semiconductors (AS) 
in two aspects: (i) the states at the fringes of valence and conduction
bands are localized by topological disorder. At moderate
temperature both localized and extended states are accessible to electronic
transport processes\cite{motda}. Fig.\ref{spectrum} is a schematic energy spectrum of AS; (ii) A carrier in a localized state $\phi
_{A}$ polarizes the nearby atoms: the static displacements $d_{p_{A}}$ of
the atoms ($p_{A}$ is the index of atoms participating in localized state $\phi _{A}$)
induced by the e-ph interaction is comparable to or even
larger than $a_{v}$,
the amplitude of thermal and zero point vibrations. In contrast, the
static displacements of atoms caused by the carriers in extended states $\xi
_{B}$ are negligible\cite{epjb}.

The imaginary time integral in the Kubo formula can be avoided in the microscopic response method (MRM)\cite{short,pre,jpcm}. If an external perturbation on a system can be expressed via additional terms in the Hamiltonian (a ``mechanical perturbation''\cite{ku57}, electromagnetic field is an example), the microscopic response 
 can be obtained from the continuity equation for charge, the time-dependent Schrodinger equation and the initial conditions. The measured macroscopic response is a coarse-grained average, and an ensemble average
 of the microscopic response\cite{short}. Thus one does not need to calculate the macroscopic response by averaging the response operator over density matrix, and the imaginary time integral is avoided\cite{pre}. The purpose of this paper is to review the MRM and to report the full formulae for the conductivity and Hall mobility.

The MRM is equivalent to and simpler than the Kubo formula\cite{pre}. Because the microscopic current density in a state is a bilinear form of the wave function of that state connected by the current operator, each transport process is a temporal evolution of a state under the driving force of the external field and the residual interactions which are time ordered in a specific way. The contribution to the microscopic current density from a process is a product of several transition amplitudes and a connector of the corresponding current operator. One may visualize the transport processes by a series of diagrams. The topology of diagrams gives us a systematic classification about various transport processes. It becomes easier to compute transport coefficients to a given order of residual interactions.

The consequence of strong polarization was first discussed by Marcus for electron transfer in polar solvents\cite{mar} and by Holstein for small polarons in molecular crystals\cite{hol59}.
Emin and others\cite{Emin74} have used small polaron theory to compute transport properties for well-localized carriers. However two related features in this model do not agree with the nature of amorphous semiconductors: (i) electronic localization in an amorphous semiconductor is caused by the static positional disorder rather than the strong e-ph interaction; (ii) the localized states caused by the geometric disorder involve several atoms rather than one atom\cite{fed}. 

Miller and Abrahams studied the carrier hopping in a doped crystalline semiconductor. They assumed that the carriers are trapped in impurity or defect states close to the Fermi level, and the e-ph interaction is weak.Then the transitions between donor sites is mainly affected by the single-phonon absorption or emission\cite{ma}.

Traditionally transport theory takes the zero order Hamiltonian $H_{0}$ as a sum of the vibrational part $H_{v}$ and the electronic part $H_{e}$ in the force field of fixed nuclei, and the e-ph interaction $H_{e-ph}$ is
viewed as a small perturbation:%
\begin{equation}
H_{0}=H_{e}+H_{vib},\text{ \ \ }H=H_{0}+H_{e-ph}.  \label{parti}
\end{equation}%
We will use $h_{e}$ and $h_{e-ph}$ to denote the corresponding single-particle quantities.
If the e-ph interaction is so weak that the static displacements of atoms
induced by the e-ph interaction is negligible compared to $a_{v}$,
partition (\ref{parti}) is reasonable. 
Both localized states and extended states are eigenstates of $h_{e}$, 
$H_{e-ph}$ is the unique residual interaction and induces\cite{mas}
all four possible transitions among the eigenstates of $h_{e}$: (i) from a
localized state to another localized state (LL), (ii) from a localized state
to an extended state (LE), (iii) from an extended state to a localized state
(EL) and (iv) from an extended state to another extended state (EE).

Both experiments\ and \textit{ab initio} simulations show that in AS the
e-ph interaction for the well-localized tail states 
is strong ($d_{p_{A}}\gtrsim a_{v}$), while the e-ph interaction
for the less localized states and for the extended states 
is weak \cite{dra,tafn}. One has to adopt different
Hamiltonian partitions for the two situations.
For a well-localized state $A$ (we use $A$ with or without subscript to
label a localized state), we take $h_{0e}^{A}=K+V_{A}$ as the
single-particle Hamiltonian, where $V_{A}$ is the effective potential energy
of the ions inside the region $D_{A}$ where $\phi _{A}^{0}$ (eigenfunction
of $h_{e}$) is nonzero. One can easily show that $\phi _{A}\thickapprox \phi
_{A}^{0}$ even under perturbation $h_{e-ph}$, where $\phi _{A}$ is the
ground state of $h_{0e}^{A}$ \cite{epjb}. The zero order Hamiltonian is $%
h_{0}=h_{0e}^{A}+(H_{v}+h_{e-ph})$. The attraction from other nuclei outside
$D_{A}$ is taken as a perturbation. In this partition, LL\ and LE transitions
are caused by the transfer integrals\cite{epjb}, cf. Sec.6 of Ref.\cite{sup}. For an extended state $\xi
_{B}$ (we use $B$ with or without subscript to label an extended state), the
zero order Hamiltonian was taken as $h_{B}=h_{e}+H_{v}$, $h_{e-ph}$ is the
perturbation. EL and EE transitions are caused by the e-ph interactions\cite{epjb}.

In this paper, we apply the MRM to compute the conductivity and Hall mobility in amorphous semiconductors.
In Sec.\ref{cur}, the
measured macroscopic response is obtained by taking a spatial and ensemble average
over the microscopic current density.
The required input is the many-body wave function $\Psi^{\prime}(t)$ of N electron + $\mathcal{N}$ nuclei in an external field.
By means of the single-electron approximation and the harmonic approximation for vibration, in Sec.\ref{sing},
one can use a simplified many-body wave function $\psi^{\prime}(t)$ to express the spatial averaged current density,
$\psi^{\prime}(t)$ describes the motion of single electron in coupling with the nuclear vibrations.
Expanding $\psi^{\prime}(t)$ with localized and extended states, the evolution equations of the transition amplitudes can be derived from the time-dependent 
Schrodinger equation. It is convenient to compute the transition amplitudes in normal coordinate representation. In Sec.\ref{evo}, we outline how to obtain $\psi^{\prime}(t)$ to required order in residual interactions and in external field.
In Sec.\ref{duc}, we compute the conductivity from both localized and
extended states.
With the help of a systematic diagrammatic perturbation expansion, one can determine conductivity to any order in external field and
small parameters of residual interactions. We point out why some important contributions have been missed in previous calculations of small polarons based on the Kubo formula.
The non-diagonal conductivity needed for Hall mobility is calculated in Sec.\ref{hall}. To describe the Hall effect, one has to expand the current density to the 2$^{nd}$ order in the external field: one order is electric field, another is magnetic field. The MRM shows that: (i) there is a new type of term in the current density; (2) there is an intrinsic interference effect between electric field and magnetic field.
A new type temperature dependence is predicted.
In the MRM, the atomic vibrations are described by quantum mechanics, the results obtained are correct for any temperature.

\section{Macroscopic current density}\label{cur}
Consider an amorphous semiconductor, with $N$ electrons and $\mathcal{N}$
nuclei in an electromagnetic field described by potentials ($\mathbf{A},\phi
$). Denote the coordinates of the $N$ electrons as $\mathbf{r}_{1},\mathbf{r}%
_{2},\cdots ,\mathbf{r}_{N}$, the coordinates of $\mathcal{N}$ nuclei as
$\mathbf{W}_{1},\mathbf{W}_{2},\cdots ,\mathbf{W}_{\mathcal{N}}$. The state $%
\Psi ^{\prime }(t)$ of the system is determined by the Schr\"{o}%
dinger equation%
\begin{equation}
i\hbar \partial \Psi ^{\prime }/\partial t=H^{\prime }\Psi ^{\prime },
\label{ts}
\end{equation}%
where $H^{\prime }=H+H_{fm}$ is the total Hamiltonian, $H$ is the
Hamiltonian of system, $H_{fm}$ is the field-matter interaction. The
arguments of $\Psi ^{\prime }$ are $(\mathbf{r}_{1},\cdots ,%
\mathbf{r}_{N};\mathbf{W}_{1},\cdots ,\mathbf{W}_{%
\mathcal{N}};t)$. In this work, we focus on the transport coefficients far below the frequency of infrared radiation. The direct contributions from nuclei will not be written out.
For an ac field with higher frequency, especially in the infrared
range, one has to take into account the direct contribution of the motion of atoms.

The microscopic current density $\mathbf{j}^{\Psi^{\prime}}$ in state $\Psi^{\prime}(t)$ is\cite{short,pre}
\begin{equation}
\mathbf{j}^{\Psi^{\prime}}(\mathbf{r};t)=-\frac{e^{2}}{m}\mathbf{A}(\mathbf{r};t)n^{\prime}(\mathbf{r};t) \label{avcm}%
\end{equation}%
\[
+\frac{i\hbar eN}{2m}\int d\tau^{\prime}(\Psi^{\prime}\nabla_{\mathbf{r}}%
\Psi^{\prime\ast}-\Psi^{\prime\ast}\nabla_{\mathbf{r}}\Psi^{\prime}),
\]
where $n^{\prime}(\mathbf{r};t)=N\int
d\tau^{\prime}\Psi^{\prime\ast}\Psi^{\prime}$ is the carrier density in state $\Psi^{\prime}(t)$, $d\tau^{\prime}%
=d\mathbf{r}_{2}\cdots d\mathbf{r}_{N}d\mathbf{W}_{1}\cdots d\mathbf{W}_{\mathcal{N}}$,
the arguments of $\Psi^{\prime}$ in Eq.(\ref{avcm}) are $(\mathbf{r}%
,\mathbf{r}_{2},\cdots,\mathbf{r}_{N};\mathbf{W}_{1}\cdots \mathbf{W}_{\mathcal{N}};t)$. The first term of Eq.(\ref{avcm}) expresses the oscillation of `free electrons'. Comparing to the second term, its contribution to transport coefficients is negligible\cite{cb}. We will not discuss it further.

By averaging over a `physical infinitesimal' volume element\cite{ku57,kubo} $\Omega _{%
\mathbf{s}}$ around point $\mathbf{s}$, the coarse-grained current density $%
\widetilde{\mathbf{j}}^{\Psi ^{\prime }}$\ is :\
\begin{equation}
\widetilde{j}_{\alpha }^{\Psi ^{\prime }}(\mathbf{s},t)=\Omega _{\mathbf{s}%
}^{-1}\int_{\mathbf{r}\in \Omega _{\mathbf{s}}}d\mathbf{r}j_{\alpha }^{\Psi
^{\prime }}(\mathbf{r},t),~\alpha=x,y,z.  \label{culi}
\end{equation}%
For a mechanical perturbation, state $\Psi ^{\prime }(t)$ is determined by
the initial condition $\Psi ^{\prime }(-\infty )$. Because one does not know
what state the system was initially in, one must average $\widetilde{j}%
_{\alpha }^{\Psi ^{\prime }}$ over all possible $\Psi ^{\prime }(-\infty )$
to compute the measured macroscopic current density $j_{\alpha }$:%
\begin{equation}
j_{\alpha }(\mathbf{s},t)=\sum_{\Psi ^{\prime }}P[\Psi ^{\prime }(-\infty )]%
\widetilde{j}_{\alpha }^{\Psi ^{\prime }}(\mathbf{s},t),  \label{ense}
\end{equation}%
where $P[\Psi ^{\prime }(-\infty )]$ is the probability that the system is
in state $\Psi ^{\prime }(-\infty )$ before the external field is
adiabatically introduced. $P[\Psi ^{\prime }(-\infty )]$ depends only on the
energy of state $\Psi ^{\prime }(-\infty )$, and may be taken to be the
canonical distribution\cite{short,pre}.

\section{Harmonic approximation and single-electron approximation}\label{sing}
In the solid state,
$\mathbf{W}_{\mathbf{n}}=\mathcal{R}_{\mathbf{n}}+\mathbf{u}_{\mathbf{n}}$,
where $\mathcal{R}_{\mathbf{n}}=(X_{3(\mathbf{n}%
-1)+1},X_{3(\mathbf{n}-1)+2},X_{3\mathbf{n}})$ and \\$\mathbf{u}_{\mathbf{n}%
}=(x_{3(\mathbf{n}-1)+1},x_{3(\mathbf{n}-1)+2},x_{3\mathbf{n}})$ are the
equilibrium position vector and the vibrational displacement vector of the $%
\mathbf{n}^{th}$ nucleus respectively. In the harmonic approximation, the vibrational Hamiltonian is
\begin{equation}
H_{v}=\sum_{j}-\frac{\hbar^{2}}{2M_{j}}\frac{\partial^{2}}{\partial x_{j}^{2}%
}+\frac{1}{2}\sum_{jk}k_{jk}x_{j}x_{k},   \label{vb}
\end{equation}
where $(k_{ij})$ is the matrix of force constants.

Because in AS the correlation between electrons is weak, one can use the single-electron approximation to $\Psi^{\prime}(t)$. The arguments of the simplified single electron wave function $\psi^{\prime}(t)$ include only the single electronic coordinate and the vibrational coordinates of nuclei.
The state $\psi^{\prime}$
of a carrier in an external field satisfies:
\begin{equation}
i\hbar\partial\psi^{\prime}/\partial t=h^{\prime}\psi^{\prime}(\mathbf{r}%
,\{x_{j}\},t),   \label{ss}
\end{equation}
where $\mathbf{r}$ is the coordinate of the carrier. $h^{\prime}=h+h_{fm}$
is the Hamiltonian of [system + external field]. In gauge $\nabla\cdot{\bf{A}}({\bf{r}},t)=0$,
\begin{equation}
h_{fm}=(i\hbar e/m)\mathbf{A}(\mathbf{r})\cdot\nabla_{\mathbf{r}}+e^{2}%
\mathbf{A}^{2}(\mathbf{r})/(2m)+e\phi(\mathbf{r}),   \label{fm}
\end{equation}
is the coupling between the carrier and the external field. Here $%
h=h_{e}+H_{v}$ is the Hamiltonian without field, where
\begin{equation}
h_{e}=\frac{-\hbar^{2}}{2m}\nabla_{\mathbf{r}}^{2}+\sum_{\mathbf{n}=1}^{%
\mathcal{N}}U(\mathbf{r}-\mathcal{R}_{\mathbf{n}}-\mathbf{u}_{\mathbf{n}}),
\label{ed}
\end{equation}
is the electronic Hamiltonian. $U(\mathbf{r}-\mathcal{R}_{\mathbf{n}%
}-\mathbf{u}_{\mathbf{n}})$ is the effective potential energy between the
electron at $\mathbf{r}$ and the $\mathbf{n}^{th}$ nucleus. Denote the $3%
\mathcal{N}$ vibrational degrees of freedom as $\{x_{j},$ $j=1,2,\cdots ,3%
\mathcal{N}\}$, then the spatially averaged current density (\ref{culi}) at $\mathbf{s}$
is reduced to\cite{short,kubo}%
\begin{equation}
\widetilde{\mathbf{j}}^{\psi^{\prime}}(\mathbf{s},t)=\frac{i\hbar eN_{e}}{%
2m\Omega_{\mathbf{s}}}\int_{\Omega_{\mathbf{s}}}d\mathbf{r}\int[\displaystyle\prod
\limits_{j=1}^{3\mathcal{N}}dx_{j}]   \label{cu}
\end{equation}%
\begin{equation*}
(\psi^{\prime}\nabla_{\mathbf{r}}\psi^{\prime\ast}-\psi^{\prime\ast}\nabla_{%
\mathbf{r}}\psi^{\prime}).
\end{equation*}

\section{Evolution equation in external field}\label{evo}
The prime ingredient required for the conductivity and Hall
mobility is the single particle state $\psi^{\prime}(t)$ of the system in an
external field\cite{short}. To find $\psi^{\prime}(t)$, we expand it using
the approximate eigenstates of the single-particle Hamiltonian\cite{epjb}:
\begin{equation}
\psi^{\prime}(t)=\sum_{A}a_{A}(t)\phi_{A}+\sum_{B}b_{B}(t)\xi_{B},
\label{ew}
\end{equation}
where the arguments of $\psi^{\prime}$ are ($\mathbf{r};x_{1},x_{2},%
\cdots,x_{3\mathcal{N}}$), $a_{A}(t)$ is the probability amplitude at time $t
$ that the carrier is in a localized state $\phi_{A}$ while the vibrational
displacements of $\mathcal{N}$ nuclei are $x_{1},x_{2},\cdots,x_{3\mathcal{N}%
}$. Here, $b_{B}(t)$ is the probability amplitude at time t that the carrier
is in an extended state $\xi_{B}$ while the vibrational displacements of $%
\mathcal{N}$ nuclei are $x_{1},x_{2},\cdots,x_{3\mathcal{N}}$. If the
external field is not too strong, it is convenient to put the change in $%
\psi^{\prime}$ caused by external field into the probability amplitudes
rather than in the zero-order eigenstates.
Substituting Eq.(\ref{ew}) into Eq.(\ref{ss}),
the time variation of probability amplitudes under the external field are determined by\cite{epjb}:%
\begin{equation}
\lbrack i\hbar\frac{\partial}{\partial t}-h_{A_{1}}]a_{A_{1}}=%
\sum_{A}a_{A}J_{A_{1}A}^{tot}+\sum_{B}b_{B}K_{A_{1}B}^{\prime tot},
\label{aev1}
\end{equation}
and%
\begin{equation}
\lbrack i\hbar\frac{\partial}{\partial t}-h_{B_{1}}]b_{B_{1}}=%
\sum_{A}J_{B_{1}A}^{\prime tot}a_{A}+\sum_{B}K_{B_{1}B}^{tot}b_{B},
\label{bev1}
\end{equation}
where%
\begin{equation}
J_{A_{1}A}^{tot}=J_{A_{1}A}+J_{A_{1}A}^{field},\text{ }K_{A_{1}B}^{\prime
tot}=K_{A_{1}B}^{\prime}+K_{A_{1}B}^{\prime field},   \label{JKP}
\end{equation}
and%
\begin{equation}
J_{B_{1}A}^{\prime tot}=J_{B_{1}A}^{\prime}+J_{B_{1}A}^{\prime field},\text{
}K_{B_{1}B}^{tot}=K_{B_{1}B}+K_{B_{1}B}^{field},   \label{JPK}
\end{equation}
and%
\begin{equation}
h_{A_{1}}=H_{v}+E_{A_{1}},\text{ }h_{B_{1}}=H_{v}+E_{B_{1}},   \label{hab}
\end{equation}
where transfer integral $J_{A_{1}A}=\int d\mathbf{r}\phi_{A_{1}}^{\ast}%
\sum_{p\notin D_{A}}U(r-\mathcal{R}_{p})\phi_{A}$ causes a transition from
localized state $\phi_{A}$ to localized state $\phi_{A_{1}}$, the transfer
integral $J_{B_{1}A}^{\prime}=\int d\mathbf{r}\xi_{B_{1}}^{\ast}\sum_{p%
\notin D_{A}}U(r-\mathcal{R}_{p})\phi_{A}$ induces a transition from
localized state $\phi_{A}$ to an extended state $\xi_{B_{1}}$. The e-ph
interaction $K_{A_{1}B}^{\prime}=\int d\mathbf{r}\phi_{A_{1}}^{\ast}%
\sum_{j}x_{j}(\partial U/\partial x_{j})\xi_{B}$ causes a transition from
extended state $\xi_{B}$ to localized state $\phi_{A_{1}}$. The e-ph
interaction $K_{B_{1}B}=\int d\mathbf{r}\xi_{B_{1}}^{\ast}\sum_{j}x_{j}(%
\partial U/\partial x_{j})\xi_{B}$ induces a transition from extended state $%
\xi_{B}$ to another extended state $\xi_{B_{1}}$. $J_{A_{1}A}^{field}=\int d%
\mathbf{r}\phi_{A_{1}}^{\ast}h_{fm}\phi_{A}$ is the coupling between two
localized states $\phi_{A}$ and $\phi_{A_{1}}$ caused by external field, etc.

The microscopic response is expressed by the state $\psi^{\prime}(t)$ of
[system + external field]. We replace the second line of Eq.(\ref{cu}) with $%
(\psi^{\prime}\nabla_{\mathbf{r}}\psi^{\prime\ast}-\psi^{\prime\ast}\nabla_{%
\mathbf{r}}\psi^{\prime})-(\psi\nabla_{\mathbf{r}}\psi^{\ast}-\psi^{\ast}%
\nabla_{\mathbf{r}}\psi)$, where $\psi(\mathbf{r},\{x_{j}\},t)$\ satisfies $%
i\hbar\partial\psi/\partial t=h\psi$, is the state of system without
external field. $(\psi\nabla_{\mathbf{r}}\psi^{\ast}-\psi ^{\ast}\nabla_{%
\mathbf{r}}\psi)$ represents the microscopic current density when no
external field is applied to the system. For carriers in localized states
and in extended states (scattered by static disorder), the contribution to
the coarse-grained current density $\widetilde{\mathbf{j}}^{\psi^{\prime}}$
from $(\psi\nabla_{\mathbf{r}}\psi^{\ast}-\psi^{\ast}\nabla_{\mathbf{r}}\psi)
$ is zero. The spatially averaged \textit{microscopic} current density to
second order in the field is\cite{short}%
\begin{equation}
\widetilde{\mathbf{j}}^{\psi^{\prime}}(\mathbf{s},t)=-\frac{N_{e}\hbar e}{%
m\Omega_{\mathbf{s}}}\int_{\mathbf{r}\in\Omega_{\mathbf{s}}}d\mathbf{r}\int[%
\displaystyle\prod \limits_{j=1}^{3\mathcal{N}}dx_{j}]   \label{1e1}
\end{equation}%
\begin{equation*}
\{\func{Im}(\psi^{(0)}\nabla_{\mathbf{r}}\psi^{(1)\ast}-\psi^{(1)\ast
}\nabla_{\mathbf{r}}\psi^{(0)})
\end{equation*}%
\begin{equation*}
+\func{Im}(\psi^{(0)}\nabla\psi^{(2)\ast}-\psi^{(2)\ast}\nabla
\psi^{(0)}+\psi^{(1)}\nabla\psi^{(1)\ast})\},
\end{equation*}
where $\psi^{(1)}$ is the change in state to first order in field, and $%
\psi^{(2)}$ is the change in state to second order in field. $\psi^{(0)}$ is
the state of carrier at time $t$ \textit{without} an external field.
Hereafter, the superscripts (0), (1) and (2) on $\psi$ indicate the order of
external field, the order of residual interactions will be denoted as
subscripts. For example $\psi_{J}^{(1)}$ represents the change in state to
first order of external field and to first order of $J$. If not explicitly
stated, we understand that $\psi^{(0)}$, $\psi^{(1)}$ and $\psi^{(2)}$ are
fully dressed by the residual interactions $J$, $J^{\prime}$, $K^{\prime}$
and $K$.

Unlike the nearly free carriers in crystalline materials or the
extended states of an amorphous semiconductor, the carriers in localized
states cannot be accelerated by an external electric field. Because the
force produced by the external field is much weaker than the binding force
from static disorder, the quantum tunneling probability\cite{bom} $\Gamma=\exp
\{-4(2m)^{1/2}\varepsilon_{b}^{3/2}/(3\hbar eE)\}$ produced by an external
field is negligible, where $\varepsilon_{b}$ is the binding energy produced
by static disorder. In addition, for a field of $10^{5}$ V$\cdot$cm$^{-1}$
and a typical distance 10\AA\ between two neighboring localized states, the
upper limit of the transition moment is $0.01$eV, the same order as $J$ and $%
J^{\prime}$. Denote $S$ as one of the small parameters $J,$ $J^{\prime},$ $%
K^{\prime},$ $K$ from the residual interactions. One must calculate $%
\psi^{(0)}(t),$ $\psi^{(1)}(t)$ and $\psi^{(2)}(t)$ to order $S^{0}$ \textit{%
and} $S^{1}$ and keep all$\mathbf{\ }$possible order $S^{0}$ \textit{and} $%
S^{1}$ contributions in current density $\mathbf{j}$.

To find $\psi^{\prime}(t)$, it is convenient to transform the
vibrational displacements $\{x\}$ of the atoms into normal coordinates $%
\{\Theta\}$\cite{epjb}. We expand probability
amplitude $a_{A_{1}}(\cdots\Theta_{\alpha}\cdots;t)$ with the eigenfunctions
of $h_{A_{1}}$:%
\begin{equation}
a_{A_{1}}=\sum_{\cdots N_{\alpha}^{\prime}\cdots}C^{A_{1}}(\cdots N_{\alpha
}^{\prime}\cdots;t)\Psi_{A_{1}}^{\{N_{\alpha}^{\prime}\}}e^{-it\mathcal{E}%
_{A_{1}}^{\{N_{\alpha}^{\prime}\}}/\hbar},   \label{az}
\end{equation}
where $C^{A_{1}}(\cdots N_{\alpha}^{\prime}\cdots;t)$ is the probability
amplitude at moment $t$ that the electron is in localized state $\phi_{A_{1}}
$ while the vibrational state of the nuclei is characterized by occupation
number $\{N_{\alpha}^{\prime},\alpha=1,2,\cdots,3\mathcal{N}\}$ in each
mode. Similarly, we expand $b_{B_{1}}(\cdots\Theta_{\alpha}\cdots;t)$ with
the eigenfunctions of $h_{B_{1}}$:%
\begin{equation}
b_{B_{1}}=\sum_{\cdots N_{\alpha}^{\prime}\cdots}F^{B_{1}}(\cdots N_{\alpha
}^{\prime}\cdots;t)\Xi_{B_{1}}^{\{N_{\alpha}^{\prime}\}}e^{-it\mathcal{E}%
_{B_{1}}^{\{N_{\alpha}^{\prime}\}}/\hbar}.   \label{bz}
\end{equation}
Eqs.(\ref{aev1},\ref{bev1}) become the evolution equations for $C^{A}$ and $F^{B}$. Using perturbation theory, one can
find the probability amplitudes $C^{A}$ and $F^{B}$ to any order of field
for any initial conditions. Substitute $C^{A}$ and $F^{B}$ into Eqs.(\ref{az}%
,\ref{bz}), and one obtains the probability amplitudes $a_{A}$ and $b_{B}$
in the normal coordinate representation. Substituting $a_{A}$ and $b_{B}$
back to Eq.(\ref{ew}), one determines the state $\psi^{\prime}(t)$ of system
in an external field. The results are listed in Sec.1 of Ref.\cite{sup}.
\begin{table}[th]
\caption{The origin of various conduction processes}%
\resizebox{8cm}{!} {
\begin{tabular}
[c]{lll}\hline\hline
order & expression & diagrams\\
\hline
S$^{0}$ & $(\psi_{S^{0}}^{(0)}\nabla_{\mathbf{r}}\psi_{S^{0}}^{(1)\ast}-\psi_{S^{0}}^{(1)\ast}\nabla_{\mathbf{r}}\psi_{S^{0}}^{(0)})$ &
Fig.\ref{A1A2}, \ref{B1B2}\\
\hline
S & $(\psi_{S}^{(0)}\nabla_{\mathbf{r}}\psi_{S^{0}}^{(1)\ast}-\psi_{S^{0}}^{(1)\ast}\nabla_{\mathbf{r}}\psi_{S}^{(0)})$ & Fig.\ref{A3A4A5A6},
\ref{B3B4B5B6}\\
\hline
S & $(\psi_{S^{0}}^{(0)}\nabla_{\mathbf{r}}\psi_{S}^{(1)\ast}-\psi
_{S}^{(1)\ast}\nabla_{\mathbf{r}}\psi_{S^{0}}^{(0)})$ & Fig.\ref{QAA},\ref{QAB},\ref{SBA},\ref{SBB}\\\hline
K$^{2}$ & $(\psi_{K^{2}}^{(0)}\nabla_{\mathbf{r}}\psi_{S^{0}}^{(1)\ast}-\psi_{S^{0}}^{(1)\ast}\nabla_{\mathbf{r}}\psi_{K^{2}}^{(0)})$ & Fig.\ref{B7}\\\hline
K$^{2}$ & $(\psi_{S^{0}}^{(0)}\nabla_{\mathbf{r}}\psi_{K^{2}}^{(1)\ast}-\psi_{K^{2}}^{(1)\ast}\nabla_{\mathbf{r}}\psi_{S^{0}}^{(0)})$ &
Fig.\ref{fKKT},\ref{KfKT},\ref{KKfT}\\\hline\hline
\end{tabular}\label{clac}
}
\end{table}
\section{Conductivity}\label{duc}
\begin{table*}[th]
\caption{States coupled by the current operator }%
\resizebox{15cm}{!} {
\begin{tabular}
[c]{lllllllll}\hline\hline
symbol & \ \ \   expression \\\hline
A$\rightsquigarrow$ A$_{2}$ &\ $\Omega
_{\mathbf{s}}^{-1}\int_{\Omega_{\mathbf{s}}}d\mathbf{r(}\phi_{A}\nabla_{\mathbf{r}}\phi_{A_{2}}^{\ast}-\phi_{A_{2}}^{\ast}\nabla_{\mathbf{r}}\phi_{A})\langle\Psi_{A_{2}}^{\{N_{\alpha}^{\prime\prime}\}}|\Psi
_{A}^{\{N_{\alpha}\}}\rangle e^{it(\mathcal{E}_{A_{2}}^{\{N_{\alpha}^{\prime\prime}\}}-\mathcal{E}_{A}^{\{N_{\alpha}\}})/\hbar}$ \\\hline
A$\rightsquigarrow$ B$_{1}$ & \ $ \Omega
_{\mathbf{s}}^{-1}\int_{\Omega_{\mathbf{s}}}d\mathbf{r(}\phi_{A}\nabla_{\mathbf{r}}\xi_{B_{1}}^{\ast}-\xi_{B_{1}}^{\ast}\nabla_{\mathbf{r}}\phi_{A})\langle\Xi_{B_{1}}^{\{N_{\alpha}^{\prime}\}}|\Psi_{A}^{\{N_{\alpha
}\}}\rangle e^{it(\mathcal{E}_{B_{1}}^{\{N_{\alpha}^{\prime}\}}-\mathcal{E}_{A}^{\{N_{\alpha}\}})/\hbar}$  \\\hline
B$\rightsquigarrow$ A$_{2}$ & \ $ \Omega
_{\mathbf{s}}^{-1}\int_{\Omega_{\mathbf{s}}}(\xi_{B}\nabla_{\mathbf{r}}\phi_{A_{2}}^{\ast}-\phi_{A_{2}}^{\ast}\nabla_{\mathbf{r}}\xi_{B})\langle
\Psi_{A_{2}}^{\{N_{\alpha}^{\prime\prime}\}}|\Xi_{B}^{\{N_{\alpha}\}}\rangle e^{it(\mathcal{E}_{A_{2}}^{\{N_{\alpha}^{\prime\prime}\}}-\mathcal{E}_{B}^{\{N_{\alpha}\}})/\hbar}$ \\\hline
B$\rightsquigarrow$ B$_{1}$ & \ $  \Omega
_{\mathbf{s}}^{-1}\int_{\Omega_{\mathbf{s}}}d\mathbf{r}(\xi_{B}\nabla
_{\mathbf{r}}\xi_{B_{1}}^{\ast}-\xi_{B_{1}}^{\ast}\nabla_{\mathbf{r}}\xi
_{B})\langle\Xi_{B_{1}}^{\{N_{\alpha}^{\prime}\}}|\Xi_{B}^{\{N_{\alpha}\}}\rangle e^{it(\mathcal{E}_{B_{1}}^{\{N_{\alpha}^{\prime}\}}-\mathcal{E}_{B}^{\{N_{\alpha}\}})/\hbar}$    \\\hline
A$_{2}$ $\dashrightarrow$ A$_{3}$ &\
$ \Omega_{s}^{-1}\int_{\mathbf{r}\in
\Omega_{s}}d\mathbf{r}\phi_{A_{2}}\nabla\phi_{A_{3}}^{\ast}\langle\Psi_{A_{3}}^{\{N_{\alpha}\}}|\Psi_{A_{2}}^{\{N_{\alpha}^{\prime\prime\prime}\}}\rangle
e^{it(\mathcal{E}_{A_{3}}^{\prime\{N_{\alpha}\}}-\mathcal{E}_{A_{2}}^{\prime\{N_{\alpha}^{\prime\prime\prime}\}})/\hbar}$   \\\hline\hline
\end{tabular}\label{current}
}
\end{table*}
For the conductivity, one
needs only the first term in Eq.(\ref{1e1}). To order $S^{0}$ and $S^{1}$,
all possible contributions to $\mathbf{j}$ are classified in Table \ref{clac}.
If the initial state is a localized state $\phi_{A}$, the expression for $\mathbf{j}$ can be found by substituting the corresponding $\psi^{(0)}(t)$ and $\psi^{(1)}(t)$ into the first term in Eq.(\ref{1e1}). There are 14 processes
contributing to conductivity. This can be understood as following. There are
three terms in $\psi^{(0)}$: (1) the free\ evolution of state $\phi_{A}$
without external field and residual interactions; (2) $J$ takes $\phi_{A}$
to another localized state; and (3) $J^{\prime}$ takes localized state $%
\phi_{A}$ to an extended state. As indicated in Eqs.(\ref{aev1},\ref{bev1}), e-ph interactions $K$ and $K^{\prime}$ do not couple a localized state to any other state. There are two order S$^{0}$
terms in $\psi^{(1)}$: the external field can bring $\phi_{A}$\ to either
another localized state or an extended state. There are 8 order S$^{1}$
terms in $\psi^{(1)}$, they differ in the time ordering of the residual
interaction S and external field. The current density operator $%
(\psi^{(0)}\nabla _{\mathbf{r}}\psi^{(1)\ast}-\psi^{(1)\ast}\nabla_{\mathbf{r%
}}\psi^{(0)})$ links all components of $\psi^{(0)}$ to all components of $%
\psi^{(1)\ast}$.
\begin{figure}[htb]%
\subfloat[]{\includegraphics[width=.2\linewidth,height=2cm]{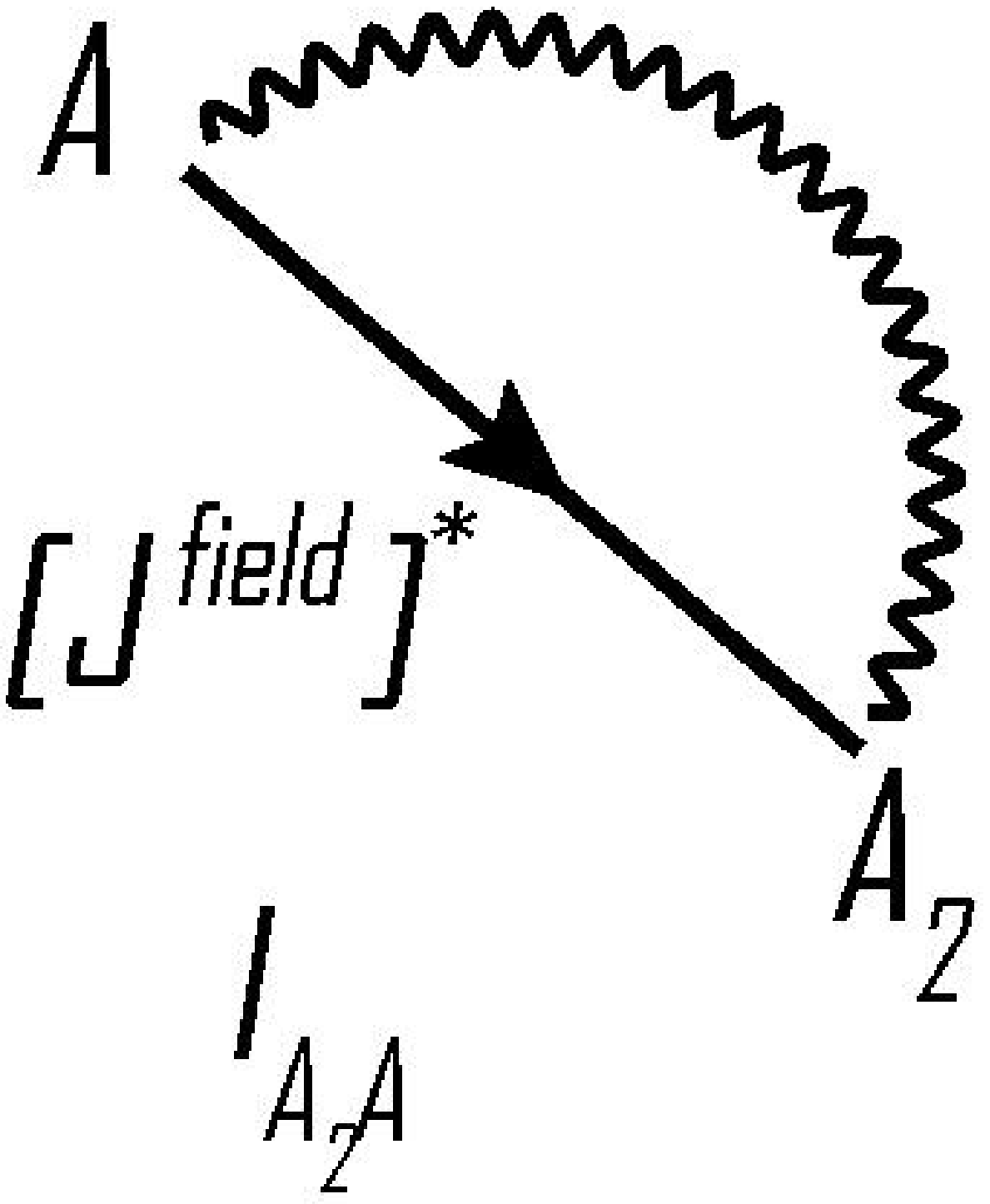}\label{A1}}\hfill
\subfloat[]{\includegraphics[width=.3\linewidth,height=2cm]{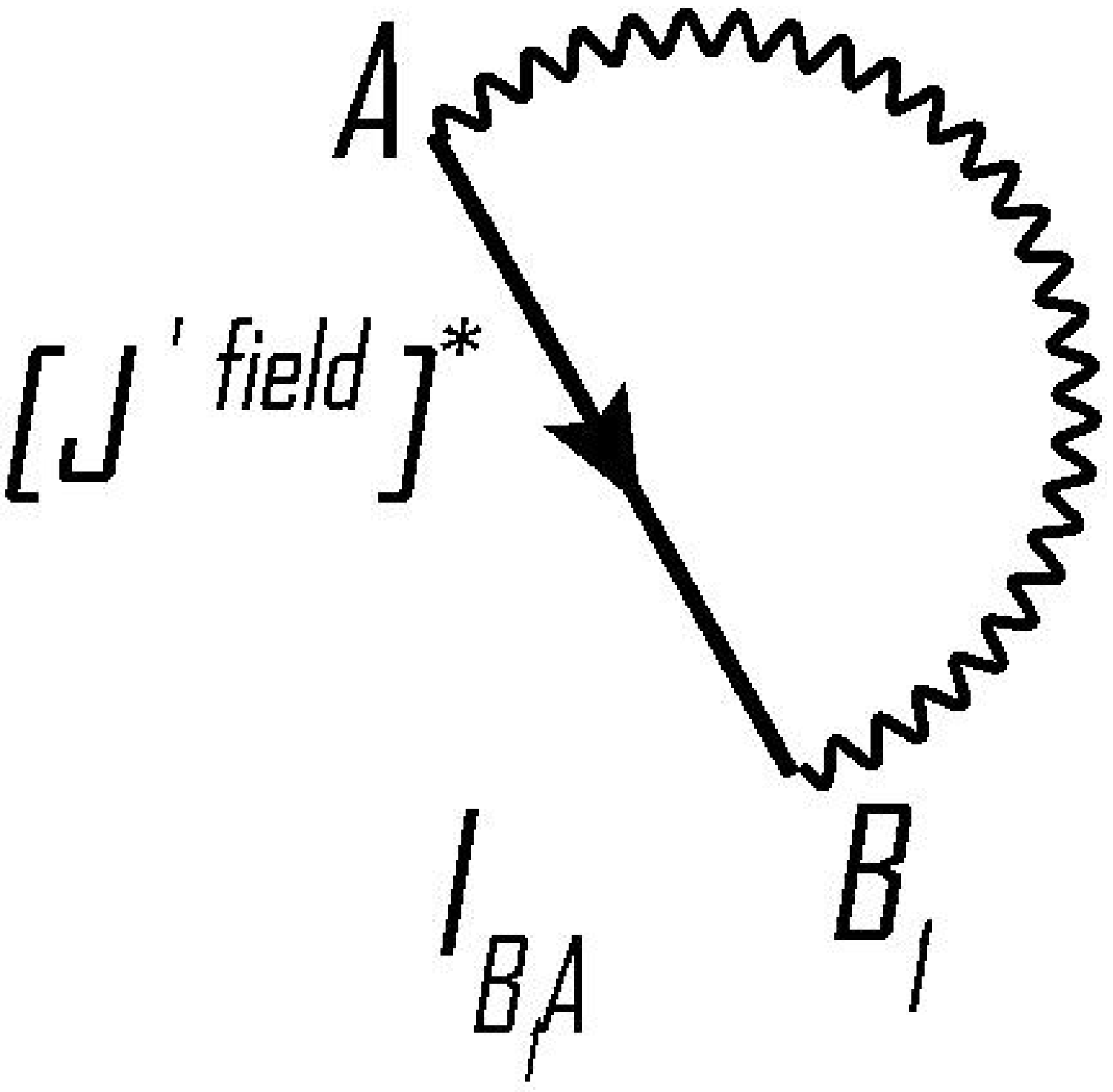}\label{A2}}\hfill
\caption{Lowest order processes contributing to conductivity. Initial state is localized state $\protect\phi_{A}$: order $J^{0}$ and $[J^{\prime}]^{0}$ contributions to conductivity  (see text).}
\label{A1A2}
\end{figure}
\begin{figure}[th]
\centering
\subfloat[] { \includegraphics[scale=0.12]{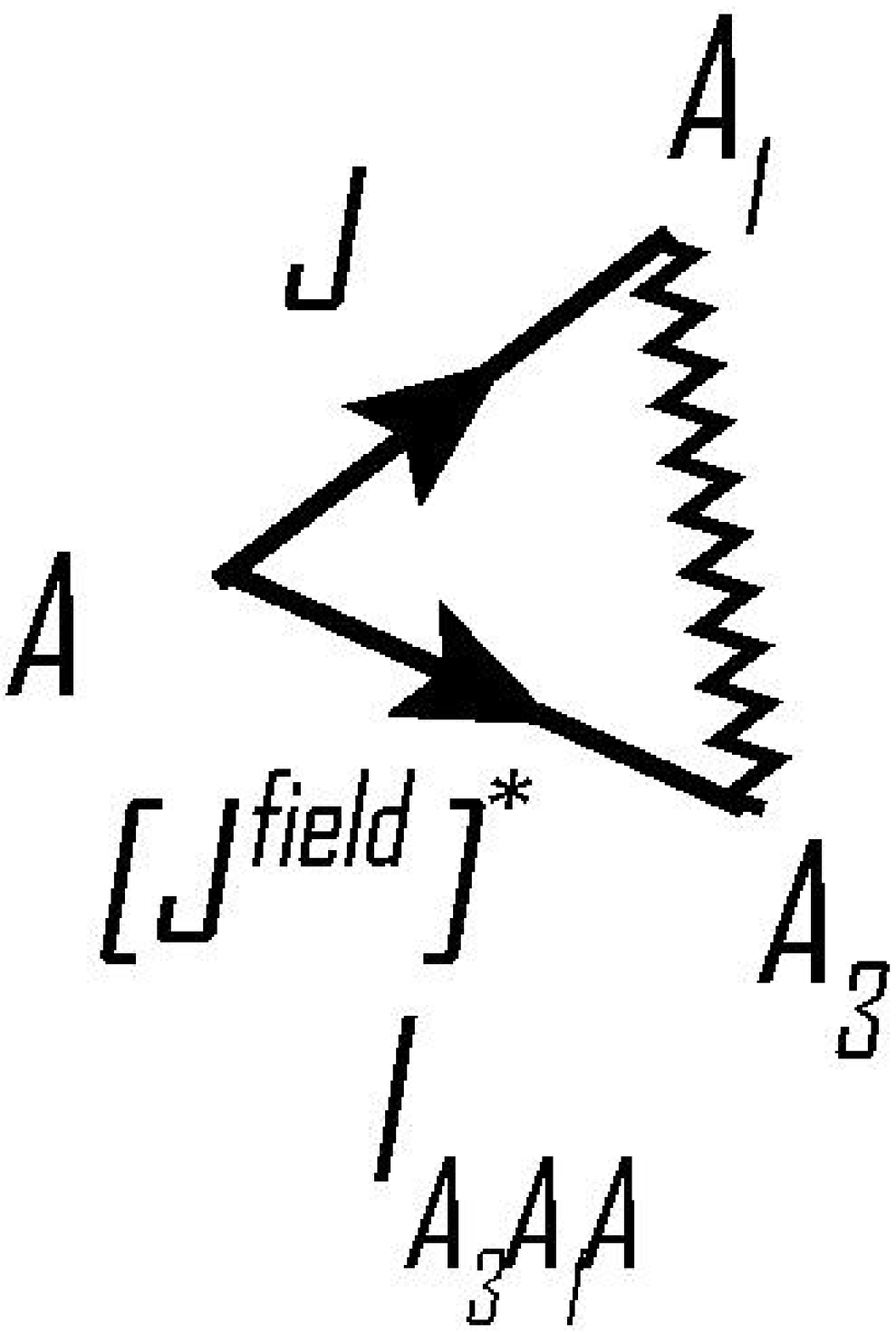} \label{A3} }\hfill
\subfloat[]{\includegraphics[scale=0.12]{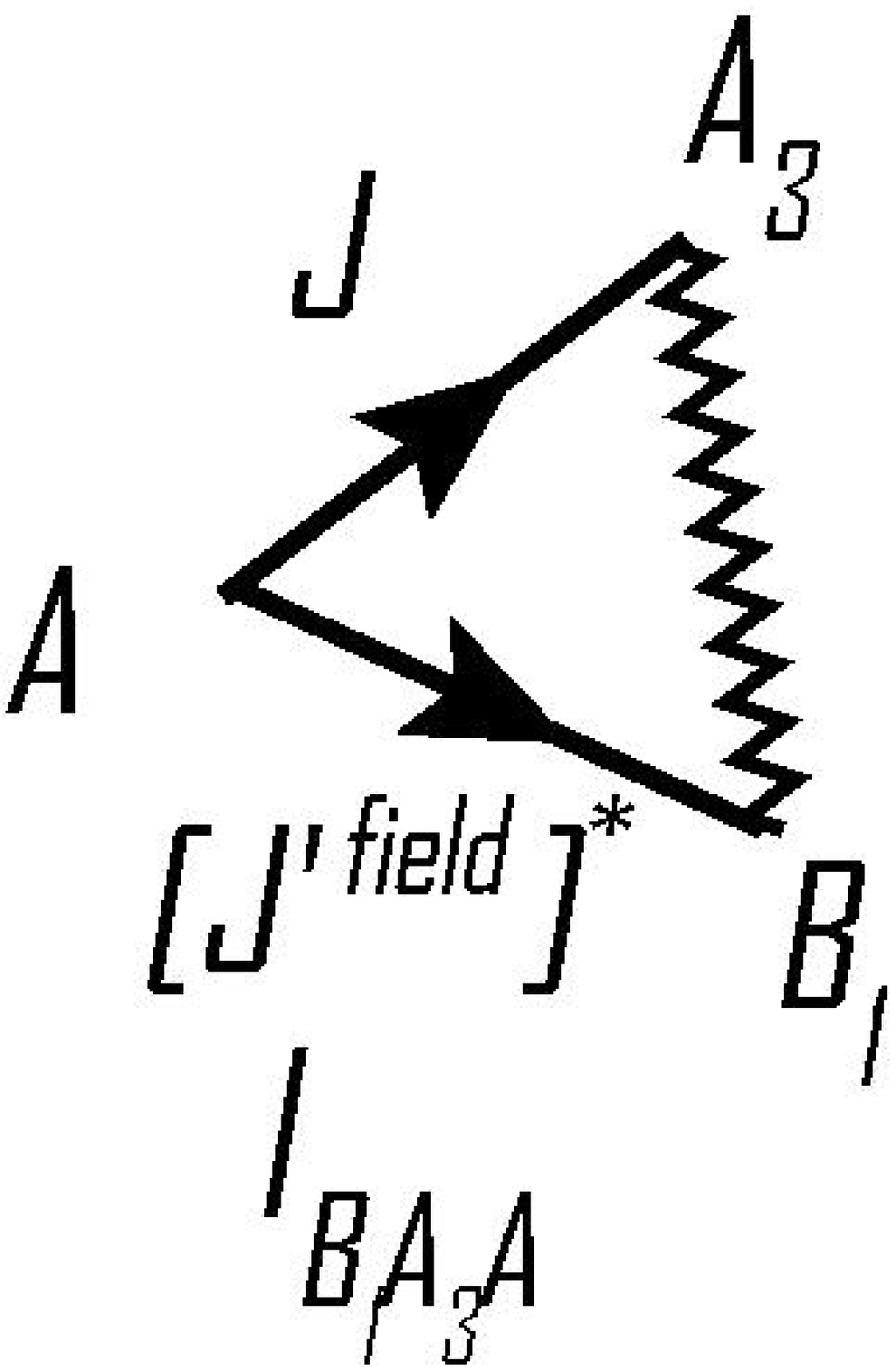}\label{A4}}\hfill
\subfloat[]{\includegraphics[scale=0.12]{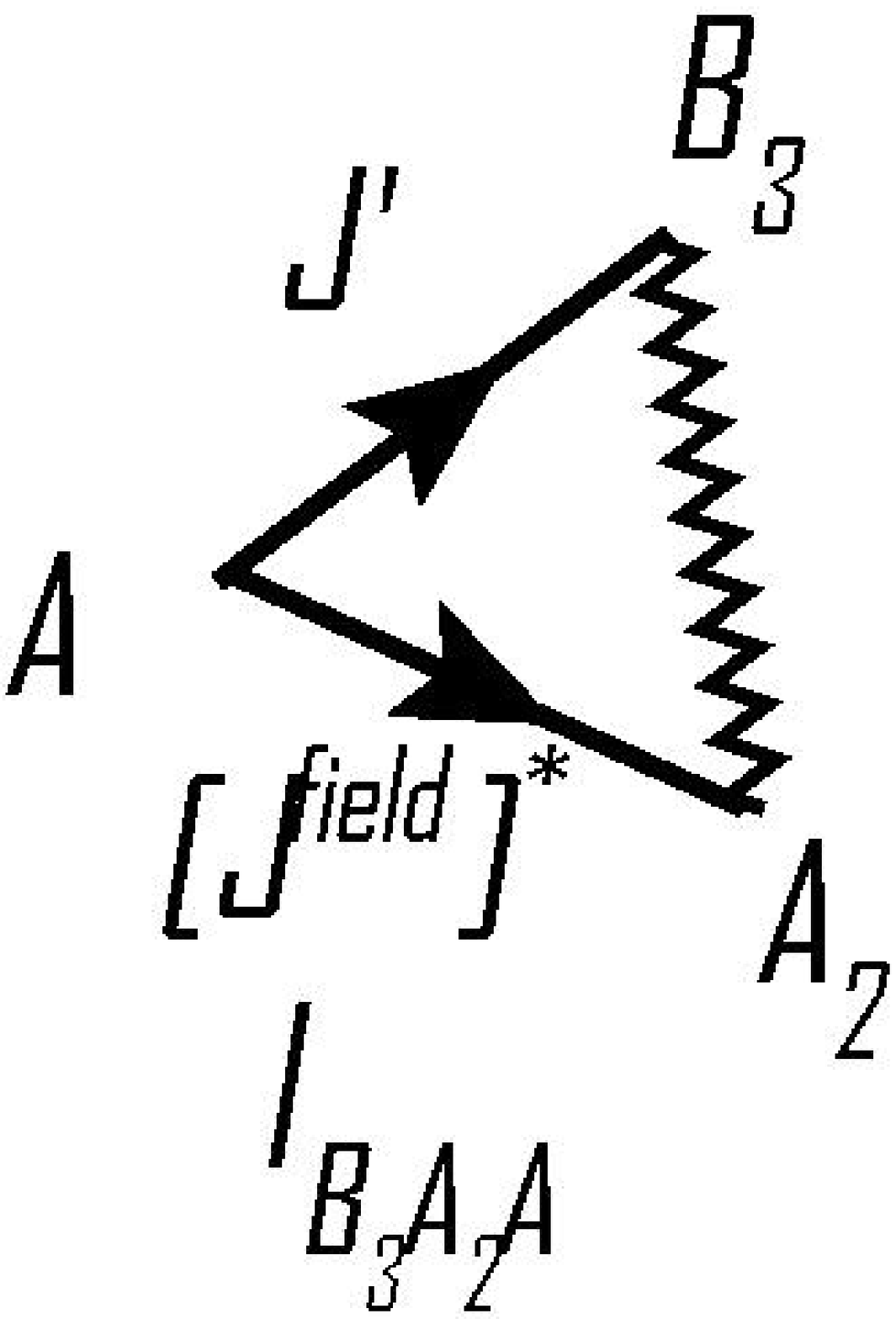}\label{A5}}\hfill
\subfloat[]{\includegraphics[scale=0.12]{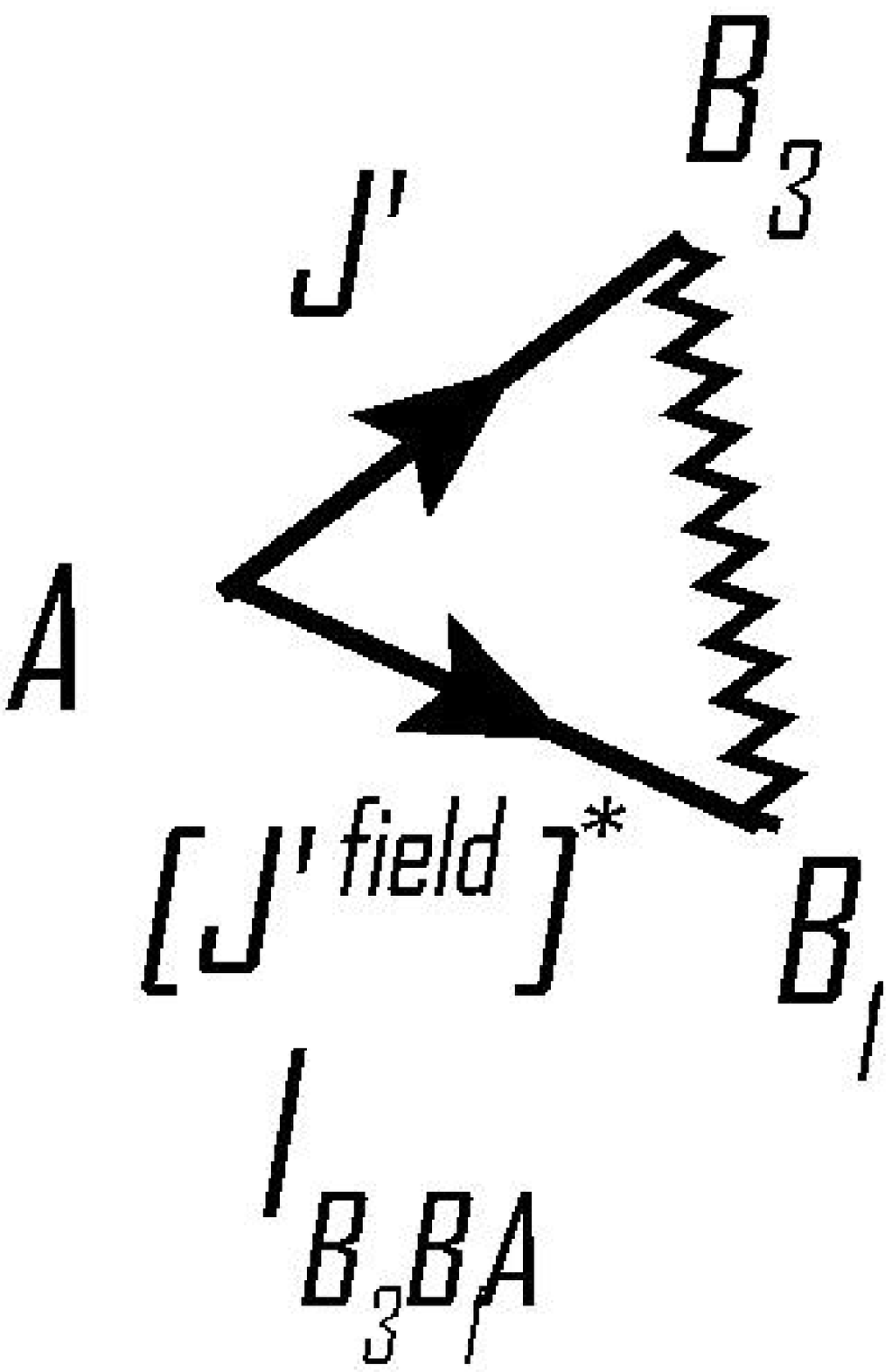}\label{A6}}\hfill
\caption{Initial state is localized state $\protect\phi_{A}$: order $J^{1}$ and order $[J^{\prime}]^{1}$ contributions to conductivity.}
\label{A3A4A5A6}
\end{figure}
\begin{figure}[th]
\centering
\subfloat[]{\includegraphics[scale=0.12]{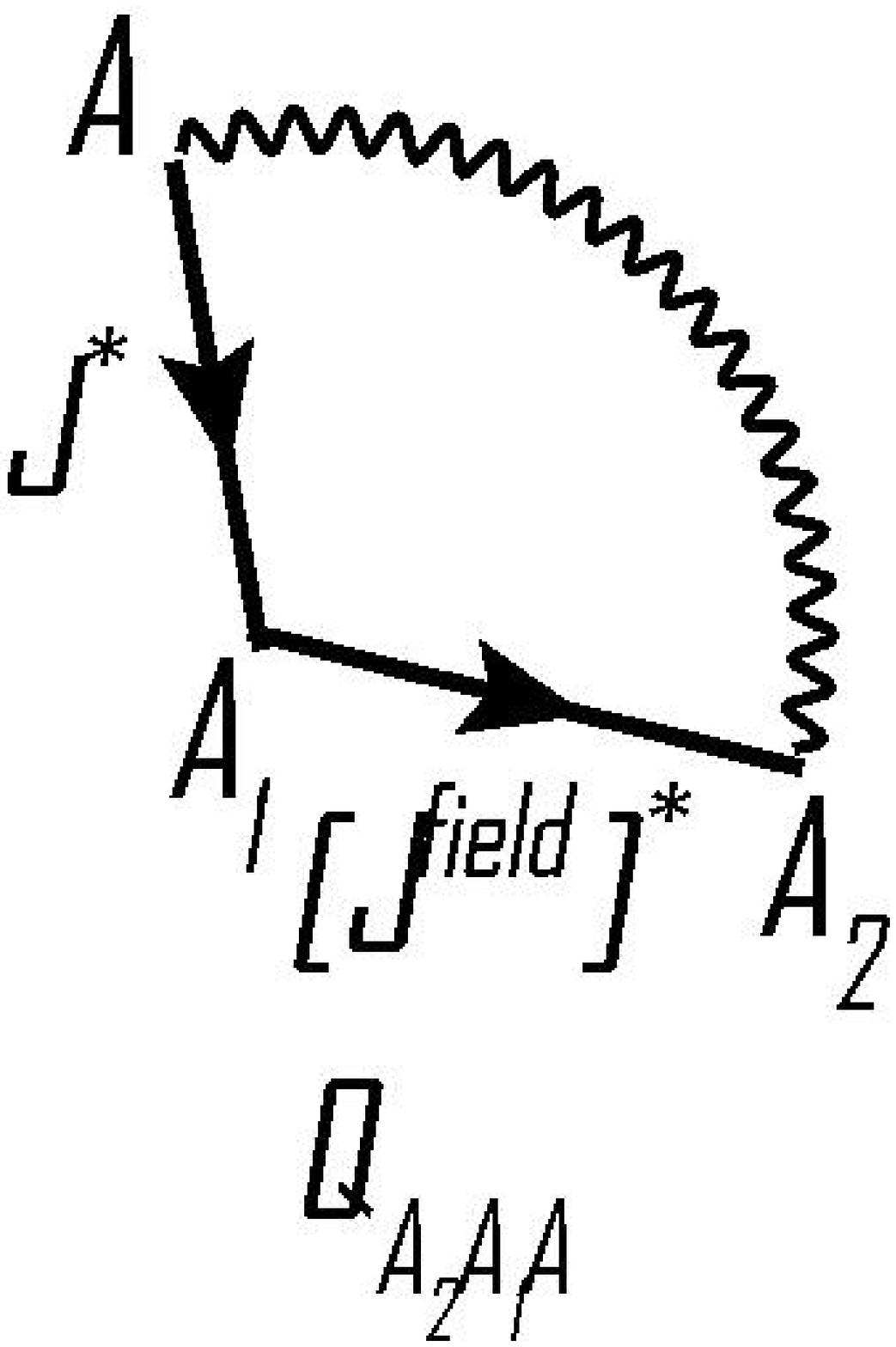}\label{QAAA1}}\hfill %
\subfloat[]{\includegraphics[scale=0.12]{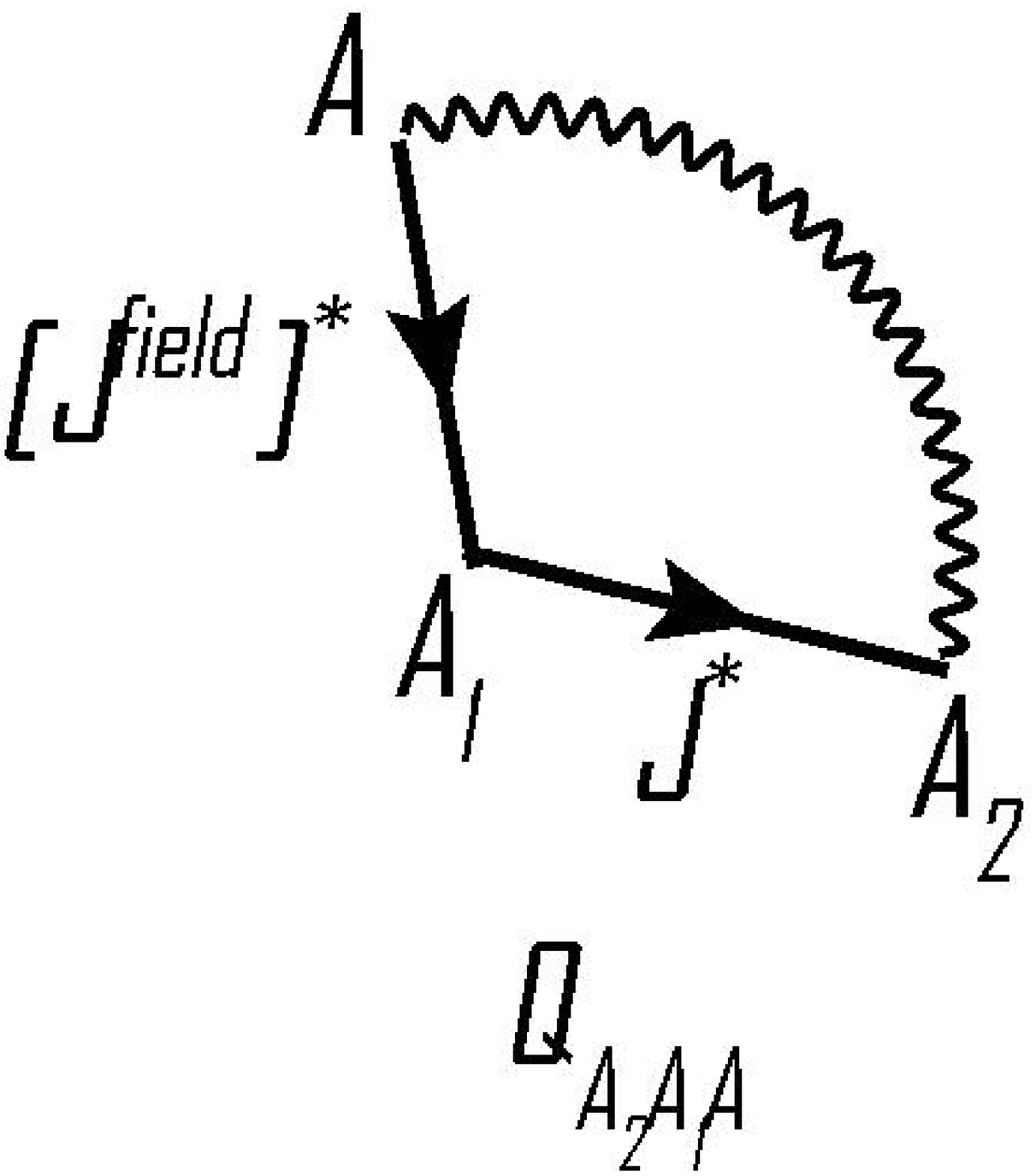}\label{QAAA2}}\hfill 
\subfloat[]{\includegraphics[scale=0.12]{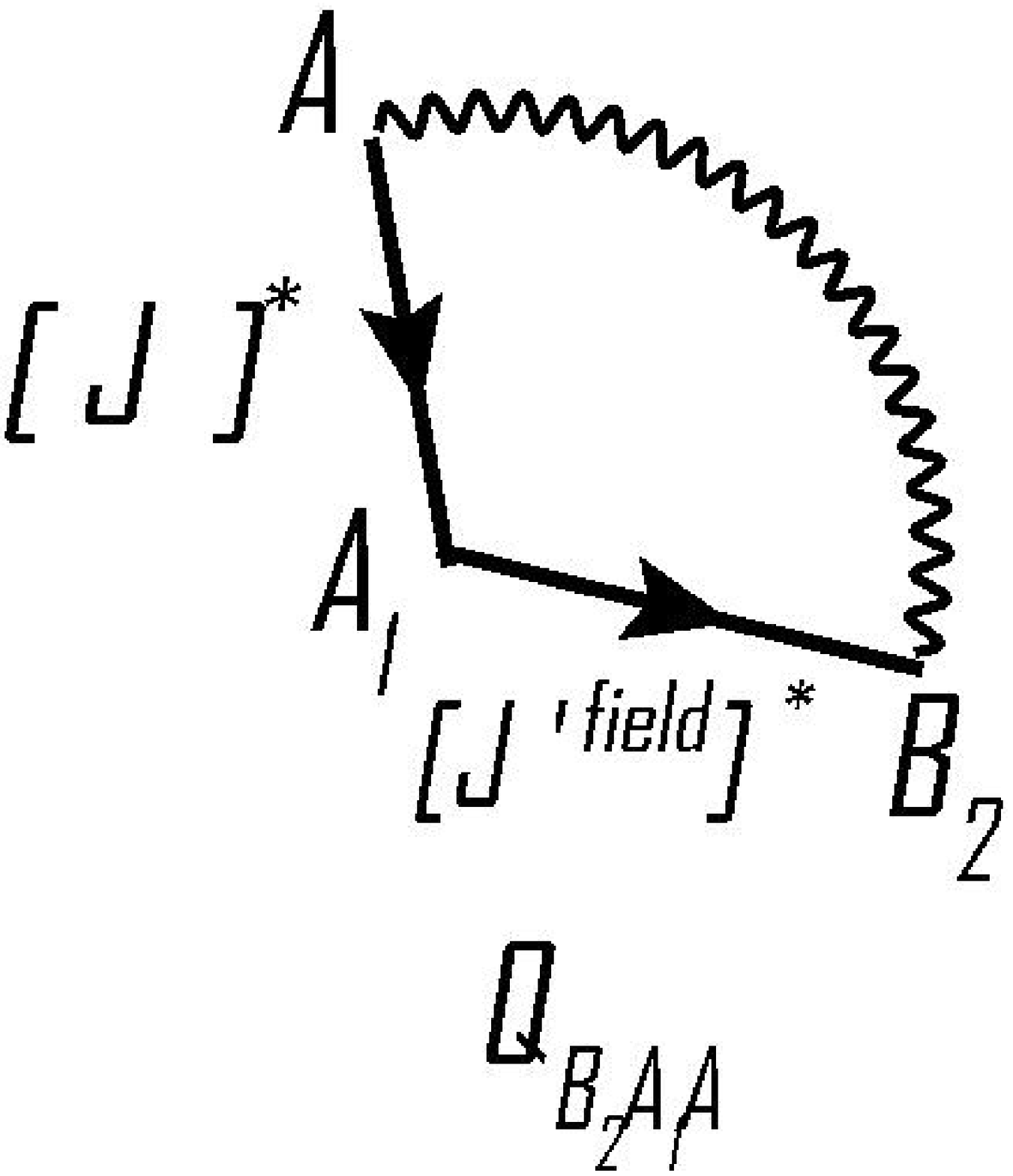}\label{QAAB1}}\hfill %
\subfloat[]{\includegraphics[scale=0.12]{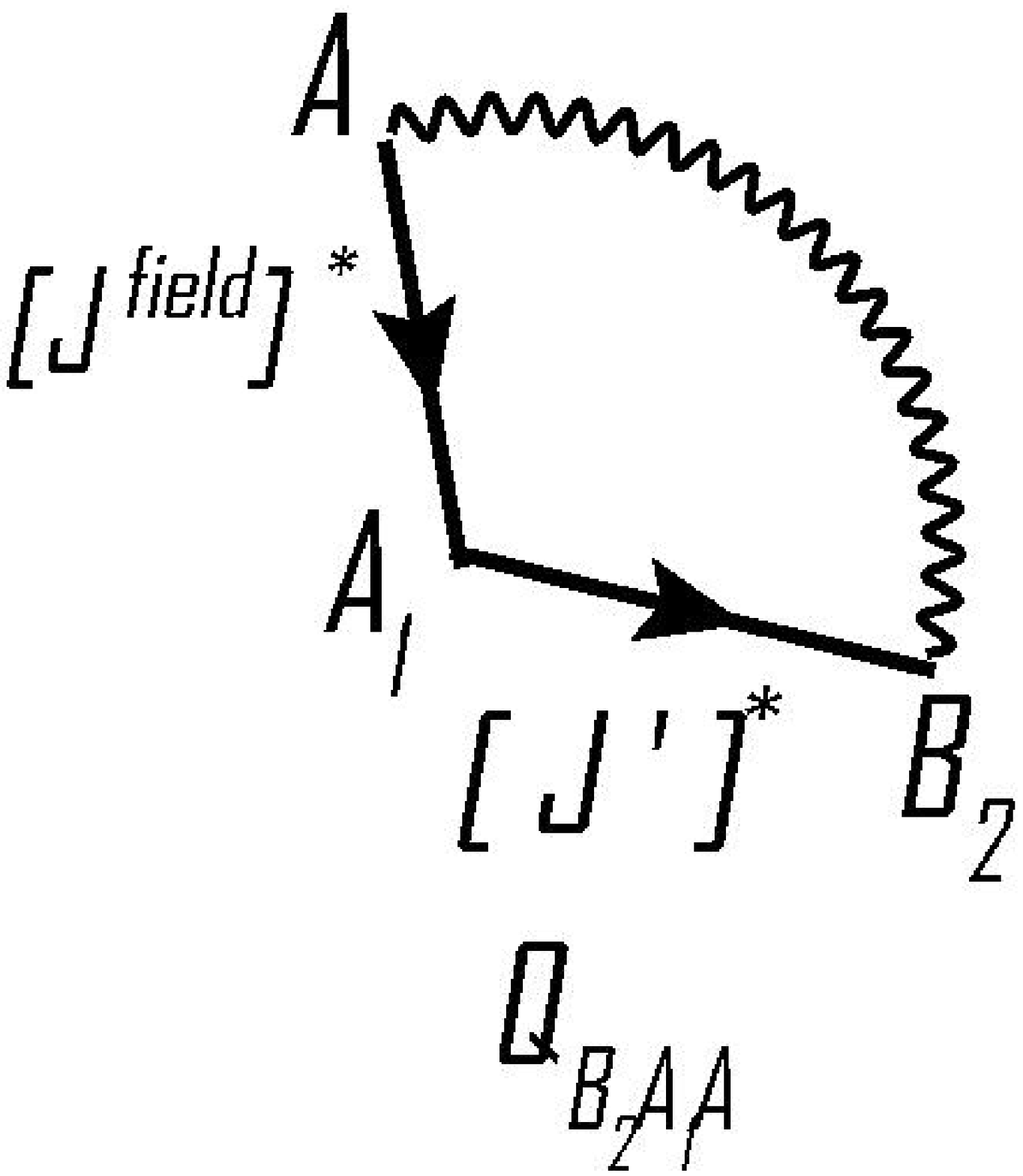}\label{QAAB2}}\hfill
\caption{Initial state is localized state $\protect\phi_{A}$: order $%
[J^{\prime}]^{1}$ and order $J^{1}$ contributions to conductivity}
\label{QAA}
\end{figure}
\begin{figure}[th]
\centering
\subfloat[]{\includegraphics[scale=0.12]{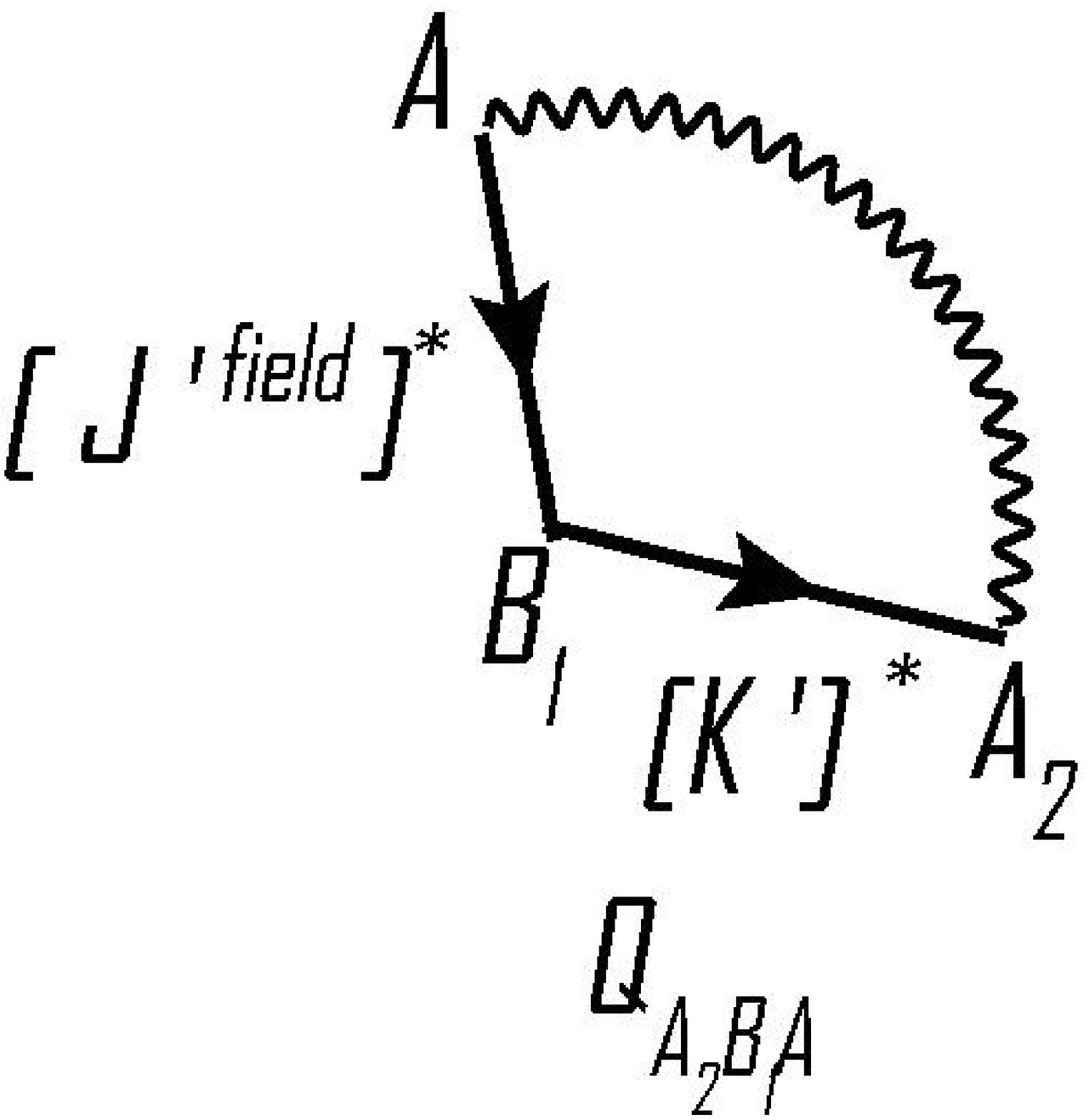}\label{QABA1}}\hfill %
\subfloat[]{\includegraphics[scale=0.12]{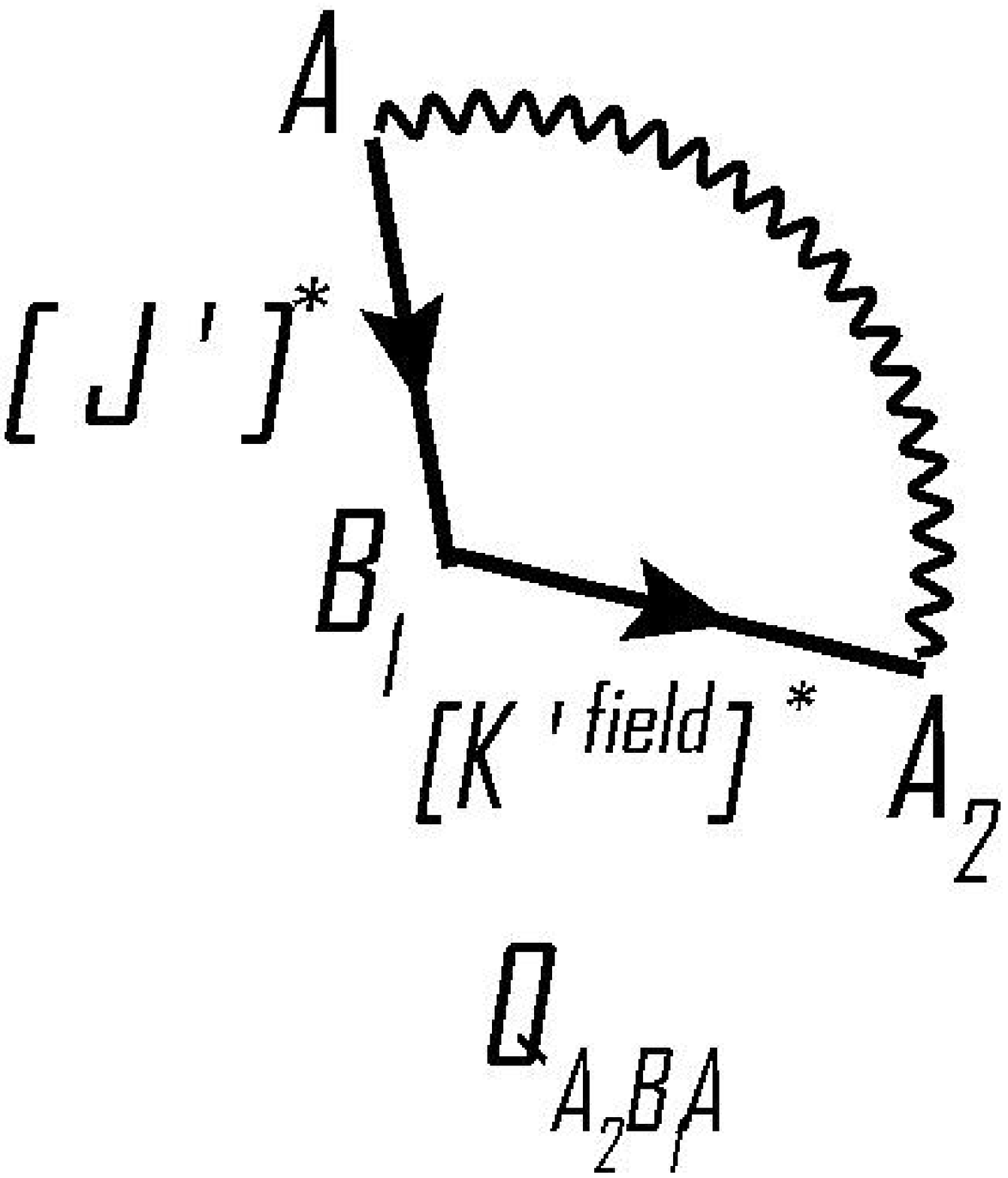}\label{QABA2}}\hfill 
\subfloat[]{\includegraphics[scale=0.12]{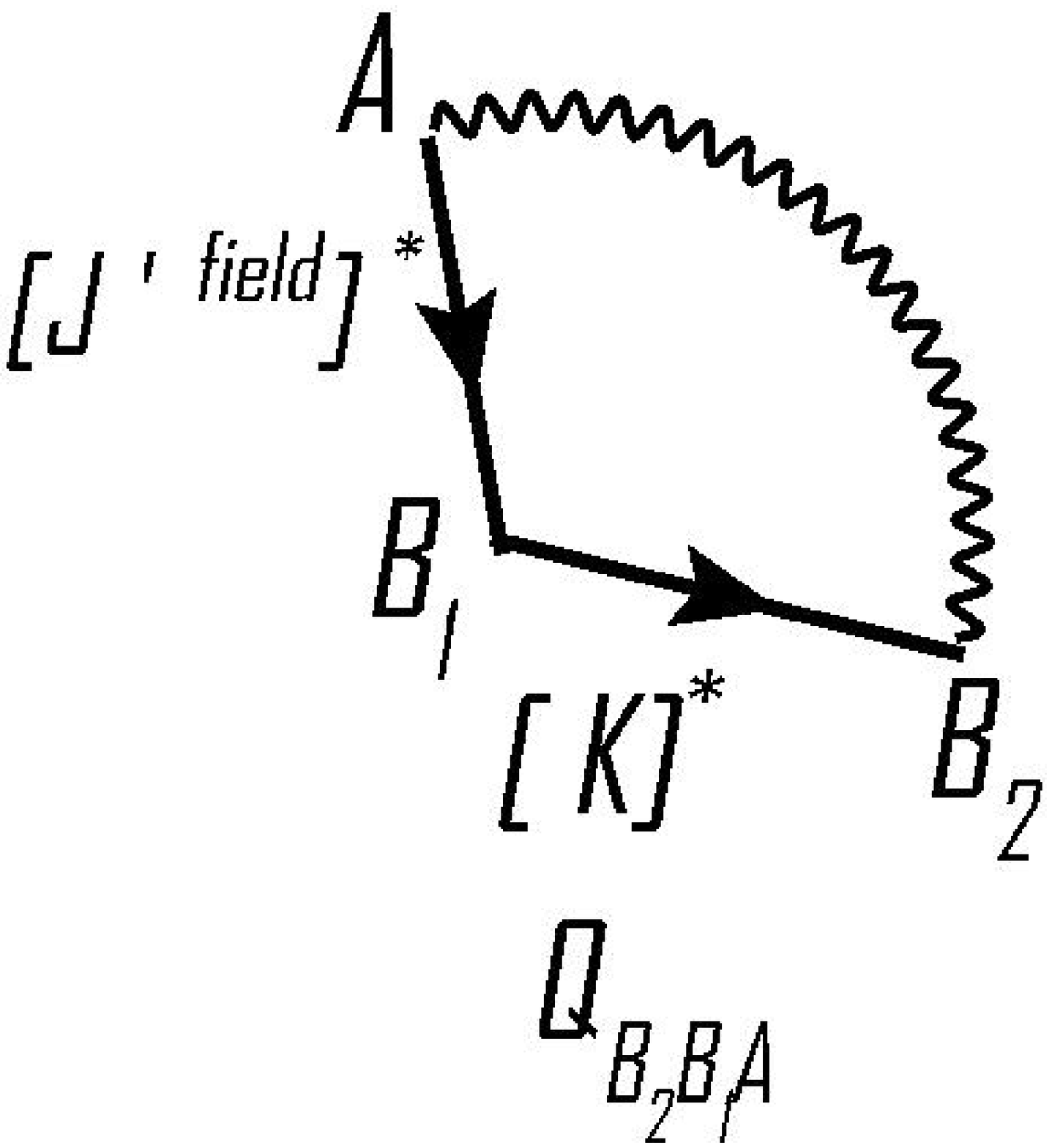}\label{QABB1}}\hfill %
\subfloat[]{\includegraphics[scale=0.12]{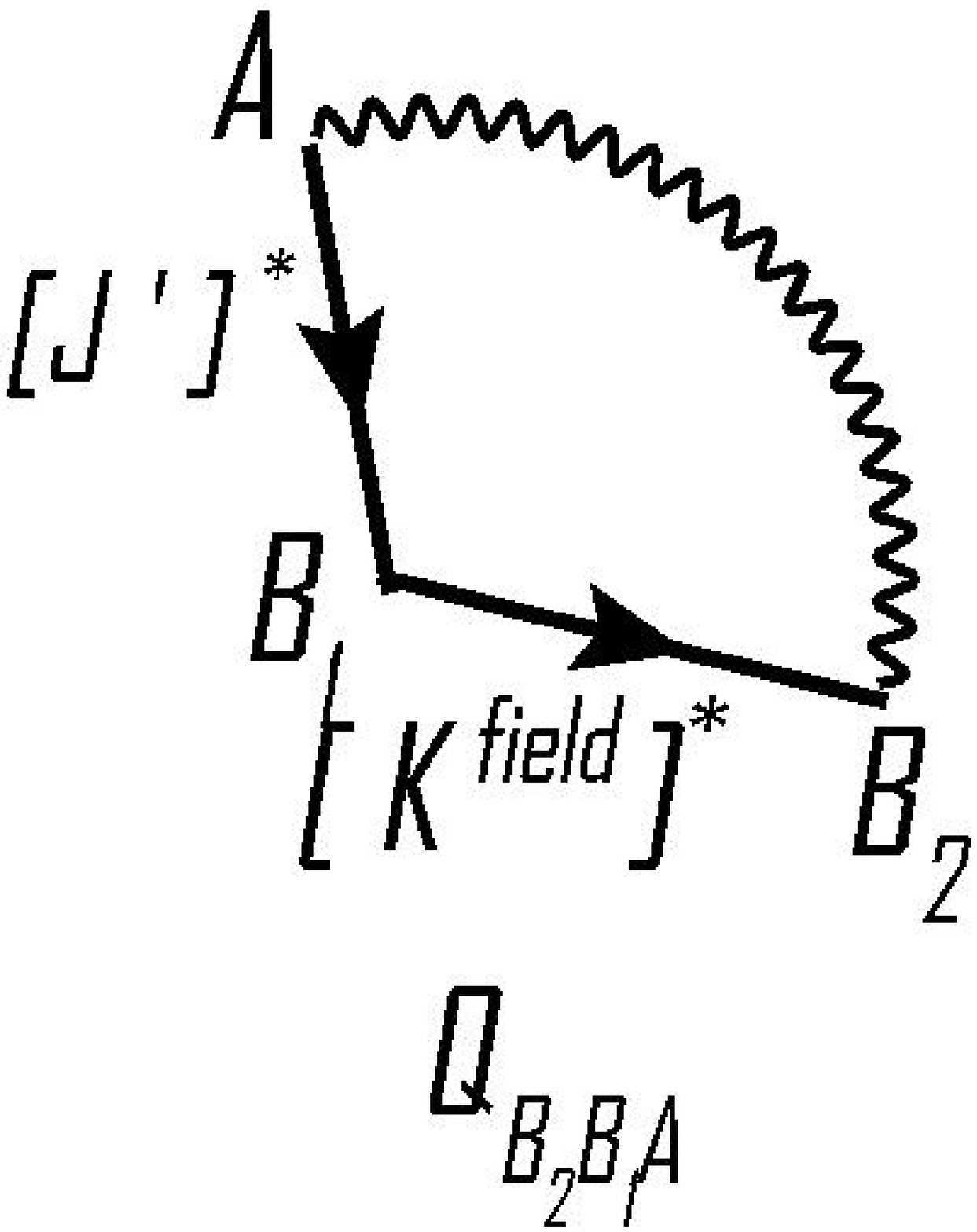}\label{QABB2}}\hfill
\caption{Initial state is localized state $\protect\phi_{A}$: order $%
[J^{\prime}]^{1}$, $K^{\prime}$ and $K$ contributions to conductivity}
\label{QAB}
\end{figure}

We may use diagrams to visualize the 14 processes (cf. Fig.\ref{A1A2}-\ref%
{QAB}). The expression for the current density Eqs.(\ref{1e1},\ref{ense})
and the perturbative solution of Eqs.(\ref{aev1},\ref{bev1}) suggest some
rules to write out the contribution to $\mathbf{j}$ for any process.
$\psi^{(0)}(t)$ and $\psi^{(1)}(t)$ are two ingredients of $\mathbf{j}$. In
Eqs.(\ref{ew},\ref{az},\ref{bz}), $\psi^{(0)}(t)$ and $\psi^{(1)}(t)$ are\
expanded with the eigenfunctions of $h_{0e}^{A}+h_{A}$ and of $h_{0e}+h_{B}
$. The expansion coefficients are transition amplitudes from a given initial
state. The current density operator $(\psi^{(0)}\nabla_{\mathbf{r}%
}\psi^{(1)\ast}-\psi^{(1)\ast}\nabla_{\mathbf{r}}\psi^{(0)})$ links one
component of $\psi^{(0)}$ to one component of $\psi^{(1)\ast}$ (connector).
Thus $\mathbf{j}$ is a sum of many terms, each term is a product of a
connector and one or several transition amplitude(s). We draw a wavy line
from one component of $\psi^{(0)}$ to one component of $\psi^{(1)\ast}$. The
spatial average produces $\Omega_{s}^{-1}\int_{\mathbf{r}\in\Omega_{s}}d%
\mathbf{r}$ before the microscopic current density, cf. the second to the fifth lines of Table \ref{current}.
Because $\psi^{(0)}$ appears in $\mathbf{j}$, the free evolution $%
\psi_{S^{0}}^{(0)}(t)$ of the initial state or one of two final components
in $\psi_{S^{1}}^{(0)}(t)$ will appear in $\mathbf{j}$. We use a solid arrow
line upward (top right or top left) to express the transition amplitude in $%
\psi_{S^{1}}^{(0)}(t)$ caused by one of residual interactions: the arrow
points from initial state to final state.
For a process in which $\psi_{S^{0}}^{(1)\ast}$ appears in $\mathbf{j}$, the
complex conjugate (cc) of the transition amplitude(s) in $\psi_{S^{0}}^{(1)}(t)$
will appear. We draw a solid arrow line downward (lower right or lower left) to
express the cc caused by an external field: the arrow points
from the initial state to the final state, see Table \ref{FHD}. For a
process in which $\psi_{S^{1}}^{(1)\ast}$ appears in $\mathbf{j}$, the
cc of two transition amplitudes in $%
\psi_{S^{1}}^{(1)}(t)$ will appear. We draw two successive solid arrow lines
downward, one represents the cc of
the transition amplitudes caused by an external field, another expresses the
cc of the transition amplitudes caused by a residual
interaction, see Table \ref{FHD}.
By retaining only order $S^{0}$ and $S^{1}$ terms in $\mathbf{j}$, Fig.\ref{A1A2}- \ref{QAB} describes all possible combinations. The two processes in Fig.\ref{A1} result from $\psi_{S^{0}}^{(0)}$ and $\psi_{S^{0}}^{(1)\ast}$. The four processes in Fig.\ref{A3A4A5A6} result from $\psi_{S^{1}}^{(0)}$ and $\psi_{S^{0}}^{(1)\ast}$. The 8 processes in Fig. \ref{QAA} and \ref{QAB} result from $\psi_{S^{0}}^{(0)}$ and $\psi_{S^{1}}^{(1)\ast}$.
\begin{table*}[ht]
\caption{Transition amplitudes induced by the residual interactions and external field}
\resizebox{15cm}{!} {
\begin{tabular}
[c]{llllllllllllllllllll}\hline\hline
symbol & \ \ \ \ \ \ \ \ \ \ \ expression &  &   \\\hline
\small A $\underrightarrow{\text{field}}$ A$_{2}$ &\ $\frac{i}{\hbar}\int_{-\infty}^{t}dt^{\prime}[J_{A_{2}A}^{field}(t^{\prime})]^{\ast}\langle\Psi_{A_{2}}^{\{N_{\alpha}^{\prime\prime}\}}|\Psi_{A}^{\{N_{\alpha}\}}\rangle
e^{-it^{\prime}(\mathcal{E}_{A_{2}}^{\{N_{\alpha}^{\prime\prime}\}}-\mathcal{E}_{A}^{\{N_{\alpha}\}})/\hbar}$ &  &  \\\hline
A $\underrightarrow{\text{field}}$ B$_{1}$ &\ $\frac{i}{\hbar}\int_{-\infty}^{t}dt^{\prime}[J_{B_{1}A}^{\prime field}(t^{\prime})]^{\ast}\langle\Xi
_{B_{1}}^{\{N_{\alpha}^{\prime}\}}|\Psi_{A}^{\{N_{\alpha}\}}\rangle
e^{-it^{\prime}(\mathcal{E}_{B_{1}}^{\{N_{\alpha}^{\prime}\}}-\mathcal{E}_{A}^{\{N_{\alpha}\}})/\hbar}$
&  & \\\hline
B $\underrightarrow{\text{field}}$ A$_{2}$ &\ $\frac{i}{\hbar}\int_{-\infty}^{t}dt^{\prime\prime\prime}[K_{A_{2}B}^{\prime field}(t^{\prime\prime\prime
})]^{\ast}\langle\Psi_{A_{2}}^{\{N_{\alpha}^{\prime\prime}\}}|\Xi
_{B}^{\{N_{\alpha}\}}\rangle e^{-it^{\prime\prime\prime}(\mathcal{E}_{A_{2}}^{\{N_{\alpha}^{\prime\prime}\}}-\mathcal{E}_{B}^{\{N_{\alpha}\}})/\hbar}$ & &  \\\hline
B $\underrightarrow{\text{field}}$ B$_{1}$ &\ $ \frac{i}{\hbar}\int_{-\infty}^{t}dt^{\prime\prime\prime}[K_{B_{1}B}^{field}(t^{\prime\prime\prime})]^{\ast
}\langle\Xi_{B_{1}}^{\{N_{\alpha}^{\prime}\}}|\Xi_{B}^{\{N_{\alpha}\}}\rangle
e^{-it^{\prime\prime\prime}(\mathcal{E}_{B_{1}}^{\{N_{\alpha}^{\prime}\}}-\mathcal{E}_{B}^{\{N_{\alpha}\}})/\hbar}$   &  & \\\hline
A $\underrightarrow{J}$ A$_{3}$ &\ $  -\frac
{i}{\hbar}J_{A_{3}A}\langle\Psi_{A_{3}}^{\{N_{\alpha}^{\prime\prime\prime}\}}|\Psi_{A}^{\{N_{\alpha}\}}\rangle\int_{-\infty}^{t}dt^{\prime}e^{it^{\prime}(\mathcal{E}_{A_{3}}^{\{N_{\alpha}^{\prime\prime\prime}\}}-\mathcal{E}_{A}^{\{N_{\alpha}\}})/\hbar}$ &  &  \\\hline
A $\underrightarrow{J^{\prime}}$ B$_{3}$ &\
$  -\frac{i}{\hbar}J_{B_{3}A}^{\prime
}\langle\Xi_{B_{3}}^{\{N_{\alpha}^{\prime\prime\prime}\}}|\Psi_{A}^{\{N_{\alpha}\}}\rangle\int_{-\infty}^{t}dt^{\prime}e^{it^{\prime
}(\mathcal{E}_{B_{3}}^{\{N_{\alpha}^{\prime\prime\prime}\}}-\mathcal{E}_{A}^{\{N_{\alpha}\}})/\hbar}$
&  & \\\hline
B $\underrightarrow{K^{\prime}}$ A$_{3}$ &\
$  -\frac{i}{\hbar}\langle\Psi_{A_{3}}^{\{N_{\alpha}^{\prime\prime\prime}\}}|K_{A_{3}B}^{\prime}|\Xi_{B}^{\{N_{\alpha}\}}\rangle\int_{-\infty}^{t}dt^{\prime}e^{it^{\prime
}(\mathcal{E}_{A_{3}}^{\{N_{\alpha}^{\prime\prime\prime}\}}-\mathcal{E}_{B}^{\{N_{\alpha}\}})/\hbar}$
&  & \\\hline
B $\underrightarrow{K}$ B$_{3}$ & \ $  -\frac
{i}{\hbar}\langle\Xi_{B_{3}}^{\{N_{\alpha}^{\prime\prime\prime}\}}|K_{B_{3}B}|\Xi_{B}^{\{N_{\alpha}\}}\rangle\int_{-\infty}^{t}dt^{\prime}e^{it^{\prime
}(\mathcal{E}_{B_{3}}^{\{N_{\alpha}^{\prime\prime\prime}\}}-\mathcal{E}_{B}^{\{N_{\alpha}\}})/\hbar}$
&  & \\\hline\hline
\end{tabular}\label{FHD}
}
\end{table*}

With the help of the diagrammatic rules listed in tables \ref{current} and \ref{FHD}, one can easily write out the corresponding macroscopic current density for each conduction processes. Taking Fig.\ref{A3} as example, the contribution to $\mathbf{j}$ is:
\begin{equation*}
(-\frac{N_{e}e\hbar}{m})\sum_{A\cdots N_{\alpha}\cdots}f(E_{A})\displaystyle\prod
\limits_{\alpha}P(N_{\alpha})\sum_{A_{1}\cdots
N_{\alpha}^{\prime}\cdots}[1-f(E_{A_{1}})]
\end{equation*}%
\begin{equation*}
\sum_{A_{3}\cdots
N_{\alpha}^{\prime\prime\prime}\cdots}[1-f(E_{A_{3}})]\Omega_{\mathbf{s}%
}^{-1}\int_{\Omega_{\mathbf{s}}}d\mathbf{r}(\phi_{A_{1}}\nabla_{\mathbf{r}%
}\phi_{A_{3}}^{\ast}-\phi_{A_{3}}^{\ast}\nabla_{\mathbf{r}}\phi_{A_{1}})
\end{equation*}%
\begin{equation}
\langle\Psi_{A_{1}}^{\{N_{\alpha}^{\prime}\}}|\Psi_{A_{3}}^{\{N_{\alpha
}^{\prime\prime\prime}\}}\rangle e^{-it(\mathcal{E}_{A_{1}}^{\{N_{\alpha
}^{\prime}\}}-\mathcal{E}_{A_{3}}^{\{N_{\alpha}^{\prime\prime\prime}\}})/%
\hbar}   \label{r1d}
\end{equation}%
\begin{equation*}
(\frac{-i}{\hbar})J_{A_{1}A}\int_{-\infty}^{t}dt^{\prime\prime}\langle
\Psi_{A_{1}}^{\{N_{\alpha}^{\prime}\}}|\Psi_{A}^{\{N_{\alpha}\}}\rangle
e^{it^{\prime\prime}(\mathcal{E}_{A_{1}}^{\{N_{\alpha}^{\prime}\}}-\mathcal{E%
}_{A}^{\{N_{\alpha}\}})/\hbar}
\end{equation*}%
\begin{equation*}
\frac{i}{\hbar}\int_{-\infty}^{t}dt^{\prime}[J_{A_{3}A}^{\text{field}%
}(t^{\prime})]^{\ast}\langle\Psi_{A_{3}}^{\{N_{\alpha}^{\prime\prime\prime}%
\}}|\Psi_{A}^{\{N_{\alpha}\}}\rangle e^{-it^{\prime}(\mathcal{E}%
_{A_{3}}^{\{N_{\alpha}^{\prime\prime\prime}\}}-\mathcal{E}%
_{A}^{\{N_{\alpha}\}})/\hbar},
\end{equation*}
where $P(N_{\alpha})=\exp[-\beta(N_{\alpha}+1/2)\hbar\omega_{\alpha }]%
/Z_{\alpha}$ is the probability that there are $N_{\alpha}$ phonons in the $%
\alpha^{th}$ mode$,$ and $Z_{\alpha}=\sum_{N_{\alpha}}\exp(-\beta(N_{\alpha
}+1/2)\hbar\omega_{\alpha})$. The second and third lines come from current operator, cf. the second line of table \ref{current}. The fourth line of Eq.(\ref{r1d}) is the transition amplitude induced by the transfer integral J, cf. the sixth line of table \ref{FHD}. The fifth line of Eq.(\ref{r1d}) is the cc of the transition amplitude induced by the external field, cf. the second line of table \ref{FHD}. Because we transformed the vibrational
displacements $\{x_{j}\}$ to normal coordinates $\{\theta_{\alpha}\}$, the
integral over displacements $\int[\displaystyle\prod \limits_{j=1}^{3\mathcal{N}}dx_{j}]$
in Eq.(\ref{1e1}) becomes an integral over normal coordinates $\int[\displaystyle\prod
\limits_{\alpha=1}^{3\mathcal{N}}d\theta_{\alpha}]$ in $\langle%
\Psi_{A_{1}}^{\{N_{\alpha}^{\prime}\}}|\Psi_{A_{3}}^{\{N_{\alpha}^{\prime%
\prime\prime}\}}\rangle$. Because $A_{1}$ is the final state of the LL transition induced by the transfer integral J, $A_{3}$ is the final state of the LL transition induced by the external field, states $A_{1}$ and $A_{3}$ must not be occupied: one has factors $[1-f(E_{A_{3}})]$ and $[1-f(E_{A_{3}})]$.
According to perturbation theory, one should (1) sum over all intermediate
states; (2) sum over all the components of the final state. Because we don't
know what the initial state is, we average over all possible initial
states. At $t=-\infty$, the external field and various residual interactions
($J,$ $J,$ $K$ and $K^{\prime}$) are not yet turned on, and the system is in
equilibrium. The probabilities of $|A\cdots N_{\alpha}\cdots\rangle$ is $f(E_{A})\displaystyle\prod
\limits_{\alpha}P(N_{\alpha})$. The first sum comes from averaging over various initial states, the second sum comes from summing over the component of final state resulted from transfer integral, the third sum comes from summing over the component of final state caused by external field. The first factor in come from Eq.(\ref{1e1}). One can similarly find the expressions of macroscopic density for other conduction processes.

In an electric field $\mathbf{E}=\mathbf{E}_{0}\cos\omega t$, the real part of
conductivity tensor Re$\sigma_{\alpha\beta}$ can be extracted from the component of $%
j_{\alpha}=\sum_{\beta }\sigma_{\alpha\beta}E_{\beta}$ with time factor $\cos\omega t$, the imaginary part Im$\sigma_{\alpha\beta}$ is extracted from $j_{\alpha}$ with time factor $\sin\omega t$.
For the conduction processes in which the initial state is a localized state, the expressions of $\func{Re}\sigma_{\alpha\beta}(\omega)$ and $\func{Im}%
\sigma_{\alpha\beta}(\omega)$ ($\alpha,\beta=x,y,z$) are given in table \ref{Acon}, the real part takes the upper sign, the imaginary part takes the lower sign. We used
\begin{equation}
w_{A_{3}A_{2}}^{\alpha}=\frac{-i\hbar}{m}\int_{\Omega_{\mathbf{s}}}d\mathbf{r%
}\phi_{A_{3}}(\mathbf{r}-\mathbf{R}_{A_{3}})\frac{\partial}{\partial
x_{\alpha}}\phi_{A_{2}}^{\ast}(\mathbf{r}-\mathbf{R}_{A_{2}}),   \label{w0}
\end{equation}
and%
\begin{equation}
v_{A_{2}A_{3}}^{\alpha}=\frac{-i\hbar}{m}\int_{\Omega_{\mathbf{s}}}d\mathbf{r%
}\phi_{A_{2}}^{\ast}(\mathbf{r}-\mathbf{R}_{A_{2}})\frac{\partial }{\partial
x_{\alpha}}\phi_{A_{3}}(\mathbf{r}-\mathbf{R}_{A_{3}}),   \label{v0}
\end{equation}
etc to denote the matrix elements of velocity operator between single particle states.
One can show that the integral over normal
coordinates $\int[\displaystyle\prod \limits_{\alpha=1}^{3\mathcal{N}}d\theta_{\alpha}]$%
, i.e. the sum over final phonon states and averaging over initial
phonon state can be carried out for all the contributions to the current density$%
\mathbf{j}$. The vibrational part of the current density $\mathbf{j}$ for each process becomes a time integral, cf. Sec.2 of Ref.\cite{sup}.  The time integrals
I$_{s}$, Q$_{s}$ and S$_{s}$ are functions of temperature $T$ and the frequency $\omega$
of the external field, are listed in in Sec.3.1 of Ref.\cite{sup}.
$\func{Re}\sigma_{\alpha\beta}(\omega)$ and $\func{Im}\sigma_{\alpha\beta}(%
\omega)$ satisfy the Kramers-Kronig relation, The reason that we list both of them is to show that (1) they are symmetric about the positive and negative frequency time integrals and (2) $\func{Im}%
\sigma_{\alpha\beta}(0)=0$.

\begin{table*}
\caption{The conduction processes and conductivity: initial state is a localized state $\phi_{A}$.}
\resizebox{13cm}{!} {
\begin{tabular}
[c]{lc}\hline\hline
$\text{diagram}$  & $~~~~~~~~~~\text{conductivity}%
$\\\hline
$\text{\ref{A1}}$ & $%
\begin{array}
[c]{c}%
{\large -}\frac{N_{e}e^{2}}{2\Omega_{s}}\sum_{AA_{2}}\operatorname{Im}%
{\large (w}_{AA_{2}}^{\alpha}{\large -v}_{A_{2}A}^{\alpha}{\large )i(E}%
_{A}^{0}{\large -E}_{A_{2}}^{0}{\large )}^{-1}{\large (v}_{A_{2}A}^{\beta
}{\large )}^{\ast}\\
{\large [I}_{A_{2}A+}{\large \pm I}_{A_{2}A-}{\large ][1-f(E}_{A_{2}%
}{\large )]f(E}_{A}{\large )}%
\end{array}
$\\\hline
$\text{\ref{A2}}$  & $%
\begin{array}
[c]{c}%
{\large -}\frac{N_{e}e^{2}}{2\Omega_{s}}\sum_{AB_{1}}\operatorname{Im}%
{\large (w}_{AB_{1}}^{\alpha}{\large -v}_{B_{1}A}^{\alpha}{\large )i(E}%
_{A}^{0}{\large -E}_{B_{1}}^{0}{\large )}^{-1}{\large (v}_{B_{1}A}^{\beta
}{\large )}^{\ast}\\
{\large [I}_{B_{1}A+}{\large \pm I}_{B_{1}A-}{\large ][1-f(E}_{B_{1}%
}{\large )]f(E}_{A}{\large )}%
\end{array}
$\\\hline
$\text{\ref{A3}}$  & $%
\begin{array}
[c]{c}%
{\large -}\frac{N_{e}e^{2}}{2\hbar\Omega_{s}}\operatorname{Im}{\large i}%
\sum_{A_{3}A_{1}A}{\large (w}_{A_{1}A_{3}}^{\alpha}{\large -v}_{A_{3}A_{1}%
}^{\alpha}{\large )i(E}_{A}^{0}{\large -E}_{A_{3}}^{0}{\large )}%
^{-1}{\large (v}_{A_{3}A}^{\beta}{\large )}^{\ast}{\large J}_{A_{1}A}\\
{\large [I}_{A_{3}A_{1}A+}{\large \pm I}_{A_{3}A_{1}A-}{\large ]f(E}_{A}%
^{0}{\large )[1-f(E}_{A_{1}}^{0}{\large )][1-f(E}_{A_{3}}^{0}{\large )]}%
\end{array}
$\\\hline
$\text{\ref{A4}}$  & $%
\begin{array}
[c]{c}%
{\large -}\frac{N_{e}e^{2}}{2\hbar\Omega_{s}}\sum_{AA_{3}B_{1}}%
\operatorname{Im}{\large (w}_{A_{3}B_{1}}^{\alpha}{\large -v}_{B_{1}A_{3}%
}^{\alpha}{\large )(E}_{A}^{0}{\large -E}_{B_{1}}^{0}{\large )}^{-1}%
{\large (v}_{B_{1}A}^{\beta}{\large )}^{\ast}{\large J}_{A_{3}A}\\
{\large [I}_{B_{1}A_{3}A+}{\large \pm I}_{B_{1}A_{3}A-}{\large ][1-f(E}%
_{B_{1}}{\large )][1-f(E}_{A_{3}}{\large )]f(E}_{A}{\large )}%
\end{array}
$\\\hline
$\text{\ref{A5}}$  & $%
\begin{array}
[c]{c}%
{\large -}\frac{N_{e}e^{2}}{2\hbar\Omega_{s}}\sum_{AA_{2}B_{3}}%
\operatorname{Im}{\large (w}_{B_{3}A_{2}}^{\alpha}{\large -v}_{A_{2}B_{3}%
}^{\alpha}{\large )(E}_{A_{2}}^{0}{\large -E}_{A}^{0}{\large )}^{-1}%
{\large (v}_{A_{2}A}^{\beta}{\large )}^{\ast}{\large J}_{B_{3}A}^{\prime}\\
{\large [I}_{B_{3}A_{2}A+}{\large \pm I}_{B_{3}A_{2}A-}{\large ][1-f(E}%
_{B_{3}}{\large )][1-f(E}_{A_{2}}{\large )]f(E}_{A}{\large )}%
\end{array}
$\\\hline
$\text{\ref{A6}}$  & $%
\begin{array}
[c]{c}%
{\large -}\frac{N_{e}e^{2}}{2\hbar\Omega_{s}}\sum_{AB_{1}B_{3}}%
\operatorname{Im}{\large (w}_{B_{3}B_{1}}^{\alpha}{\large -v}_{B_{1}B_{3}%
}^{\alpha}{\large )J}_{B_{3}A}^{\prime}{\large (E}_{A}^{0}{\large -E}_{B_{1}%
}^{0}{\large )}^{-1}{\large (v}_{B_{1}A}^{\beta}{\large )}^{\ast}\\
{\large [I}_{B_{3}B_{1}A+}{\large \pm I}_{B_{3}B_{1}A-}{\large ][1-f(E}%
_{B_{3}}{\large )][1-f(E}_{B_{1}}{\large )]f(E}_{A}{\large )}%
\end{array}
$\\\hline
\ref{QAAA1} & $%
\begin{array}
[c]{c}%
{\large +}\frac{N_{e}e^{2}}{2\hbar\Omega_{s}}\sum_{A_{2}A_{1}A}%
\operatorname{Im}{\large (w}_{AA_{2}}^{\beta}{\large -v}_{A_{2}A}^{\beta
}{\large )(E}_{A_{1}}{\large -E}_{A_{2}}{\large )}^{-1}{\large (v}_{A_{2}%
A_{1}}^{\alpha}{\large )}^{\ast}{\large J}_{A_{1}A}^{\ast}\\
{\large (Q}_{1A_{2}A_{1}A+}{\large \pm Q}_{1A_{2}A_{1}A-}{\large )f(E}%
_{A}{\large )[1-f(E}_{A_{2}}{\large )]}%
\end{array}
$\\\hline
\ref{QAAA2}  & $%
\begin{array}
[c]{c}%
{\large +}\frac{N_{e}e^{2}}{2\hbar\Omega_{\mathbf{s}}}\sum_{A_{2}A_{1}%
A}\operatorname{Im}{\large (w}_{AA_{2}}^{\beta}{\large -v}_{A_{2}A}^{\beta
}{\large )J}_{A_{2}A_{1}}^{\ast}{\large (E}_{A}^{0}{\large -E}_{A_{1}}%
^{0}{\large )}^{-1}{\large (v}_{A_{1}A}^{\alpha}{\large )}^{\ast}\\
{\large (Q}_{2A_{2}A_{1}A+}{\large \pm Q}_{2A_{2}A_{1}A-}{\large )f(E}%
_{A}{\large )[1-f(E}_{A_{2}}{\large )]}%
\end{array}
$\\\hline
\ref{QAAB1}  & $%
\begin{array}
[c]{c}%
{\large +}\frac{N_{e}e^{2}}{2\hbar\Omega_{\mathbf{s}}}\sum_{B_{2}A_{1}%
A}\operatorname{Im}{\large (w}_{AB_{2}}^{\beta}{\large -v}_{B_{2}A}^{\beta
}{\large )(E}_{A_{1}}^{0}{\large -E}_{B_{2}}^{0}{\large )}^{-1}{\large (v}%
_{B_{2}A_{1}}^{\alpha}{\large )}^{\ast}{\large J}_{A_{1}A}^{\ast}\\
{\large (Q}_{1B_{2}A_{1}A+}{\large \pm Q}_{1B_{2}A_{1}A-}{\large )f(E}%
_{A}{\large )[1-f(E}_{B_{2}}{\large )]}%
\end{array}
$\\\hline
\ref{QAAB2}  & $%
\begin{array}
[c]{c}%
{\Large +}\frac{N_{e}e^{2}}{2\hbar\Omega_{\mathbf{s}}}\sum_{A_{2}A_{1}%
A}\operatorname{Im}{\Large (w}_{AB_{2}}^{\beta}{\Large -v}_{B_{2}A}^{\beta
}{\Large )J}_{B_{2}A_{1}}^{\prime\ast}{\Large (E}_{A}^{0}{\Large -E}_{A_{1}%
}^{0}{\Large )}^{-1}{\Large (v}_{A_{1}A}^{\alpha}{\Large )}^{\ast}\\
{\Large (Q}_{2B_{2}A_{1}A+}{\Large \pm Q}_{2B_{2}A_{1}A-}{\Large )f(E}%
_{A}{\Large )[1-f(E}_{B_{2}}{\Large )]}%
\end{array}
$\\\hline
\ref{QABA1}  & $%
\begin{array}
[c]{c}%
{\large +}\frac{N_{e}e^{2}}{2\hbar\Omega_{\mathbf{s}}}\sum_{A_{2}B_{1}%
A}\operatorname{Im}{\large (w}_{AA_{2}}^{\beta}{\large -v}_{A_{2}A}^{\beta
}{\large )(E}_{A}^{0}{\large -E}_{B_{1}}^{0}{\large )}^{-1}{\large (v}%
_{B_{1}A}^{\alpha}{\large )}^{\ast}\\
{\large (Q}_{1A_{2}B_{1}A+}^{K^{\prime}}{\large \pm Q}_{1A_{2}B_{1}%
A-}^{K^{\prime}}{\large )f(E}_{A}{\large )[1-f(E}_{A_{2}}{\large )]}%
\end{array}
$\\\hline
\ref{QABA2}  & $%
\begin{array}
[c]{c}%
{\large +}\frac{N_{e}e^{2}}{2\hbar\Omega_{\mathbf{s}}}\sum_{A_{2}A_{1}%
A}\operatorname{Im}{\large (w}_{AA_{2}}^{\beta}{\large -v}_{A_{2}A}^{\beta
}{\large )(E}_{B_{1}}^{0}{\large -E}_{A_{2}}^{0}{\large )}^{-1}{\large (v}%
_{A_{2}B_{1}}^{\alpha}{\large )}^{\ast}{\large J}_{B_{1}A}^{\prime\ast}\\
{\large (Q}_{2A_{2}B_{1}A+}{\large \pm Q}_{2A_{2}B_{1}A-}{\large )f(E}%
_{A}{\large )[1-f(E}_{A_{2}}{\large )]}%
\end{array}
$\\\hline
\ref{QABB1}  & $%
\begin{array}
[c]{c}%
{\large +}\frac{N_{e}e^{2}}{2\hbar\Omega_{\mathbf{s}}}\sum_{B_{2}B_{1}%
A}\operatorname{Im}{\large (w}_{AB_{2}}^{\beta}{\large -v}_{B_{2}A}^{\beta
}{\large )(E}_{A}^{0}{\large -E}_{B_{1}}^{0}{\large )}^{-1}{\large (v}%
_{B_{1}A}^{\alpha}{\large )}^{\ast}\\
{\large (Q}_{1B_{2}B_{1}A+}^{K}{\large \pm Q}_{1B_{2}B_{1}A-}^{K}%
{\large )f(E}_{A}{\large )[1-f(E}_{B_{2}}{\large )]}%
\end{array}
$\\\hline
\ref{QABB2}  & $%
\begin{array}
[c]{c}%
{\large +}\frac{N_{e}e^{2}}{2\hbar\Omega_{\mathbf{s}}}\sum_{B_{2}B_{1}%
A}\operatorname{Im}{\large (w}_{AB_{2}}^{\beta}{\large -v}_{B_{2}A}^{\beta
}{\large )(E}_{B_{1}}^{0}{\large -E}_{B_{2}}^{0}{\large )}^{-1}{\large (v}%
_{B_{2}B_{1}}^{\alpha}{\large )}^{\ast}{\large J}_{B_{1}A}^{\prime\ast}\\
{\large (Q}_{2A_{2}B_{1}A+}{\large \pm Q}_{2A_{2}B_{1}A-}{\large )f(E}%
_{A}{\large )[1-f(E}_{A_{2}}{\large )]}%
\end{array}
$\\ \hline\hline
\end{tabular}
}
\label{Acon}
\end{table*}

The conductivity from LL transitions\cite{short} derives from the processes depicted in Fig. \ref{A1}, \ref{A3}, \ref{QAAA1} and \ref{QAAA2}.
For small polarons, the Kubo formula was used in both Ref.\cite{lan62} and Ref.%
\cite{sch65} to obtain the second and fourth terms in table \ref{Acon},
but not the eighth  and ninth terms. The reason is the
following. If one views both $J$ and $J^{field}$ as small parameters, the eighth
 and ninth terms in table \ref{Acon} result from a second order
change in state, one in $J$ and one in $J^{field}$.
To obtain conductivity in Kubo formulation, the change in the density matrix of system is computed to first order in external field. By substituting this first order change into the macroscopic current density, one obtains the conductivity by factoring out the
external field\cite{ku57}. In this way, the new combinations of $J$ and $%
J^{field}$ (Fig.\ref{QAAA1} and \ref{QAAA2}) are excluded. In the present
work, we apply linear response to external field
at the last step, so that there are various
combinations between $S$ and $J^{field}$. From Eqs.(21, 25,27) of Ref.\cite{sup}, we see that the $(\omega,T)$ dependence of the fourth term
is different from those of the eighth and ninth terms in table \ref{Acon}. It will be
interesting to see if new features of the eighth and ninth terms can be
observed experimentally.

\begin{figure}[th]
\centering
\subfloat[]{\includegraphics[scale=0.15]{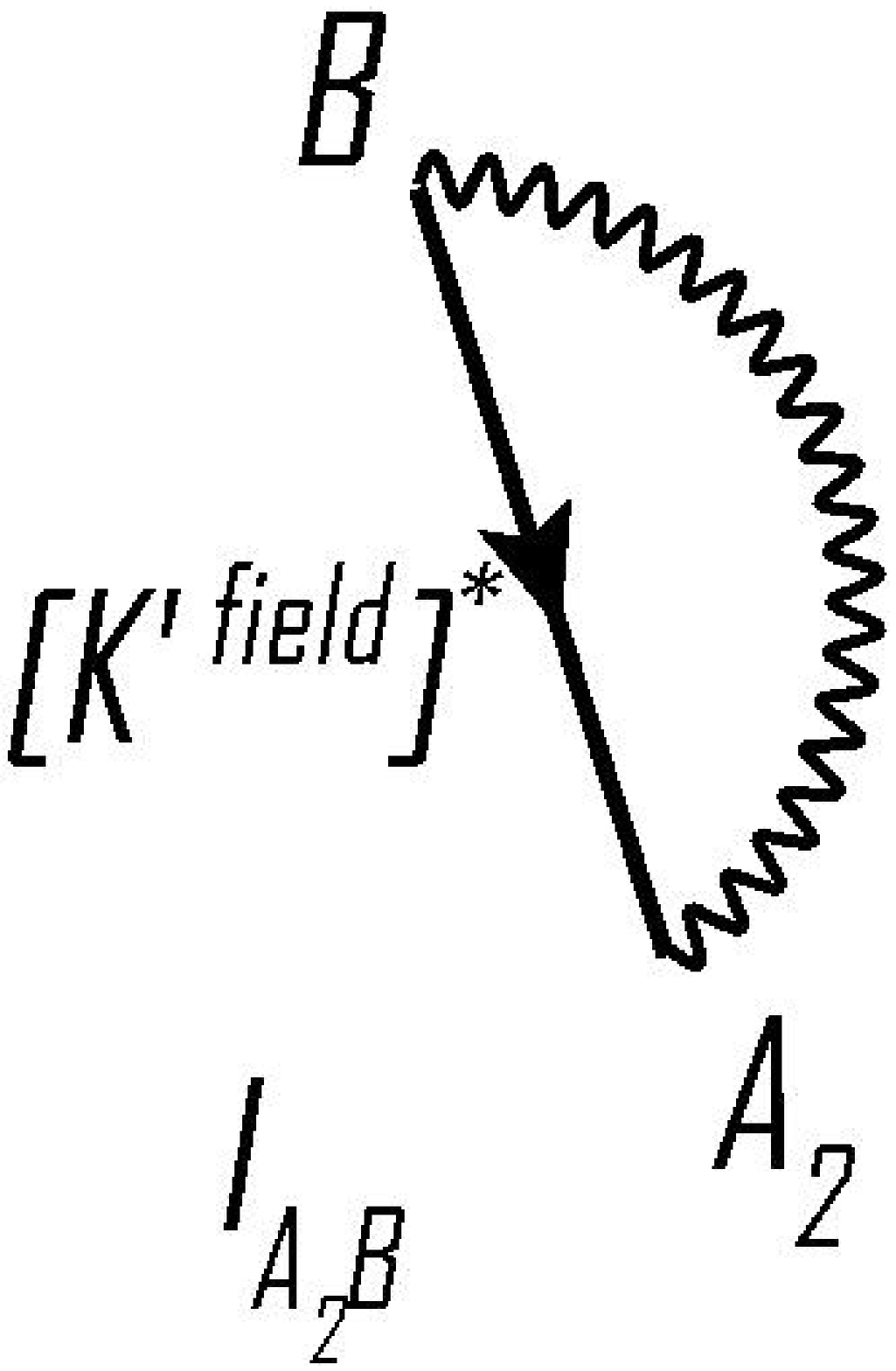}\label{B1}}\hfill
\subfloat[]{\includegraphics[scale=0.17]{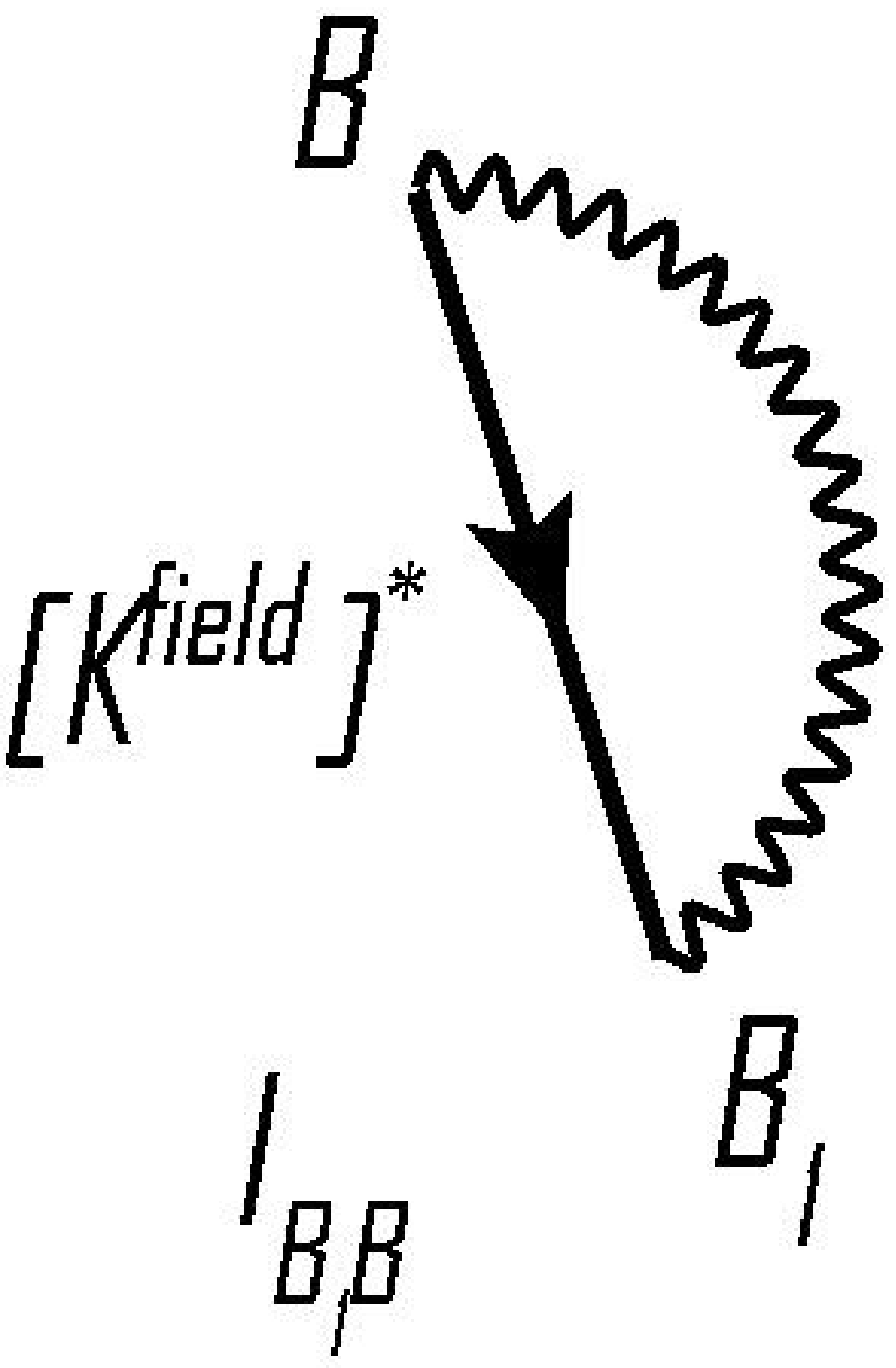}\label{B2}}\hfill
\caption{Initial state is extended state $\protect\xi_{B}$: order $%
[K^{\prime}]^{0}$ and order $K^{0}$ contributions to conductivity}
\label{B1B2}
\end{figure}

\begin{figure}[th]
\centering
\subfloat[]{\includegraphics[scale=0.11]{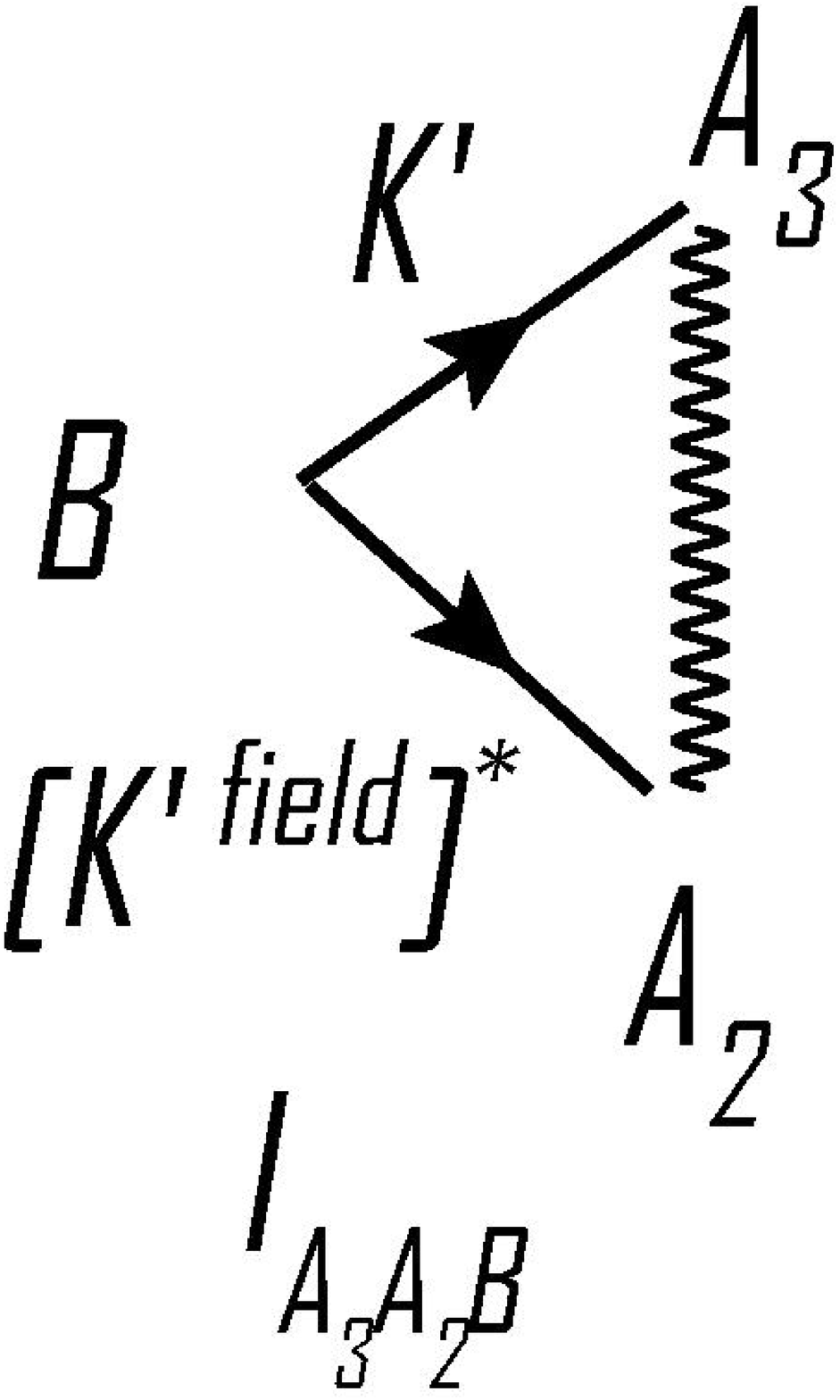} \label{B3}}\hfill
\subfloat[]{\includegraphics[scale=0.11]{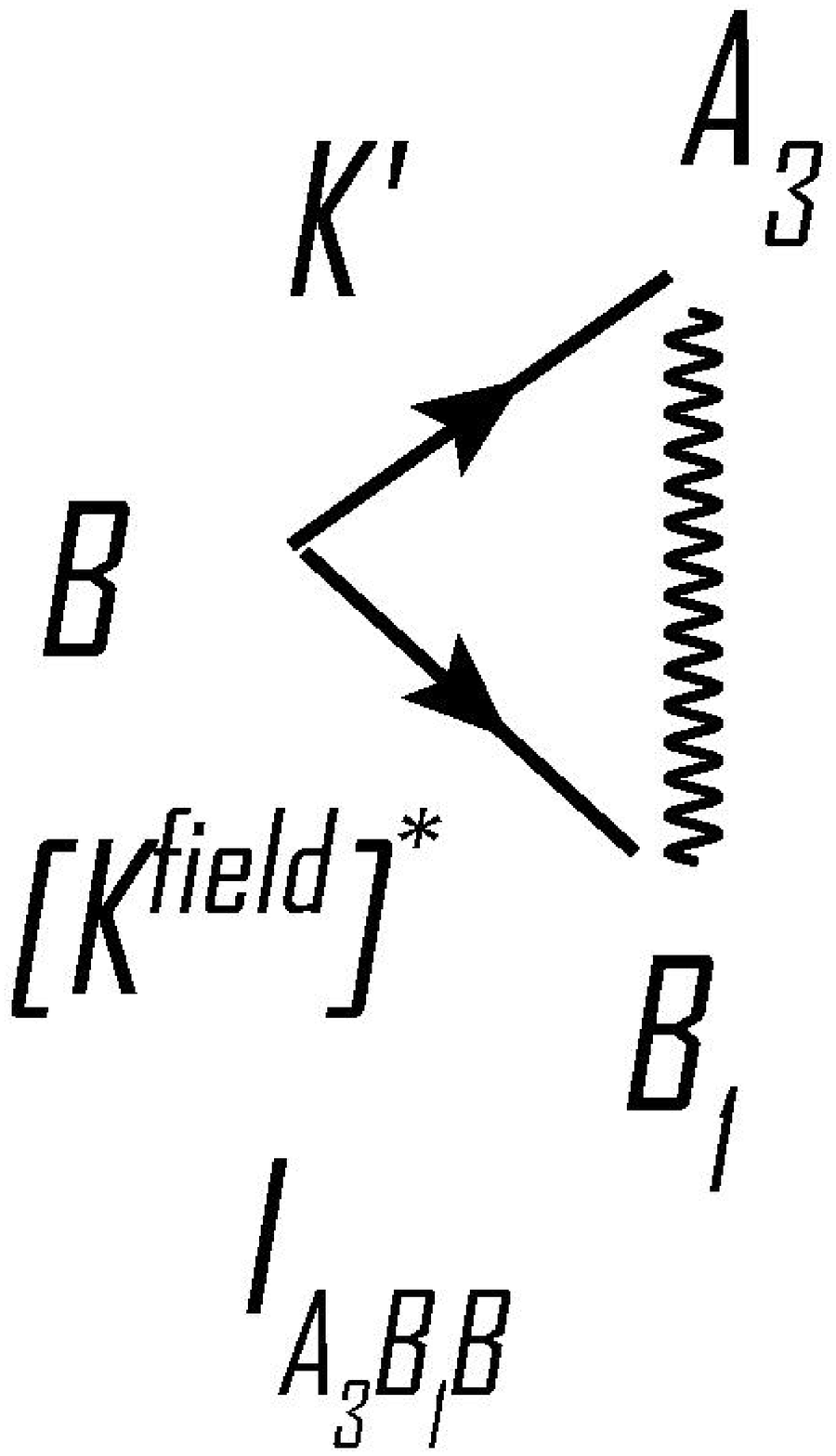}\label{B4}}\hfill
\subfloat[]{\includegraphics[scale=0.11]{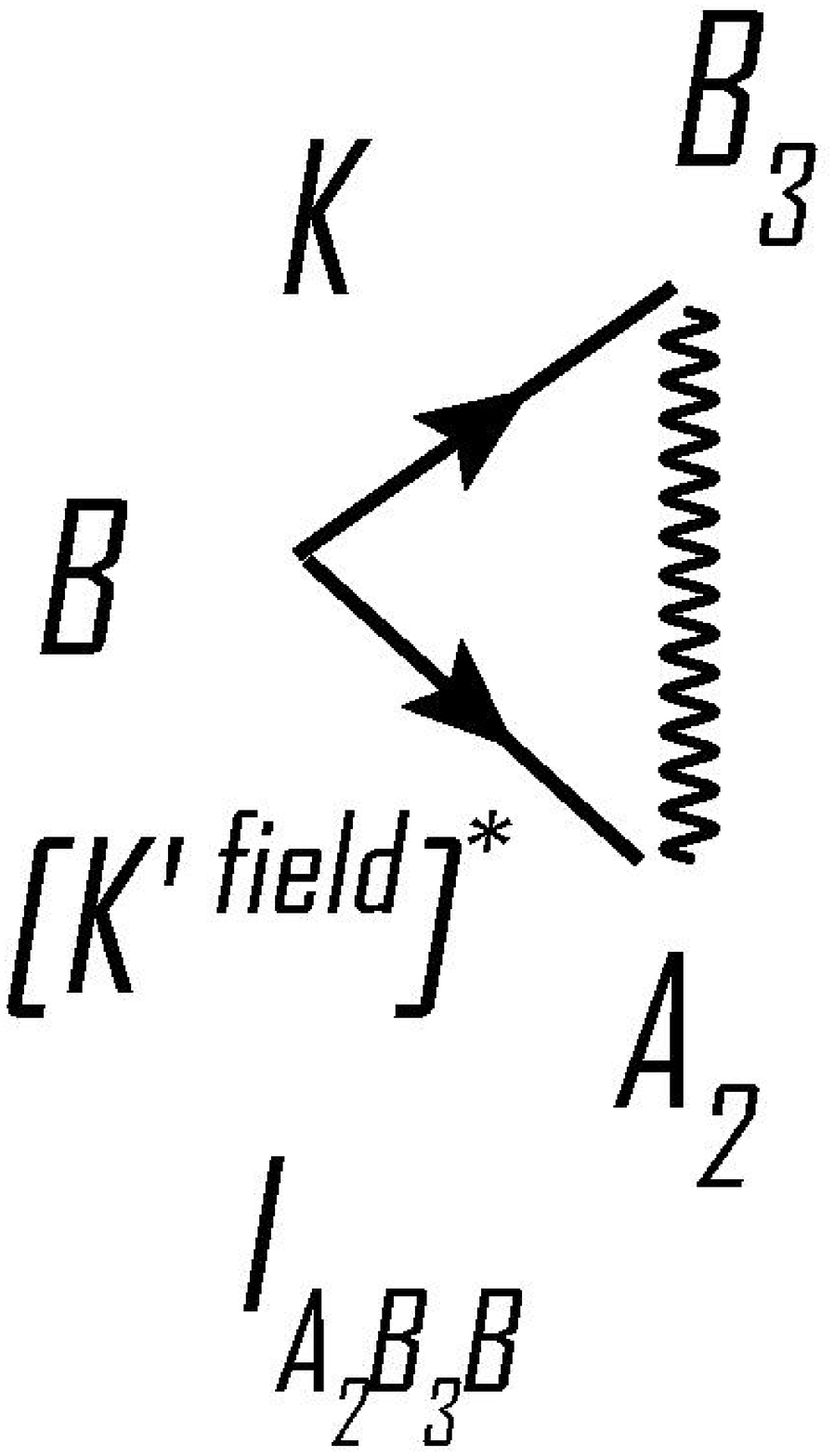}\label{B5}}\hfill
\subfloat[]{\includegraphics[scale=0.11]{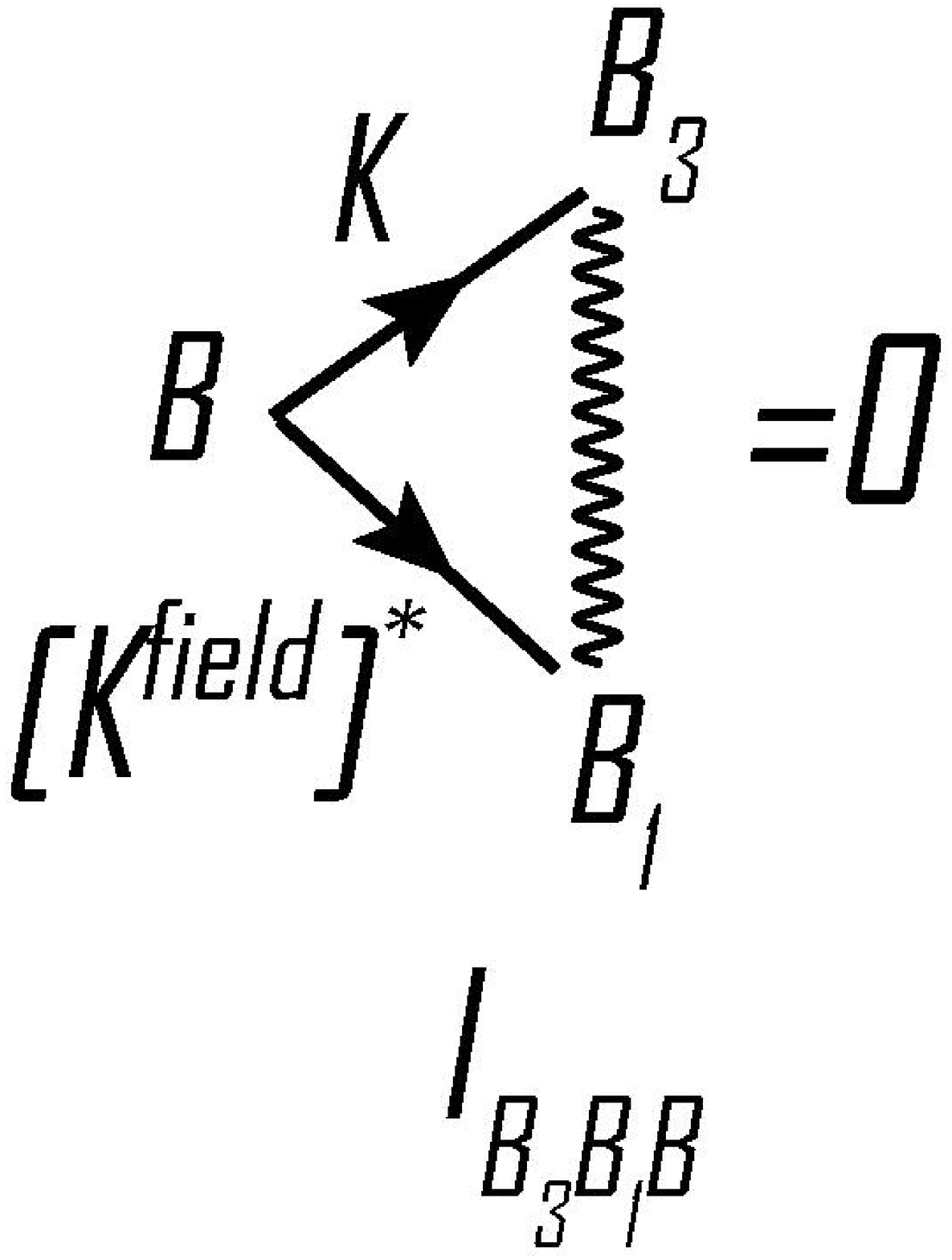}\label{B6}}\hfill
\caption{Initial state is extended state $\protect\xi_{B}$: order $%
[K^{\prime}]^{1}$ and order $K^{1}$ contributions to conductivity}
\label{B3B4B5B6}
\end{figure}

\begin{figure}[th]
\centering
\subfloat[]{\includegraphics[scale=0.12]{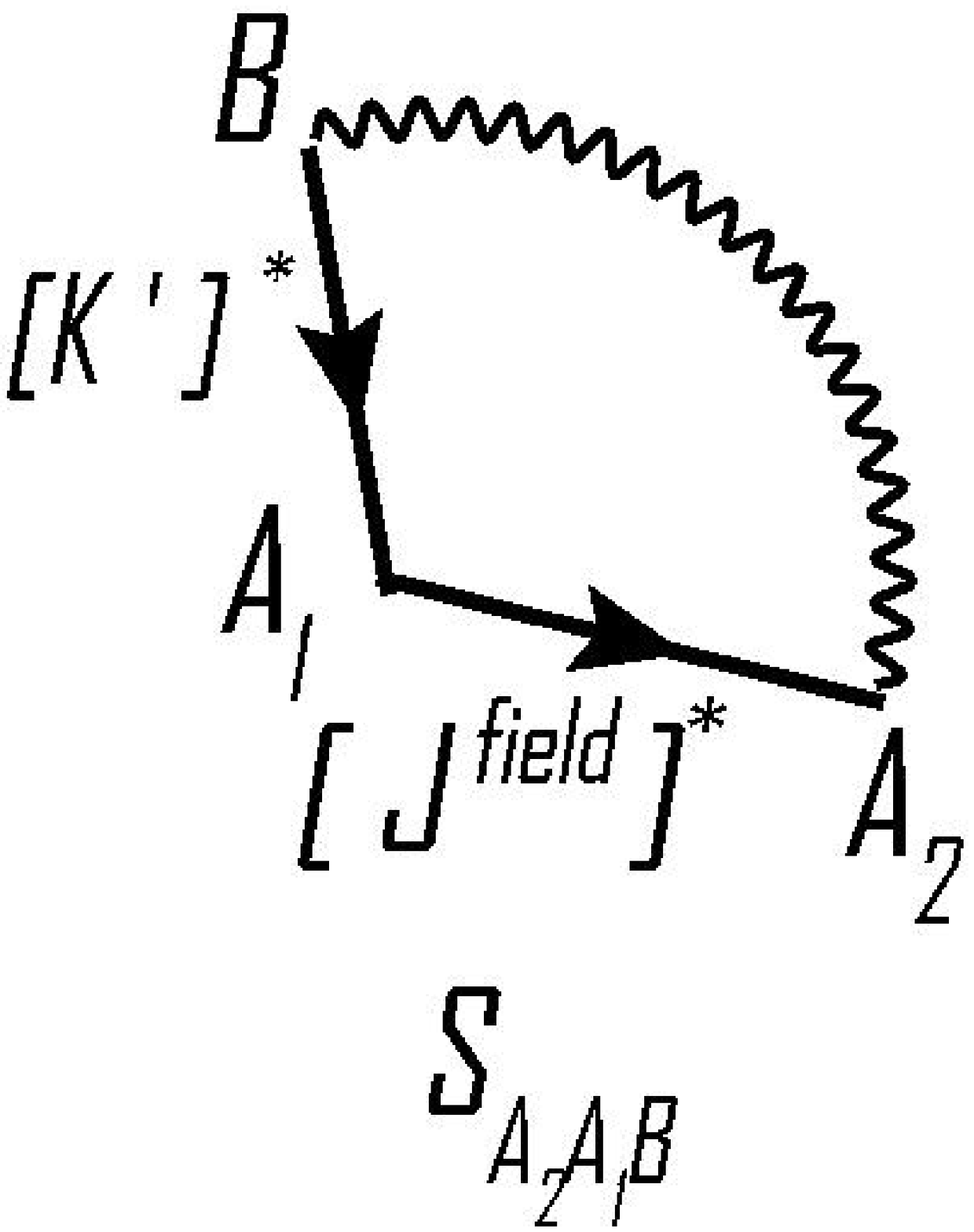}\label{SBAA1}}\hfill %
\subfloat[]{\includegraphics[scale=0.12]{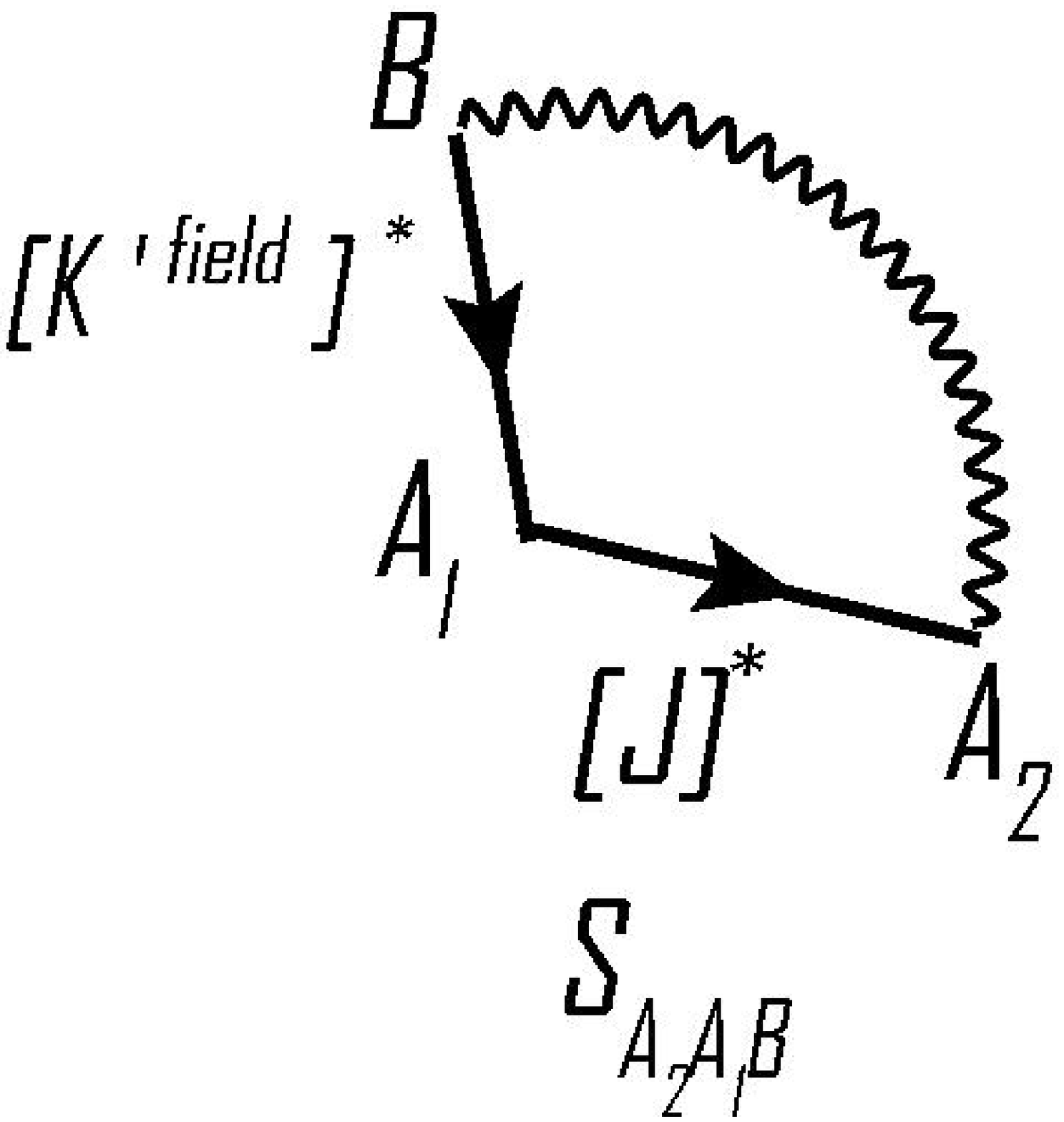}\label{SBAA2}}\hfill 
\subfloat[]{\includegraphics[scale=0.12]{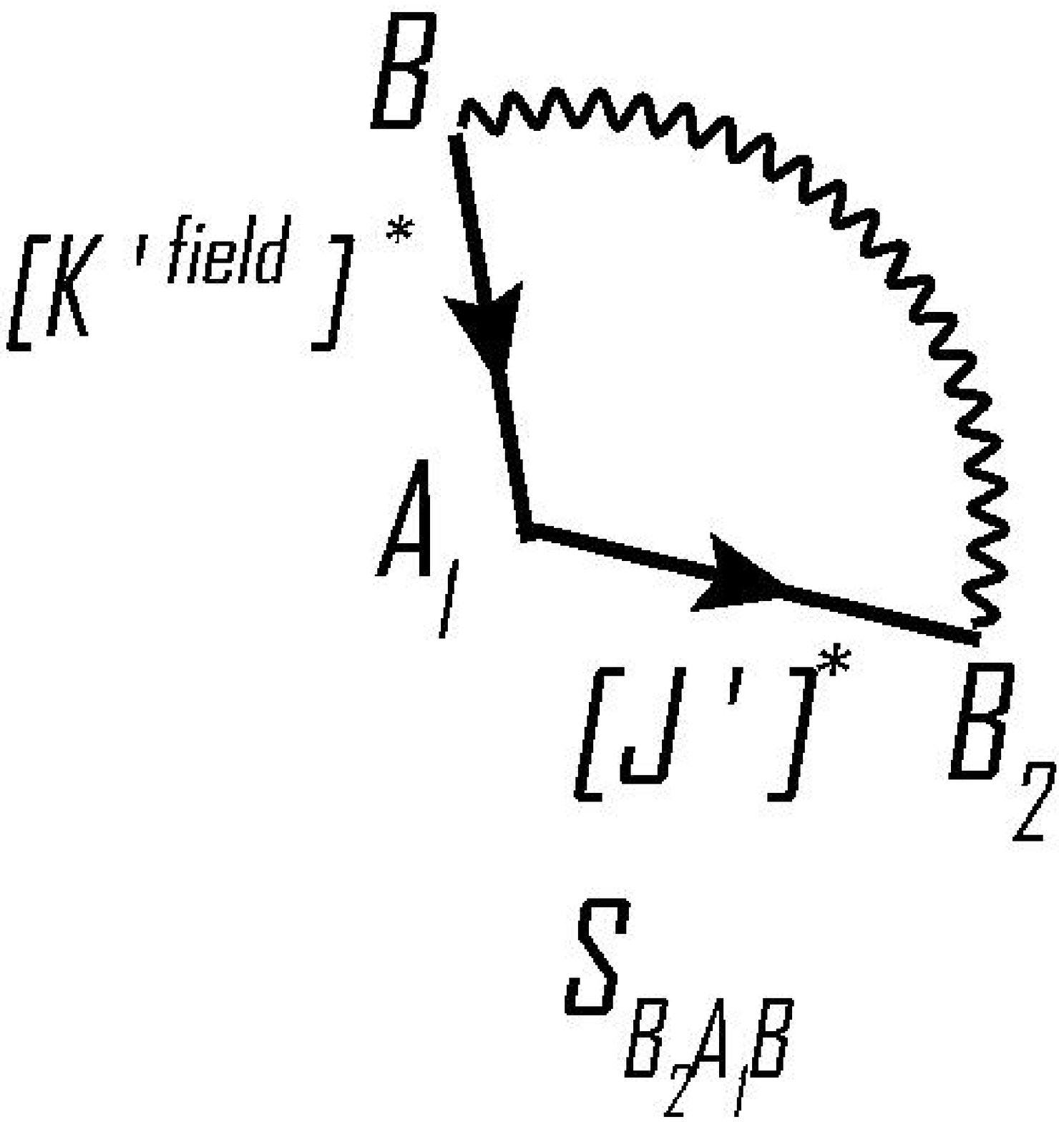}\label{SBAB1}}\hfill %
\subfloat[]{\includegraphics[scale=0.12]{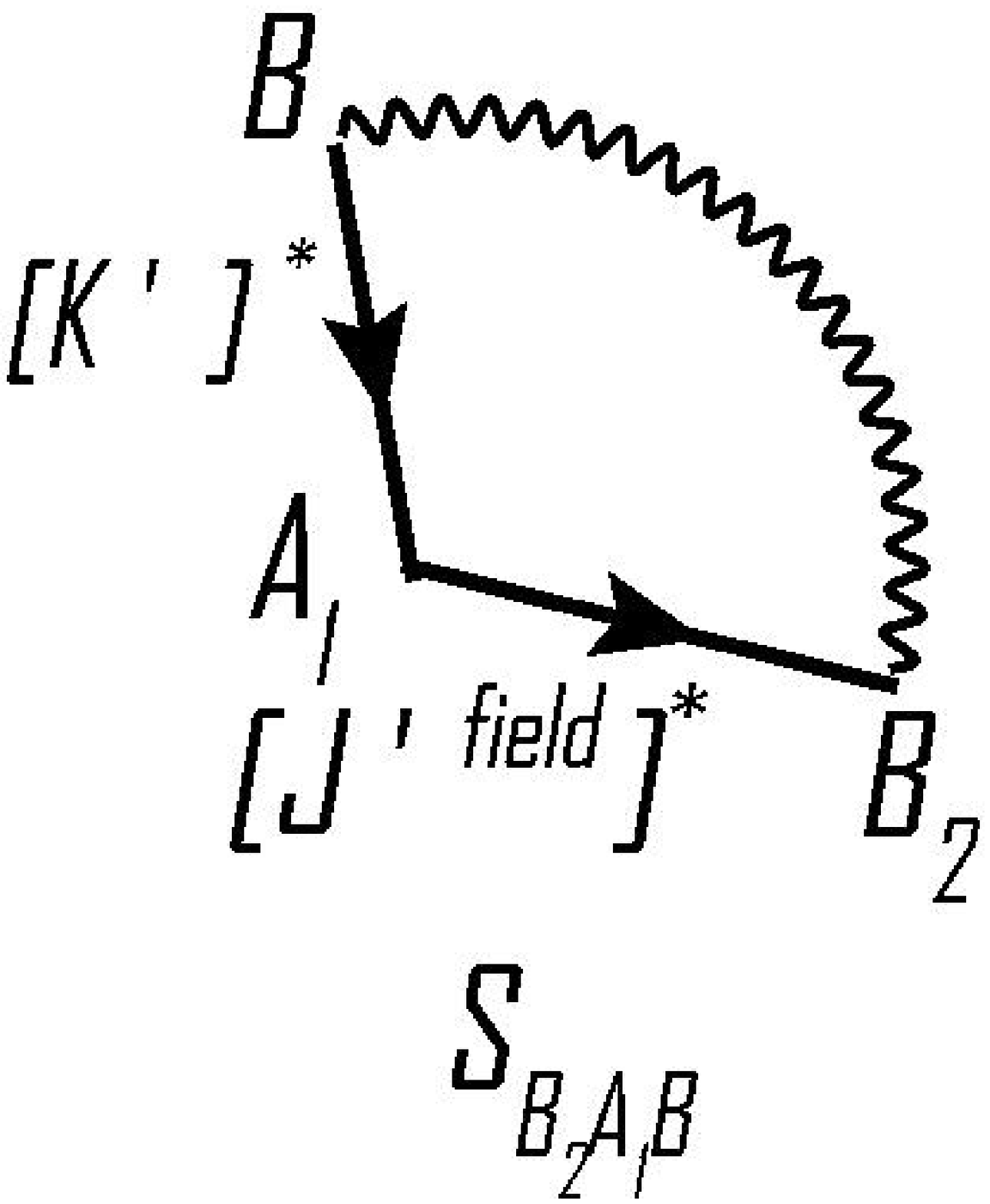}\label{SBAB2}}\hfill
\caption{Initial state is extended state $\protect\xi_{B}$: order$K^{\prime}$%
, $[J^{\prime}]^{1}$ and order $J^{1}$ contributions to conductivity}
\label{SBA}
\end{figure}

\begin{figure}[th]
\centering
\subfloat[]{\includegraphics[scale=0.12]{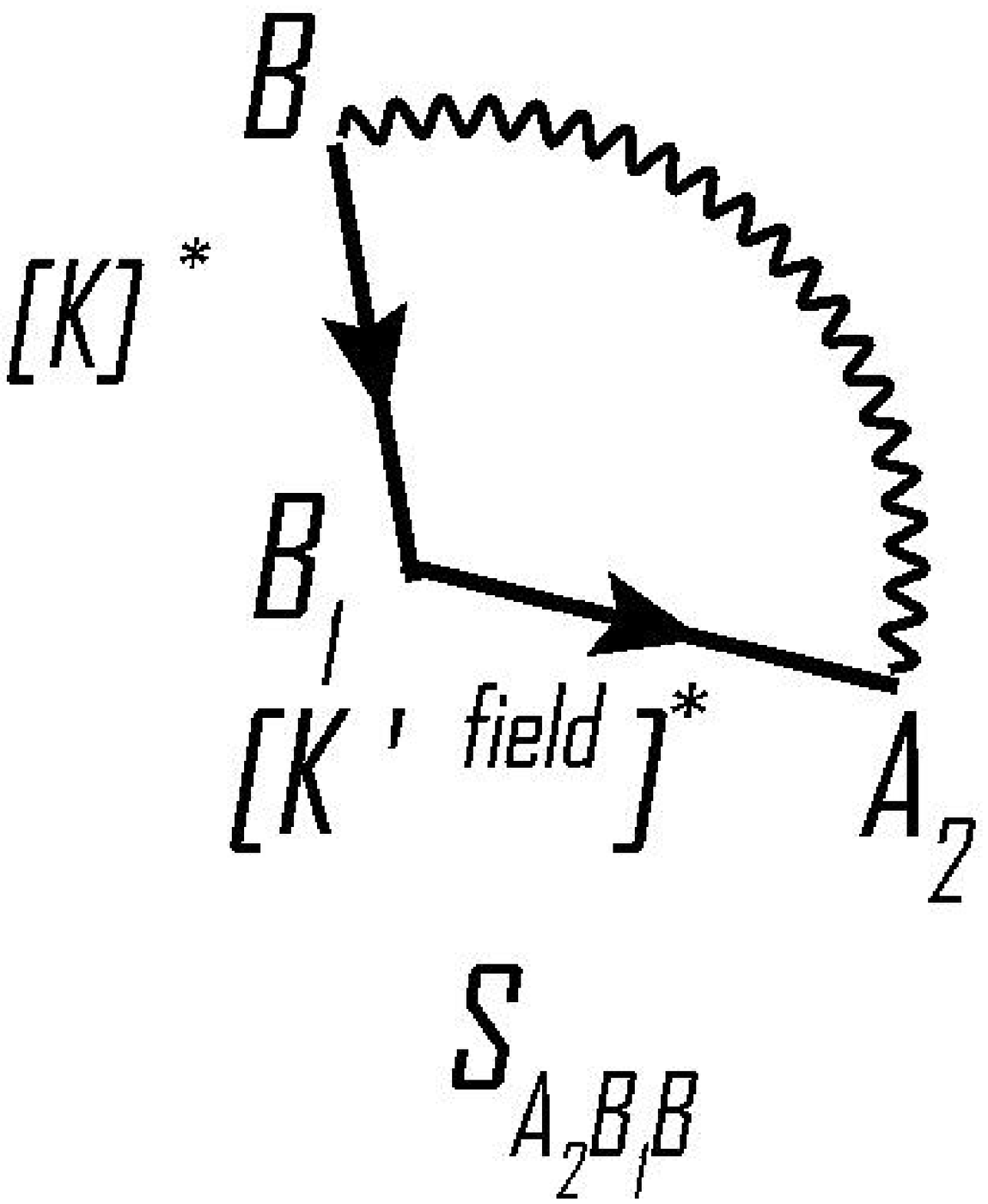}\label{SBBA1}}\hfill %
\subfloat[]{\includegraphics[scale=0.12]{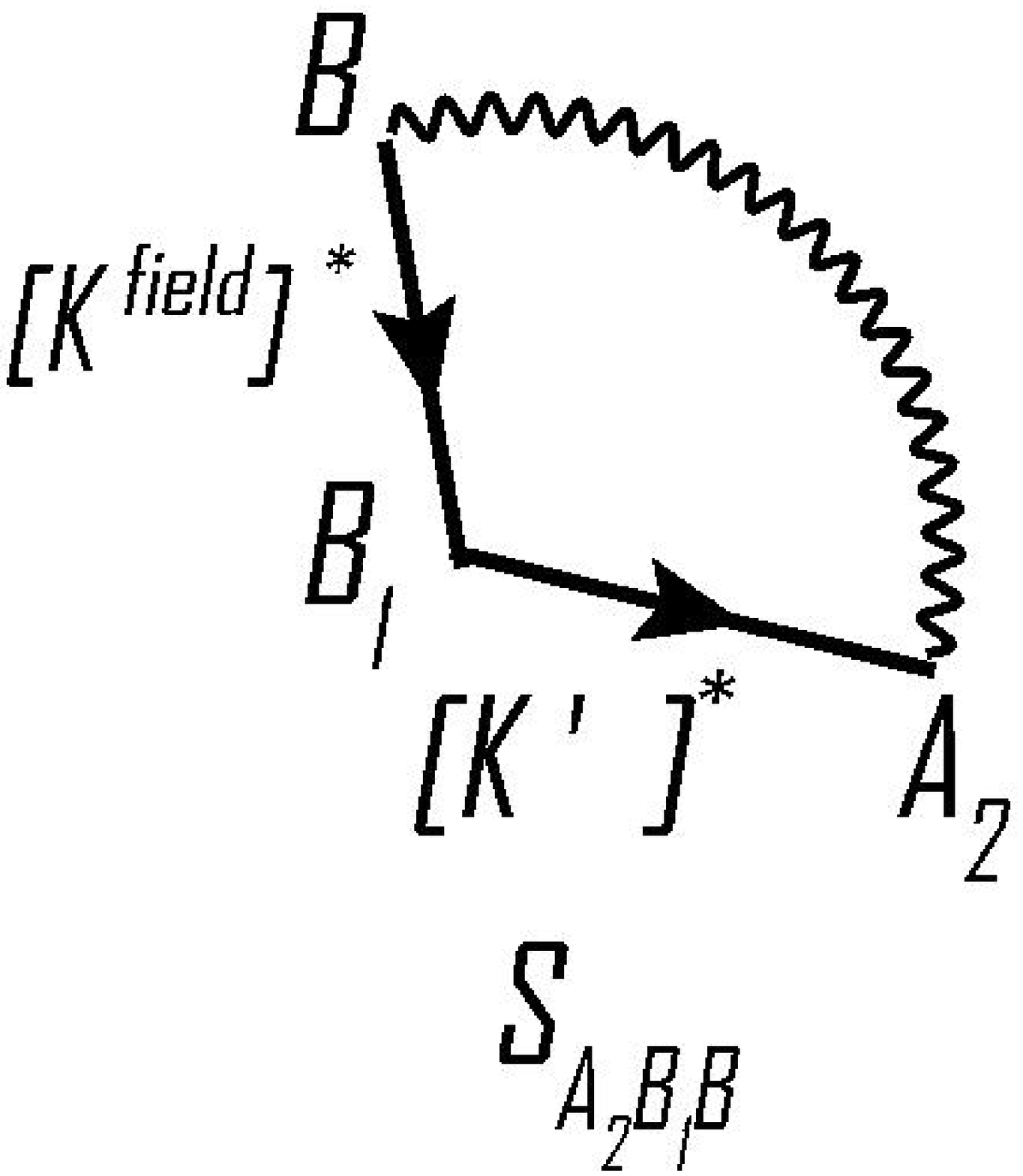}\label{SBBA2}}\hfill 
\subfloat[]{\includegraphics[scale=0.12]{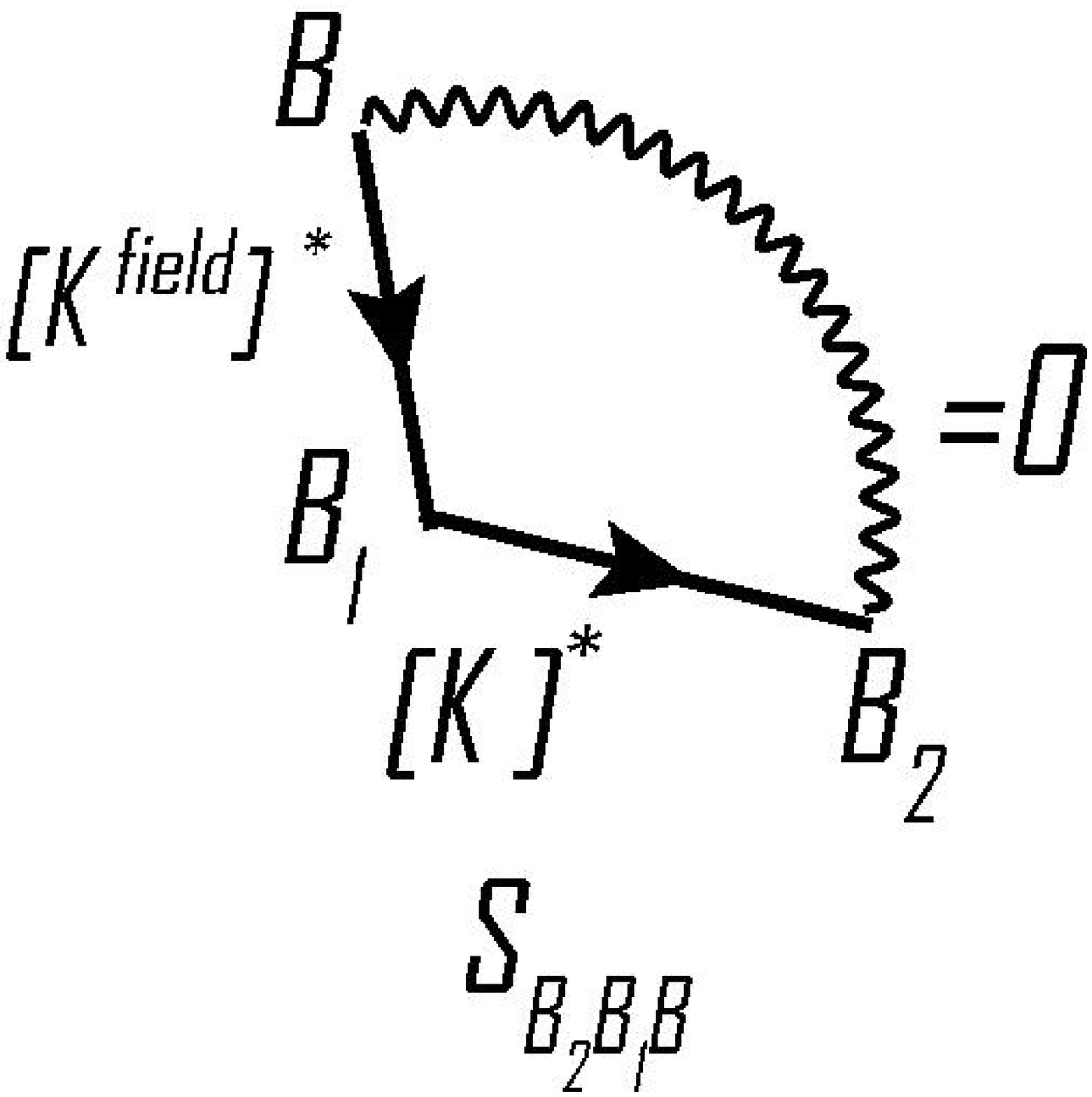}\label{SBBB1}}\hfill %
\subfloat[]{\includegraphics[scale=0.12]{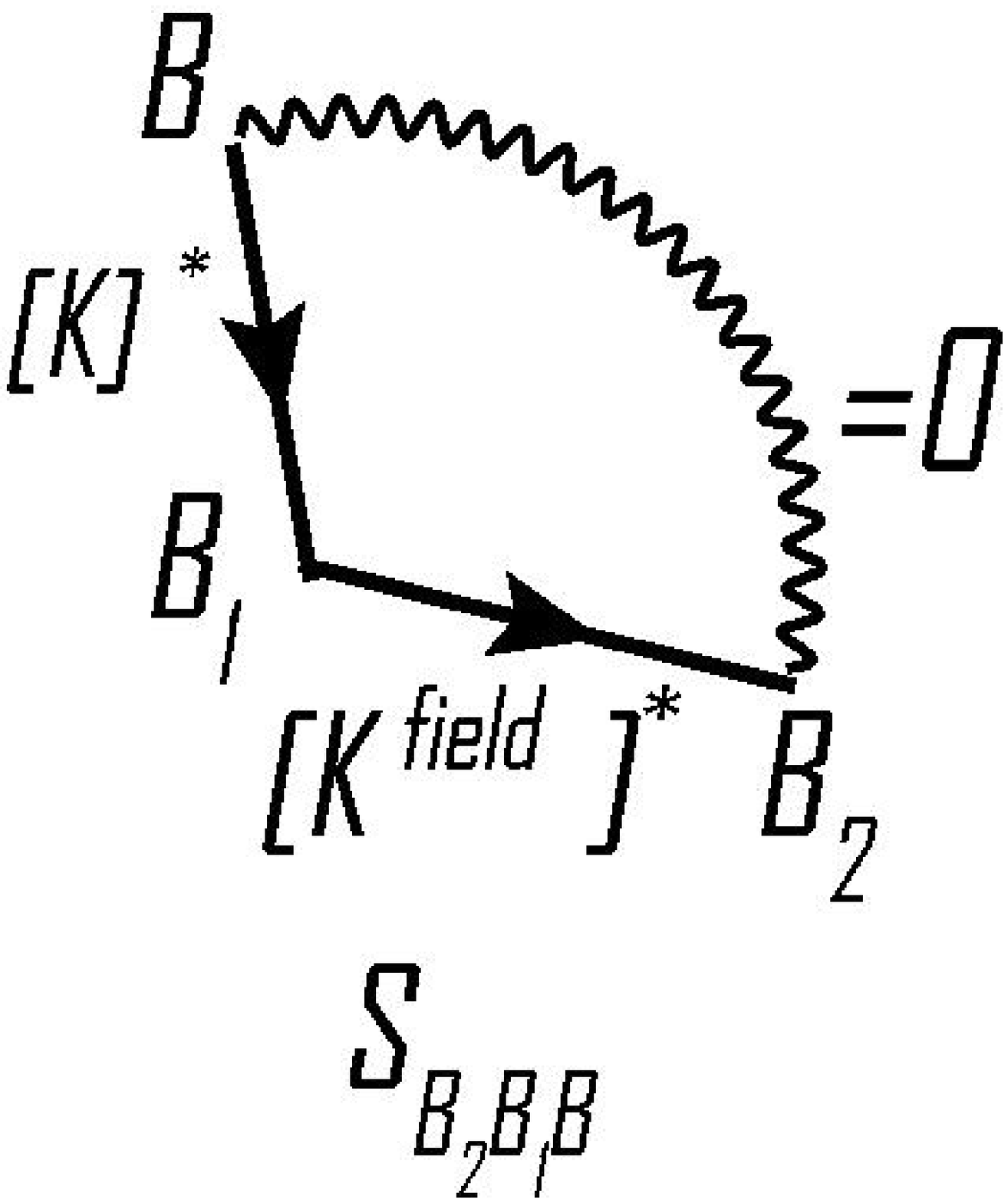}\label{SBBB2}}\hfill
\caption{Initial state is extended state $\protect\xi_{B}$: order $%
[K^{\prime}]^{1}$ and order $K^{1}$ contributions to conductivity}
\label{SBB}
\end{figure}

\begin{figure}[th]
\centering
\subfloat[]{\includegraphics[scale=0.12]{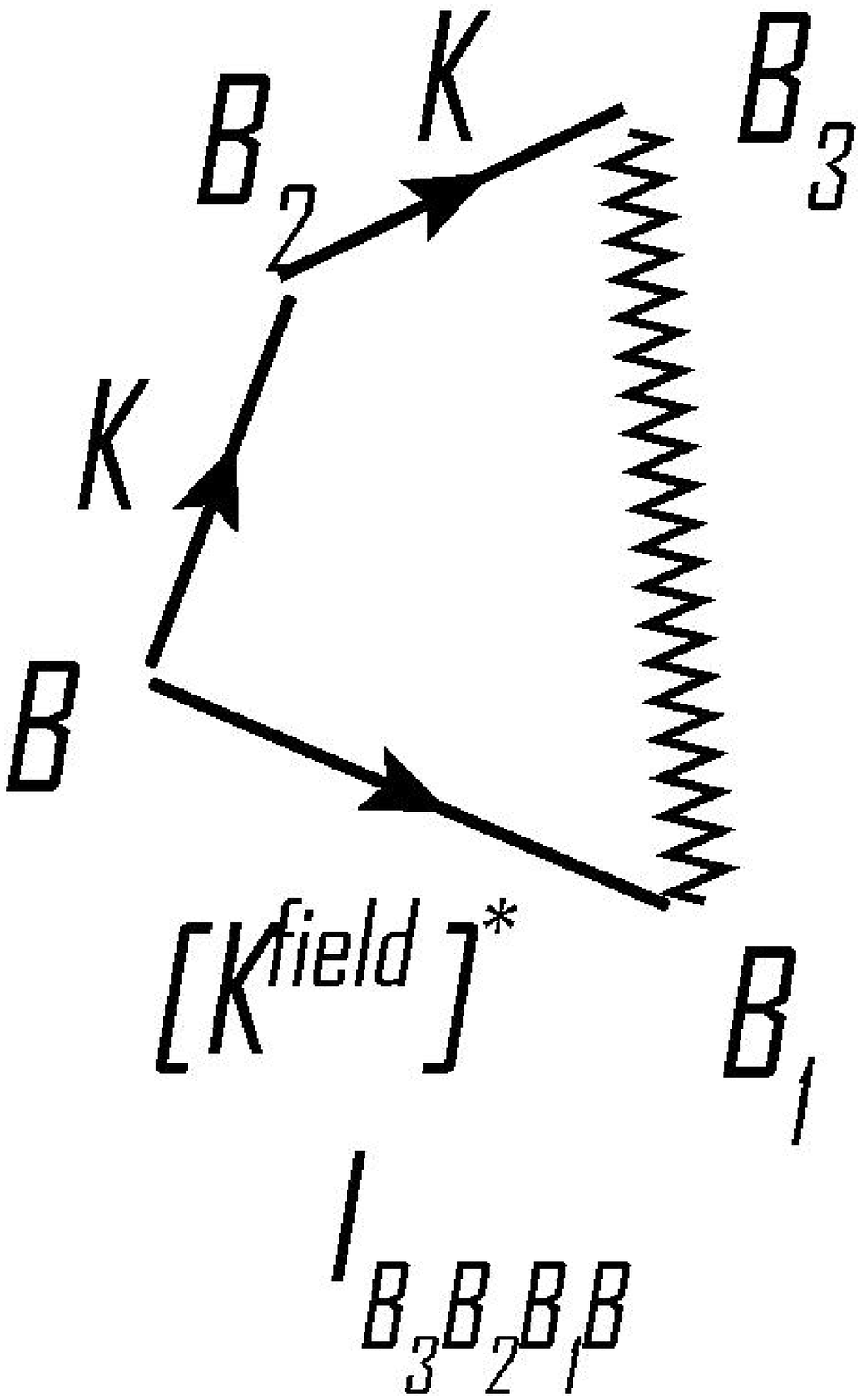}\label{B7}}\hfill
\subfloat[]{\includegraphics[scale=0.12]{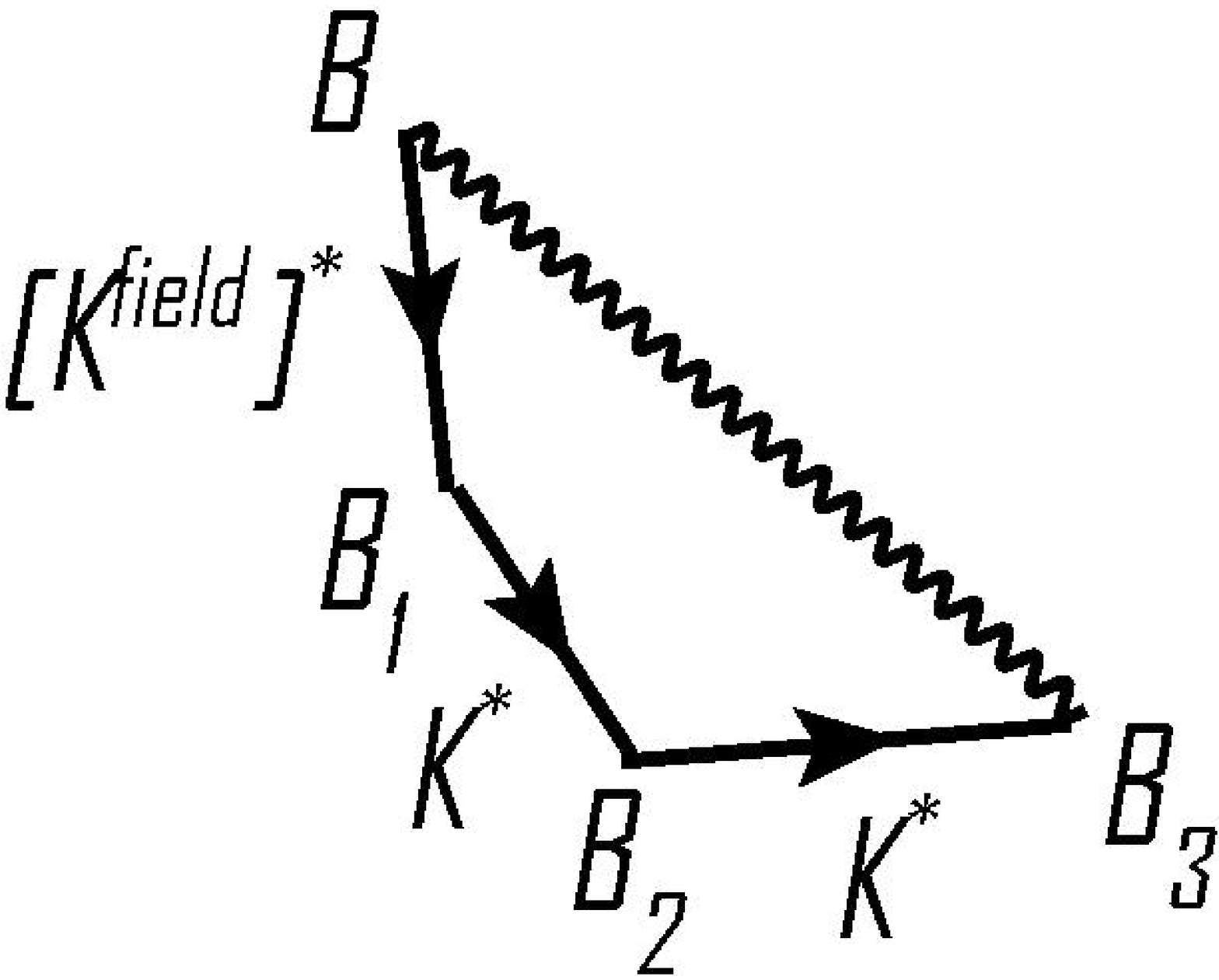}\label{fKKT}}\hfill\newline
\subfloat[]{\includegraphics[scale=0.12]{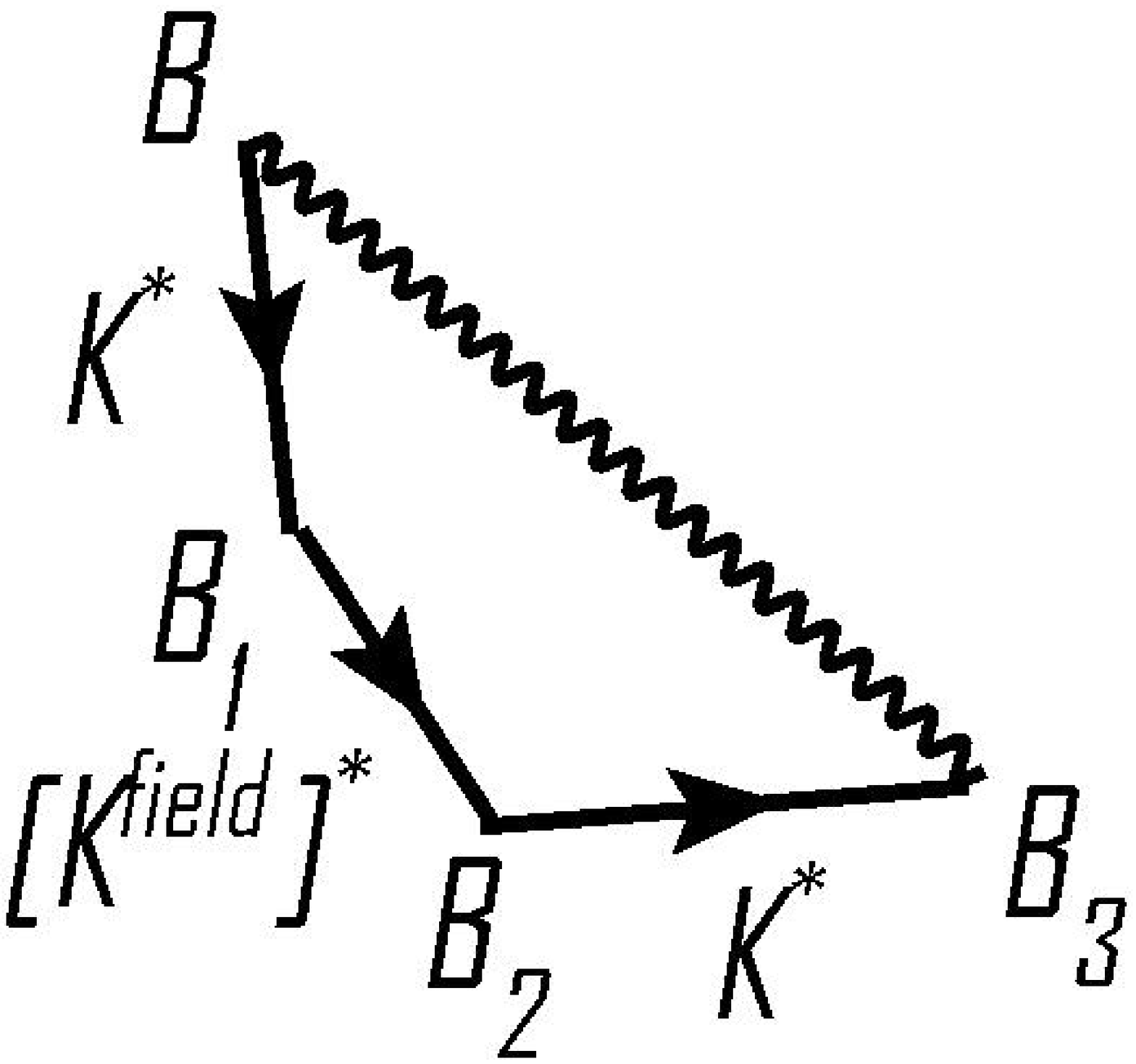}\label{KfKT}}\hfill
\subfloat[]{\includegraphics[scale=0.12]{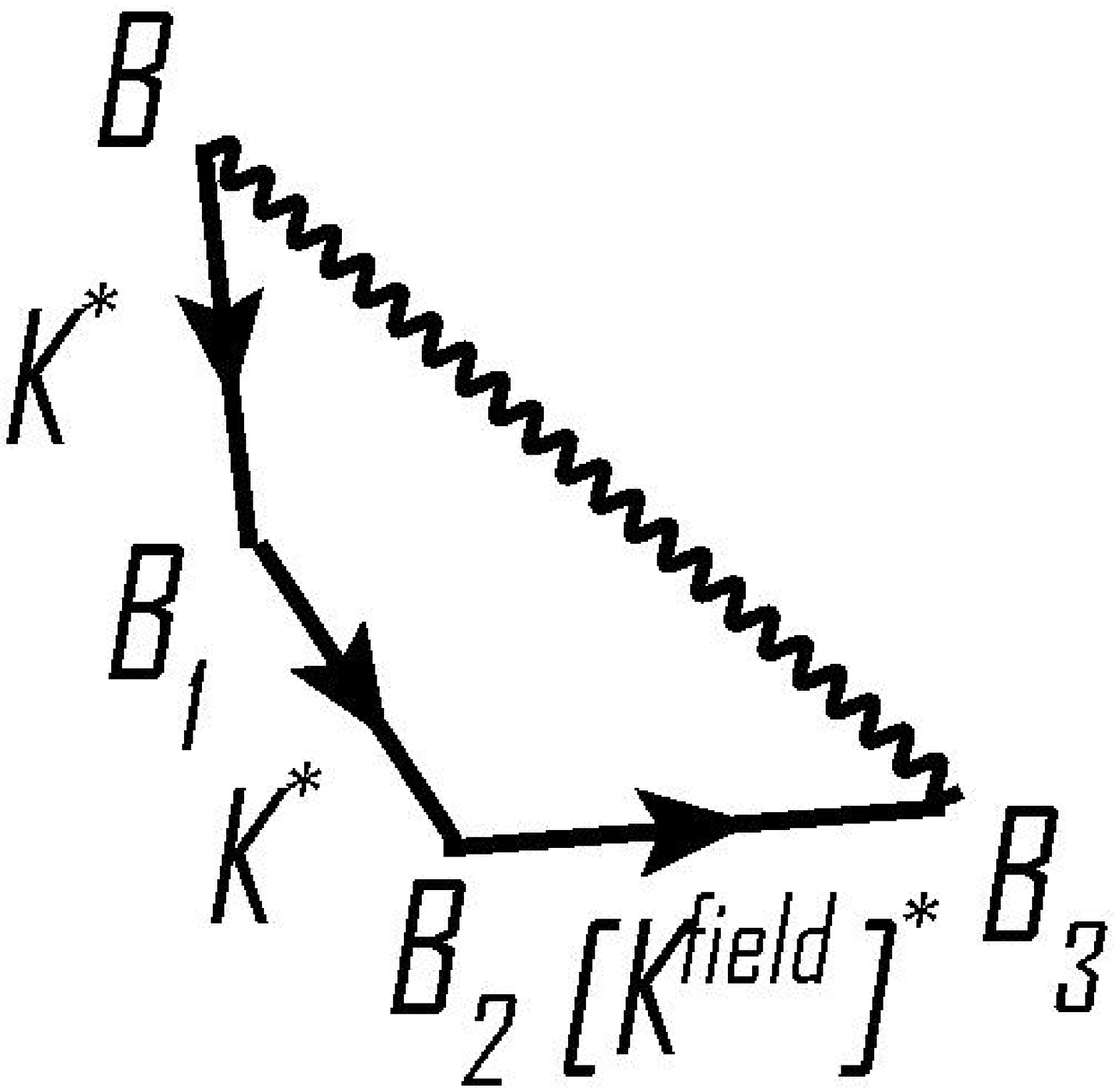}\label{KKfT}\hfill
} 
\caption{Initial state is extended state $\protect\xi_{B}$: order $K^{2}$
contribution to conductivity}
\label{2EE}
\end{figure}

If the initial state is an extended state $\xi_{B}$, substituting the corresponding $\psi^{(0)}(t)$ and $\psi^{(1)}(t)$ into the first term of Eq.(\ref{1e1}), one can similarly derive the
corresponding relation between the contribution to $\mathbf{j}$ and the
diagram of a conduction process, see tables \ref{current} and \ref{FHD}. There are 14
processes contributing to the conductivity, cf. Fig.\ref{B1B2}-\ref{SBB}.
However three order K$^{1}$ EE transition processes (Fig.\ref{B6}, \ref%
{SBBB1}, \ref{SBBB2}) are zero due to the e-ph interaction selection rule.
For processes involving only extended states, order K$^{2}$ processes are
the first nonzero contributions. The first term in Eq.(\ref{1e1}) indicates
there are 4 such terms (Fig.\ref{2EE}). Thus if the initial state is an extended state $\xi_{B}$, to lowest order self-consistent approximation, there are 15 conduction processes. The corresponding expressions for conductivity are listed in table \ref{Econ}. The time integrals are given in Sec. 3.2 of Ref.\cite{sup}.

For EE transition, the order $K^{1}$ contributions to conduction
are zero (Fig.\ref{B6}, \ref{SBBB1}, \ref{SBBB2}), so that one has to take
into account order $K^{2}$ contribution (Fig.\ref{B7}, \ref{fKKT}, \ref{KfKT}%
, \ref{KKfT}). The time integrals are given in Sec. 3.3 of Ref.\cite{sup}.
 The conductivity from the pure EE transitions derives from the second (Fig.\ref{B2}: carriers are scattering by the static disorder), and the thirteenth to the sixteenth terms (Fig.\ref{2EE}: the carriers in extended states are inelastically scattered by phonons) in table \ref{Econ}. This is consistent with the Boltzmann theory: in the lowest
order approximation, the scattering probability is second order in the e-ph
coupling constant $K$ while the distribution function $f^{(0)}$ is order $%
K^{0}$ (the non-interacting elementary excitations)\cite{pita}. Thus the
collision integral is proportional to K$^{2}$ and the change $f^{(1)}$ in
distribution function and the resulting conductivity is proportional to K$%
^{2}$.
\begin{table*}
\caption{The conduction processes and conductivity: initial state is an extended state $\xi_{B}$.}
\resizebox{15cm}{!} {
\begin{tabular}[c]{lc}\hline\hline
diagram  & conductivity\\\hline
\ref{B1}  & $%
\begin{array}
[c]{c}%
{\large -}\frac{N_{e}e^{2}}{2\Omega_{s}}\sum_{BA_{2}}\operatorname{Im}%
{\large i(E}_{A_{2}}^{0}{\large -E}_{B}^{0}{\large )}^{-1}{\large v}_{BA_{2}%
}^{\beta}{\large (w}_{BA_{2}}^{\alpha}{\large -v}_{A_{2}B}^{\alpha}%
{\large )}\\
{\large [I}_{A_{2}B+}{\large (\omega)\pm I}_{A_{2}B-}{\large (\omega
)][1-f(E}_{A_{2}}{\large )]f(E}_{B}{\large )}%
\end{array}
$\\\hline
\ref{B2}  & $%
\begin{array}
[c]{c}%
{\large -}\frac{N_{e}e^{2}}{2\Omega_{s}}\sum_{BB_{1}}\operatorname{Im}%
{\large i(E}_{B_{1}}^{0}{\large -E}_{B}^{0}{\large )}^{-1}{\large v}_{BB_{1}%
}^{\beta}{\large (w}_{BB_{1}}^{\alpha}{\large -v}_{B_{1}B}^{\alpha}%
{\large )}\\
{\large [I}_{B_{1}B+}{\large (\omega)\pm I}_{B_{1}B-}{\large (\omega
)][1-f(E}_{B_{1}}{\large )]f(E}_{B}{\large )}%
\end{array}
$\\\hline
\ref{B3}  & $%
\begin{array}
[c]{c}%
{\large -}\frac{N_{e}e^{2}}{2\hbar\Omega_{s}}\sum_{BA_{3}A_{2}}%
\operatorname{Im}{\large (E}_{A_{2}}^{0}{\large -E}_{B}^{0}{\large )}%
^{-1}{\large v}_{BA_{2}}^{\beta}{\large (w}_{A_{3}A_{2}}^{\alpha}%
{\large -v}_{A_{2}A_{3}}^{\alpha}{\large )[I}_{A_{3}A_{2}B+}{\large (\omega
)\pm I}_{A_{3}A_{2}B-}{\large (\omega)]}\\
{\large [1-f(E}_{A_{3}}{\large )][1-f(E}_{A_{2}}{\large )]f(E}_{B}{\large )}%
\end{array}
$\\\hline
\ref{B4}  & $%
\begin{array}
[c]{c}%
{\large -}\frac{N_{e}e^{2}}{2\hbar\Omega_{s}}\sum_{BB_{1}A_{3}}%
\operatorname{Im}{\large (E}_{B_{1}}^{0}{\large -E}_{B}^{0}{\large )}%
^{-1}{\large v}_{BB_{1}}^{\beta}{\large (w}_{A_{3}B_{1}}^{\alpha}%
{\large -v}_{B_{1}A_{3}}^{\alpha}{\large )[I}_{A_{3}B_{1}B+}{\large (\omega
)\pm I}_{A_{3}B_{1}B-}{\large (\omega)]}\\
{\large [1-f(E}_{A_{3}}{\large )][1-f(E}_{B_{1}}{\large )]f(E}_{B}{\large )}%
\end{array}
$\\\hline
\ref{B5}  & $%
\begin{array}
[c]{c}%
{\large -}\frac{N_{e}e^{2}}{2\hbar\Omega_{s}}\sum_{BB_{3}A_{2}}%
\operatorname{Im}{\large (E}_{A_{2}}^{0}{\large -E}_{B}^{0}{\large )}%
^{-1}{\large v}_{BA_{2}}^{\beta}{\large (w}_{B_{3}A_{2}}^{\alpha}%
{\large -v}_{A_{2}B_{3}}^{\alpha}{\large )[I}_{A_{2}B_{3}B+}{\large (\omega
)\pm I}_{A_{2}B_{3}B-}{\large (\omega)]}\\
{\large [1-f(E}_{A_{2}}{\large )][1-f(E}_{B_{3}}{\large )]f(E}_{B}{\large )}%
\end{array}
$\\\hline
\ref{SBAA1}  & $%
\begin{array}
[c]{c}%
{\large +}\frac{N_{e}e^{2}}{2\Omega_{\mathbf{s}}\hbar}\sum_{A_{2}A_{1}%
B}\operatorname{Im}{\large (w}_{BA_{2}}^{\alpha}{\large -v}_{A_{2}B}^{\alpha
}{\large )(E}_{A_{1}}^{0}{\large -E}_{A_{2}}^{0}{\large )}^{-1}{\large (v}%
_{A_{2}A_{1}}^{\beta}{\large )}^{\ast}\\
{\large (S}_{1A_{2}A_{1}B+}^{K^{\prime}}{\large \pm S}_{1A_{2}A_{1}%
B-}^{K^{\prime}}{\large )f(E}_{B}{\large )[1-f(E}_{A_{2}}{\large )]}%
\end{array}
$\\
\hline
\ref{SBAA2}  & $%
\begin{array}
[c]{c}%
{\large +}\frac{N_{e}e^{2}}{2\Omega_{\mathbf{s}}\hbar}\sum_{A_{2}A_{1}%
B}\operatorname{Im}{\large (w}_{BA_{2}}^{\alpha}{\large -v}_{A_{2}B}^{\alpha
}{\large )J}_{A_{2}A_{1}}^{\ast}{\large (E}_{B}^{0}{\large -E}_{A_{1}}%
^{0}{\large )}^{-1}{\large (v}_{A_{1}B}^{\beta}{\large )}^{\ast}\\
{\large (S}_{2A_{2}A_{1}B+}{\large \pm S}_{2A_{2}A_{1}B-}{\large )f(E}%
_{B}{\large )[1-f(E}_{A_{2}}{\large )]}%
\end{array}
$\\
\hline
\ref{SBAB1}  & $%
\begin{array}
[c]{c}%
{\large +}\frac{N_{e}e^{2}}{2\Omega_{\mathbf{s}}\hbar}\sum_{A_{2}A_{1}%
B}\operatorname{Im}{\large (w}_{BB_{2}}^{\alpha}{\large -v}_{B_{2}B}^{\alpha
}{\large )J}_{B_{2}A_{1}}^{\prime\ast}{\large (E}_{B}^{0}{\large -E}_{A_{1}%
}^{0}{\large )}^{-1}{\large (v}_{A_{1}B}^{\beta}{\large )}^{\ast}\\
{\large (S}_{1B_{2}A_{1}B+}{\large \pm S}_{1B_{2}A_{1}B-}{\large )f(E}%
_{B}{\large )[1-f(E}_{B_{2}}{\large )]}%
\end{array}
$\\
\hline
\ref{SBAB2}  & $%
\begin{array}
[c]{c}%
{\large +}\frac{N_{e}e^{2}}{2\Omega_{\mathbf{s}}\hbar}\sum_{B_{2}A_{1}%
B}\operatorname{Im}{\large (w}_{BB_{2}}^{\alpha}{\large -v}_{B_{2}B}^{\alpha
}{\large )(E}_{A_{1}}^{0}{\large -E}_{B_{2}}^{0}{\large )}^{-1}{\large (v}%
_{B_{2}A_{1}}^{\beta}{\large )}^{\ast}\\
{\large (S}_{2B_{2}A_{1}B+}^{K^{\prime}}{\large \pm S}_{2B_{2}A_{1}%
B-}^{K^{\prime}}{\large )f(E}_{B}{\large )[1-f(E}_{B_{2}}{\large )]}%
\end{array}
$\\
\hline
\ref{SBBA1}  & $%
\begin{array}
[c]{c}%
{\large +}\frac{N_{e}e^{2}}{2\Omega_{\mathbf{s}}\hbar}\sum_{B_{2}A_{1}%
B}\operatorname{Im}{\large (w}_{BA_{2}}^{\alpha}{\large -v}_{A_{2}B}^{\alpha
}{\large )(E}_{B_{1}}^{0}{\large -E}_{A_{2}}^{0}{\large )}^{-1}{\large (v}%
_{A_{2}B_{1}}^{\beta}{\large )}^{\ast}\\
{\large (S}_{1A_{2}B_{1}B+}^{K}{\large \pm S}_{1A_{2}B_{1}B-}^{K}%
{\large )f(E}_{B}{\large )[1-f(E}_{A_{2}}{\large )]}%
\end{array}
$\\%
\hline
\ref{SBBA2} & $%
\begin{array}
[c]{c}%
{\large +}\frac{N_{e}e^{2}}{2\Omega_{\mathbf{s}}\hbar}\sum_{A_{2}B_{1}%
B}\operatorname{Im}{\large (w}_{BA_{2}}^{\alpha}{\large -v}_{A_{2}B}^{\alpha
}{\large )(E}_{B}^{0}{\large -E}_{B_{1}}^{0}{\large )}^{-1}{\large (v}%
_{B_{1}B}^{\beta}{\large )}^{\ast}\\
{\large (S}_{2A_{2}B_{1}B+}^{K^{\prime}}{\large \pm S}_{2A_{2}B_{1}%
B-}^{K^{\prime}}{\large )f(E}_{B}{\large )[1-f(E}_{A_{2}}{\large )]}%
\end{array}
$\\
\hline\hline
\ref{B7}  & $%
\begin{array}
[c]{c}%
{\large +}\frac{N_{e}e^{2}}{2\Omega_{s}\hbar^{2}}\sum_{BB_{1}B_{2}B_{3}%
}\operatorname{Im}{\large (w}_{B_{3}B_{1}}^{\alpha}{\large -v}_{B_{1}B_{3}%
}^{\alpha}{\large )i(E}_{B}^{0}{\large -E}_{B_{1}}^{0}{\large )}^{-1}\\
{\large (v}_{B_{1}B}^{\beta}{\large )}^{\ast}{\large [I}_{B_{3}B_{2}B_{1}%
B+}{\large (\omega)\pm I}_{B_{3}B_{2}B_{1}B-}{\large (\omega)]}\\
{\large [1-f(E}_{B_{1}}{\large )][1-f(E}_{B_{3}}{\large )]f(E}_{B}{\large )}%
\end{array}
$\\
\hline
\ref{fKKT}  & $%
\begin{array}
[c]{c}%
{\large -}\frac{N_{e}e^{2}}{2\Omega_{\mathbf{s}}\hbar^{2}}\sum_{B_{3}%
B_{2}B_{1}B}\operatorname{Im}{\large (w}_{BB_{3}}^{\alpha}{\large -v}_{B_{3}%
B}^{\alpha}{\large )i(E}_{B}^{0}{\large -E}_{B_{1}}^{0}{\large )}%
^{-1}{\large (v}_{B_{1}B}^{\beta}{\large )}^{\ast}\\
{\large (K}_{1B_{3}B_{2}B_{1}B+}{\large \pm K}_{1B_{3}B_{2}B_{1}%
B-}{\large )f(E}_{B}{\large )[1-f(E}_{B_{3}}{\large )]}%
\end{array}
$\\
\hline
\ref{KfKT}  & $%
\begin{array}
[c]{c}%
{\large -}\frac{N_{e}e^{2}}{2\Omega_{\mathbf{s}}\hbar^{2}}\sum_{B_{3}%
B_{2}B_{1}B}\operatorname{Im}{\large (w}_{BB_{3}}^{\alpha}{\large -v}_{B_{3}%
B}^{\alpha}{\large )i(E}_{B_{1}}^{0}{\large -E}_{B_{2}}^{0}{\large )}%
^{-1}{\large (v}_{B_{2}B_{1}}^{\beta}{\large )}^{\ast}\\
{\large (K}_{2B_{3}B_{2}B_{1}B+}{\large \pm K}_{2B_{3}B_{2}B_{1}%
B-}{\large )f(E}_{B}{\large )[1-f(E}_{B_{3}}{\large )]}%
\end{array}
$\\
\hline
\ref{KKfT}  & $%
\begin{array}
[c]{c}%
{\large -}\frac{N_{e}e^{2}}{2\Omega_{\mathbf{s}}\hbar^{2}}\sum_{B_{3}%
B_{2}B_{1}B}\operatorname{Im}{\large (w}_{BB_{3}}^{\alpha}{\large -v}_{B_{3}%
B}^{\alpha}{\large )i(E}_{B_{2}}^{0}{\large -E}_{B_{3}}^{0}{\large )}%
^{-1}{\large (v}_{B_{3}B_{2}}^{\beta}{\large )}^{\ast}\\
{\large (K}_{3B_{3}B_{2}B_{1}B+}{\large \pm K}_{3B_{3}B_{2}B_{1}%
B-}{\large )f(E}_{B}{\large )[1-f(E}_{B_{3}}{\large )]}%
\end{array}
$\\%
\hline\hline
\end{tabular}
}\label{Econ}
\end{table*}

The expression for the conductivity is a sum of 29 terms listed in tables \ref{Acon} and \ref{Econ}.


\section{Hall mobility}

\begin{table}[th]
\caption{Contributions to $\protect\sigma_{yx}$ from $\protect\psi^{(1)}%
\protect\nabla\protect\psi^{(1)\ast}$}
\label{11H}%
\resizebox{6cm}{!} {
\begin{tabular}
[c]{lll}\hline\hline
J$^{0}$ & $\psi_{J^{0}}^{(1)}\nabla\psi_{J^{0}}^{(1)\ast}$ & Fig.\ref{Up}\\
\hline
J & $\psi_{J^{0}}^{(1)}\nabla\psi_{J}^{(1)\ast}$ & Fig.\ref{Z1}, Fig.\ref{Z3}\\
\hline
J & $\psi_{J}^{(1)}\nabla\psi_{J^{0}}^{(1)\ast}$ & Fig.\ref{Zp1},
Fig.\ref{Zp2}\\
\hline\hline
\end{tabular}
}
\end{table}

\begin{table}[th]
\caption{Contribution to $\protect\sigma_{yx}$ from $(\protect\psi^{(0)}%
\protect\nabla\protect\psi^{(2)\ast }-\protect\psi^{(2)\ast}\protect\nabla%
\protect\psi^{(0)})$}
\label{02H}%
\resizebox{8cm}{!} {
\begin{tabular}
[c]{lll}\hline\hline
J$^{0}$ & $\psi_{J^{0}}^{(0)}\nabla\psi_{J^{0}}^{(2)\ast}-\psi_{J^{0}}^{(2)\ast}\nabla\psi_{J^{0}}^{(0)}$ & Fig.\ref{U}\\
\hline
J & $\psi_{J}^{(0)}\nabla\psi_{J^{0}}^{(2)\ast}-\psi_{J^{0}}^{(2)\ast}\nabla\psi_{J}^{(0)}$ & Fig.\ref{Z1}\\
\hline
J & $\psi_{J^{0}}^{(0)}\nabla\psi_{J}^{(2)\ast}-\psi_{J}^{(2)\ast}\nabla
\psi_{J^{0}}^{(0)}$ & Fig.\ref{Z3}, \ref{Zp1}, \ref{Zp2}\\
\hline\hline
\end{tabular}
}
\end{table}

\label{hall}If an amorphous semiconductor is placed in an external magnetic
field, the conductivity is still defined through the current density $%
j_{\alpha}=\sum_{\beta}\sigma_{\alpha\beta}E_{\beta}$. If we apply a static
electric field along the $x$ direction and a static magnetic field along the
$z$ direction, the Hall voltage is along the $y$ direction and proportional
to both $B_{z}$ and $E_{x}$. Amorphous semiconductors are isotropic, and the
Hall mobility is given by\cite{hol68}:
\begin{equation}
\mu_{H}=B_{z}^{-1}\sigma_{yx}/\sigma_{xx}.   \label{hm}
\end{equation}
The non-diagonal conductivity in a magnetic field is described by the second
term of Eq.(\ref{1e1}).

In general, both the carriers in localized states and carriers in
extended states contribute to the Hall effect. There are too many terms in
the full expression for non-diagonal conductivity to reproduce here. We
restrict ourselves to the LL transitions. This is a reasonable approximation
for intrinsic and lightly-doped amorphous semiconductors, where carriers in
extended states are rare. At the end of this Section, we will indicate how
to obtain the full expression of Hall mobility.

\begin{figure}[th]
\centering
\subfloat[]{\includegraphics[scale=0.2]{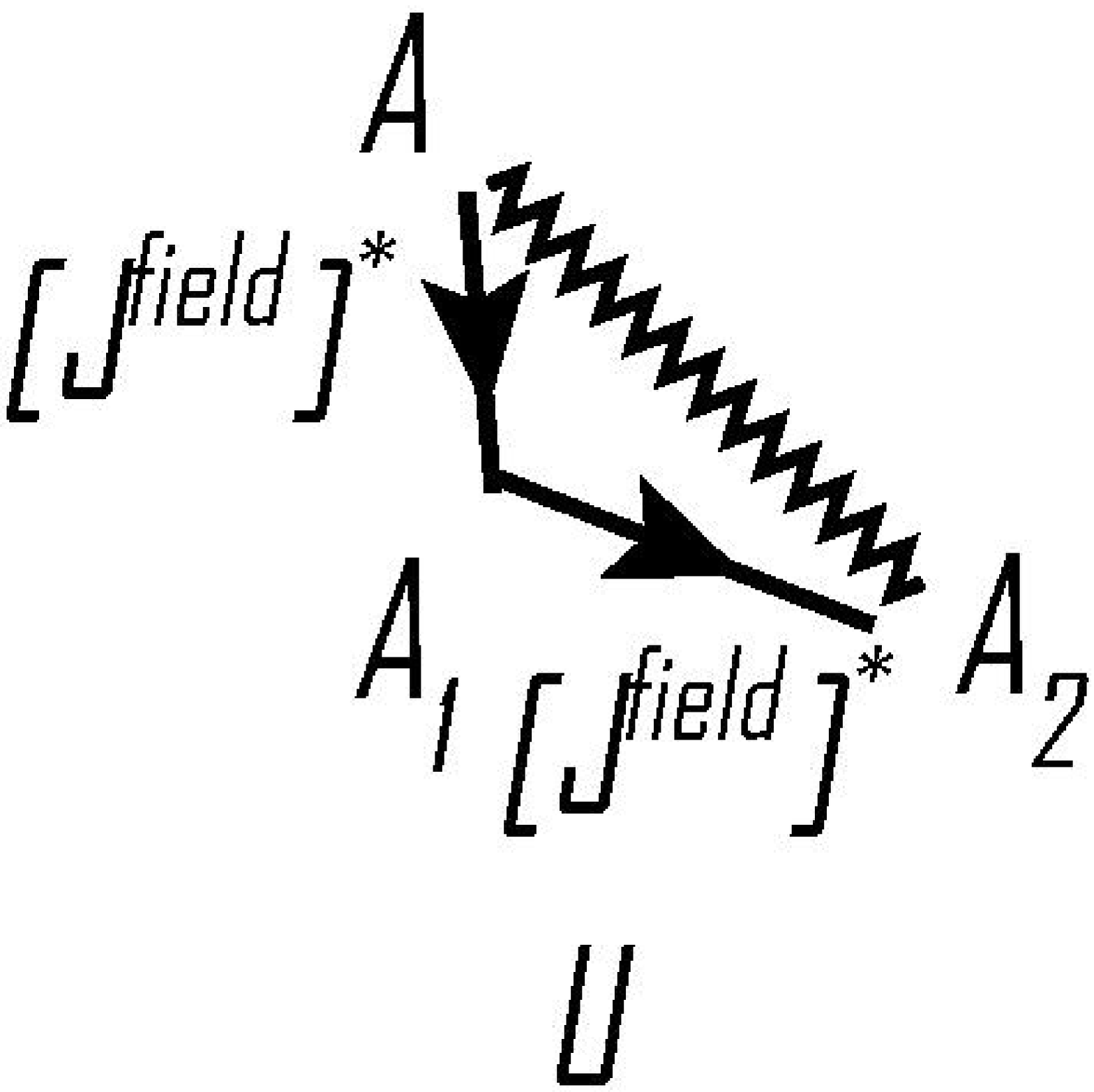}\label{U}}\hfill
\subfloat[]{\includegraphics[scale=0.2]{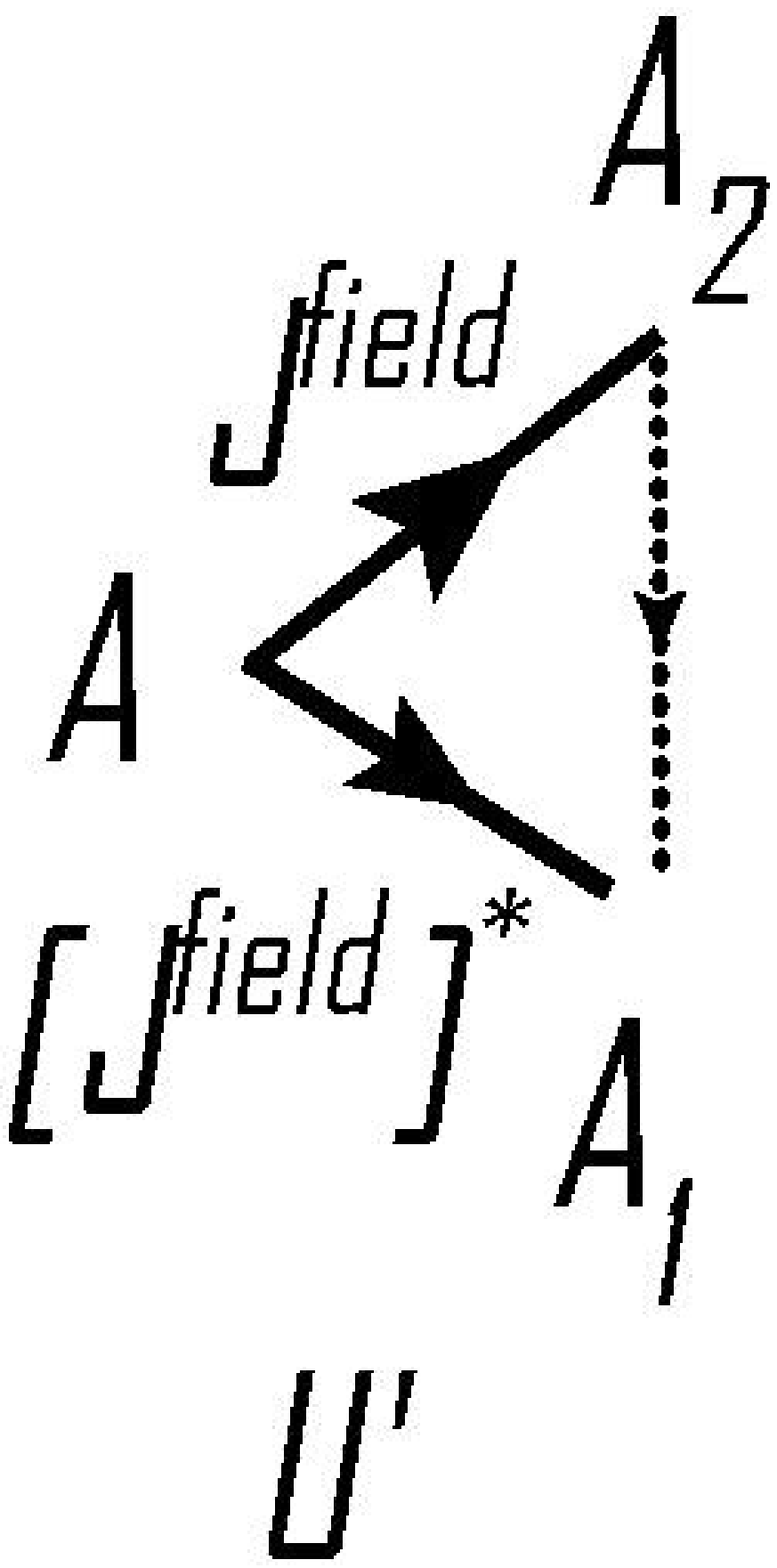}\label{Up}}\hfill
\caption{Order $J^{0}$ contributions to $\protect\sigma_{yx}$. \protect\ref%
{U} results from $(\protect\psi_{J^{0}}^{(0)}\protect\nabla\protect\psi%
_{J^{0}}^{(2)\ast}-\protect\psi_{J^{0}}^{(2)\ast}\protect\nabla\protect\psi%
_{J^{0}}^{(0)})$. \protect\ref{Up} results from $\protect\psi_{J^{0}}^{(1)}%
\protect\nabla\protect\psi_{J^{0}}^{(1)\ast}$.}
\label{UUp}
\end{figure}

\begin{figure}[th]
\centering
\subfloat[]{\includegraphics[scale=0.15]{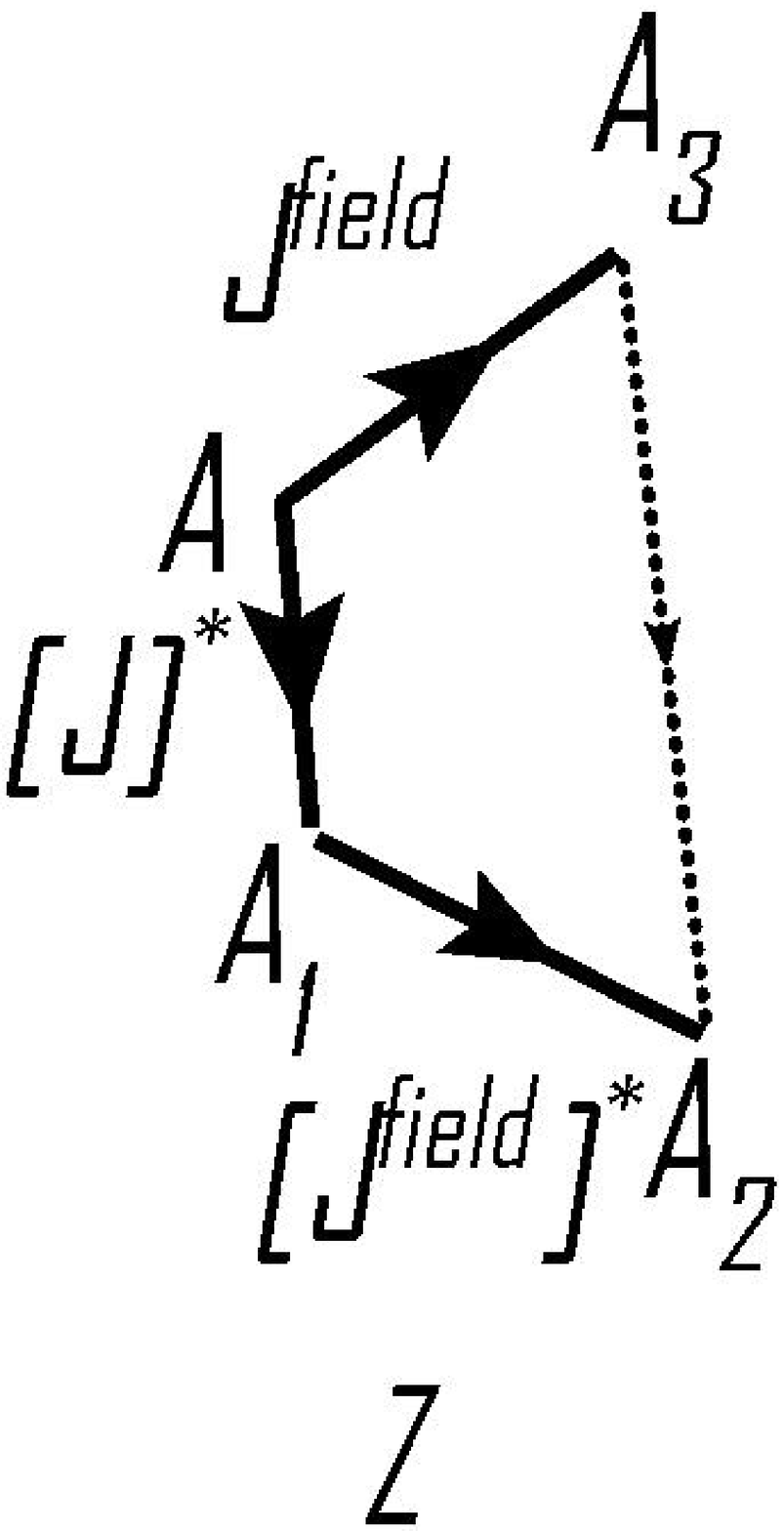}\label{Z1}}\hfill %
\subfloat[]{\includegraphics[scale=0.15]{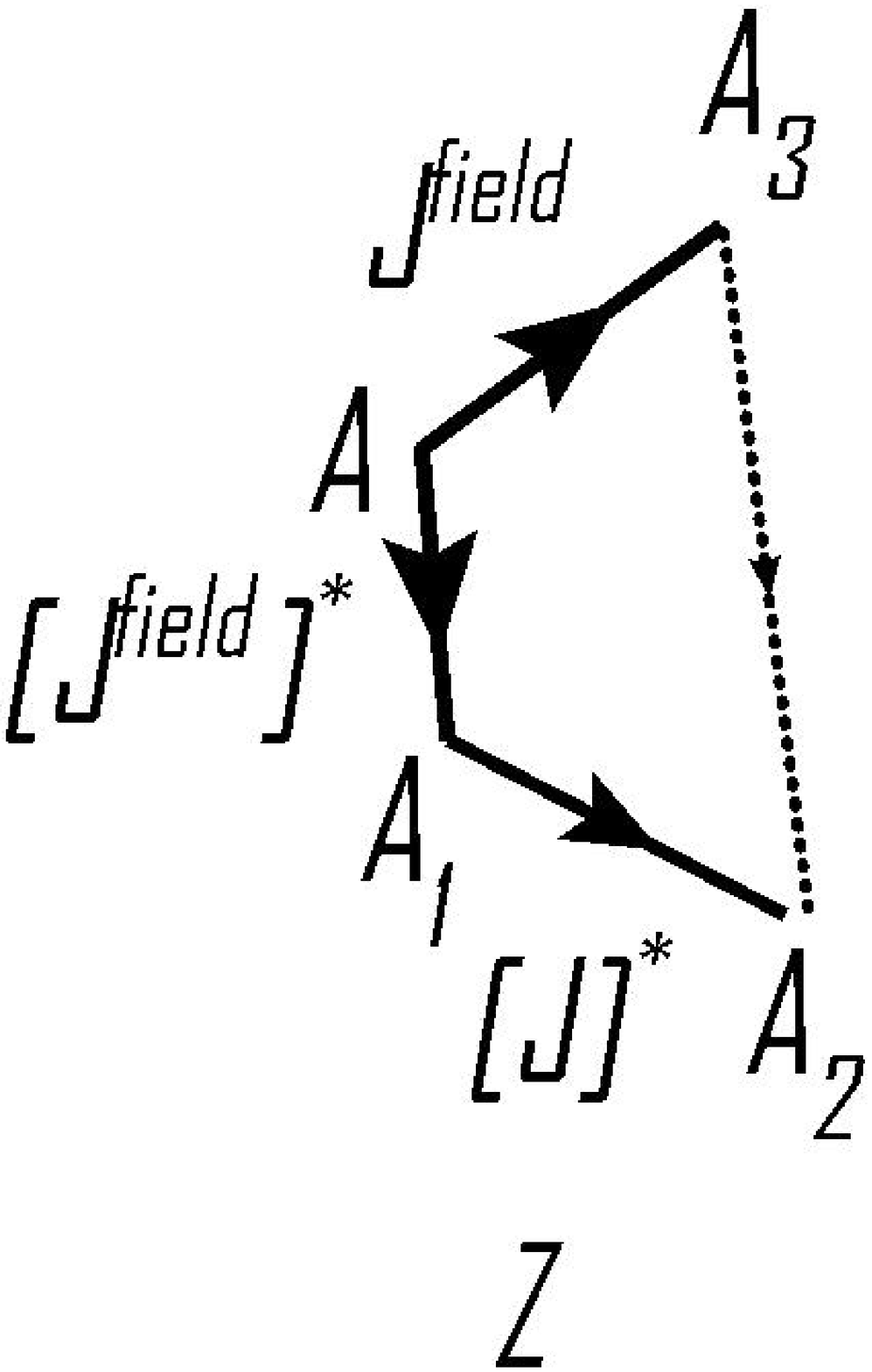}\label{Z3}}\hfill %
\subfloat[]{\includegraphics[scale=0.15]{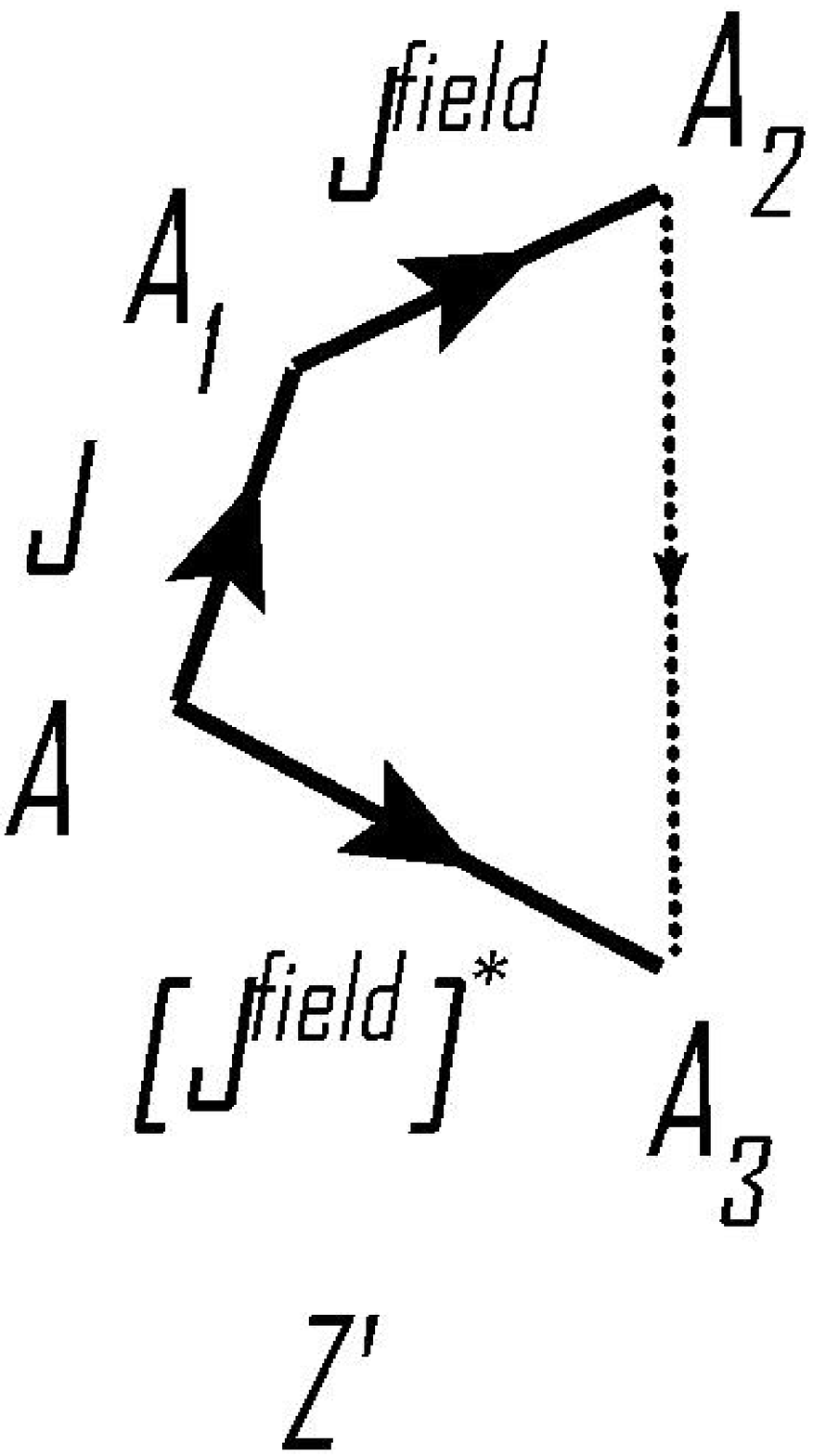}\label{Zp1}}\hfill %
\subfloat[]{\includegraphics[scale=0.15]{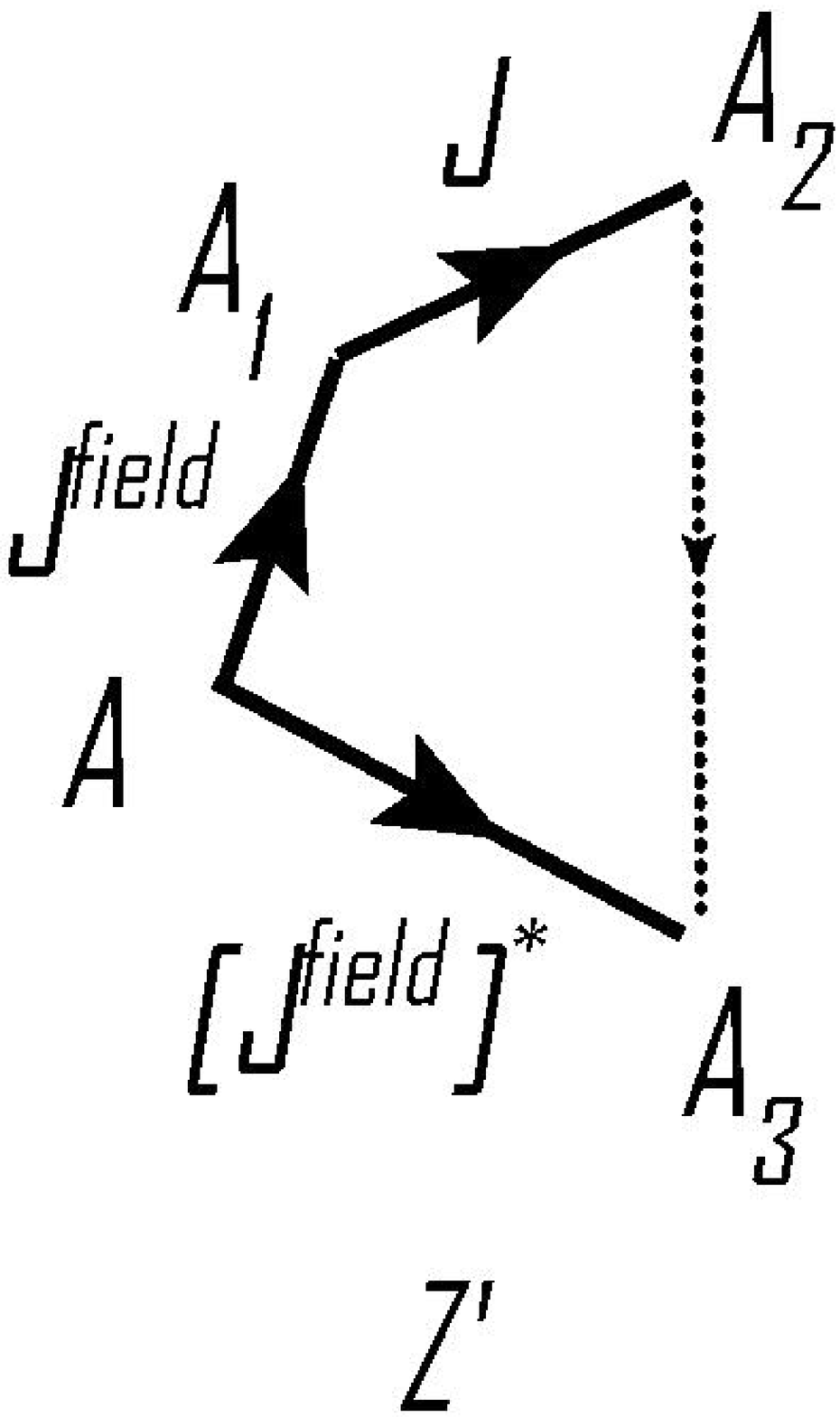}\label{Zp2}}\hfill
\caption{Order $J^{1}$ contributions to $\protect\sigma_{yx}$: \protect\ref%
{Z1} results from $\protect\psi_{J^{0}}^{(1)}\protect\nabla\{\protect\psi%
^{(1)}[$A $\protect\overset{J^{1}}{\rightarrow}$ A$_{1}$ $\protect\overset{%
J^{field}}{\rightarrow}$ A$_{2}]\}^{\ast}$. \protect\ref{Z3} results from $%
\protect\psi_{J^{0}}^{(1)}\protect\nabla\{\protect\psi^{(1)}[$A $%
\protect\overset{J^{field}}{\rightarrow}$ A$_{1}$ $\protect\overset{J^{1}}{%
\rightarrow}$ A$_{2}]\}^{\ast}$. \protect\ref{Zp1} results from $\protect\psi%
^{(1)}[$A $\protect\overset{J^{1}}{\rightarrow}$ A$_{1}$ $\protect\overset{%
J^{field}}{\rightarrow}$ A$_{2}]\protect\nabla\protect\psi_{J^{0}}^{(1)\ast}$%
. \protect\ref{Zp2} results from $\protect\psi^{(1)}$[A$\protect\overset{%
J^{field}}{\rightarrow}$$A_{1}$ $\protect\overset{J^{1}}{\rightarrow}$ A$_{2}
$]$\protect\nabla\protect\psi_{J^{0}}^{(1)\ast}$.}
\label{Zp1Zp2}
\end{figure}

\begin{figure}[th]
\centering
\par
\subfloat[]{\includegraphics[width=.2\linewidth,height=3cm]{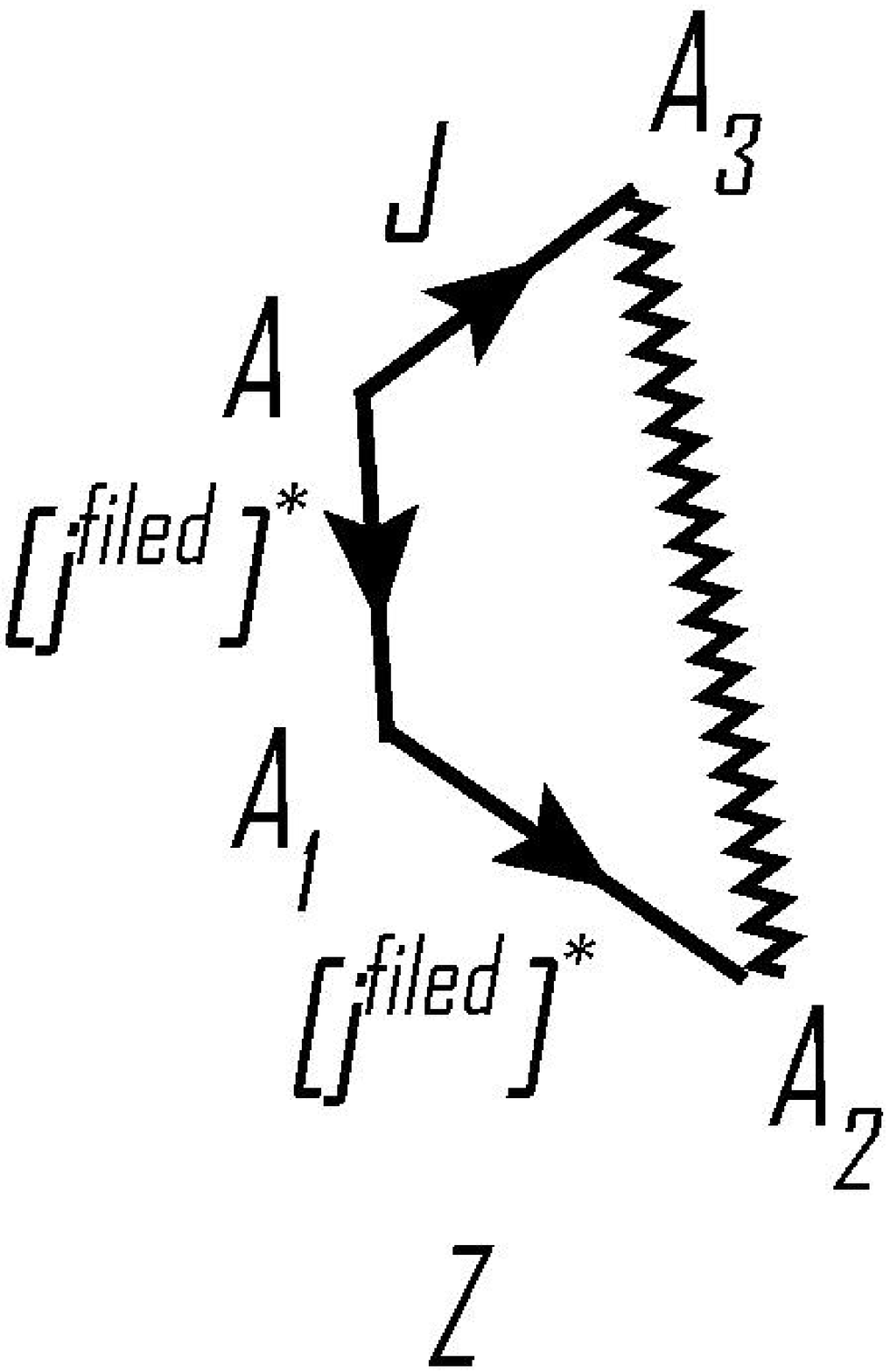}\label{Z2}}\hfill %
\subfloat[]{\includegraphics[width=.2\linewidth,height=2.5cm]{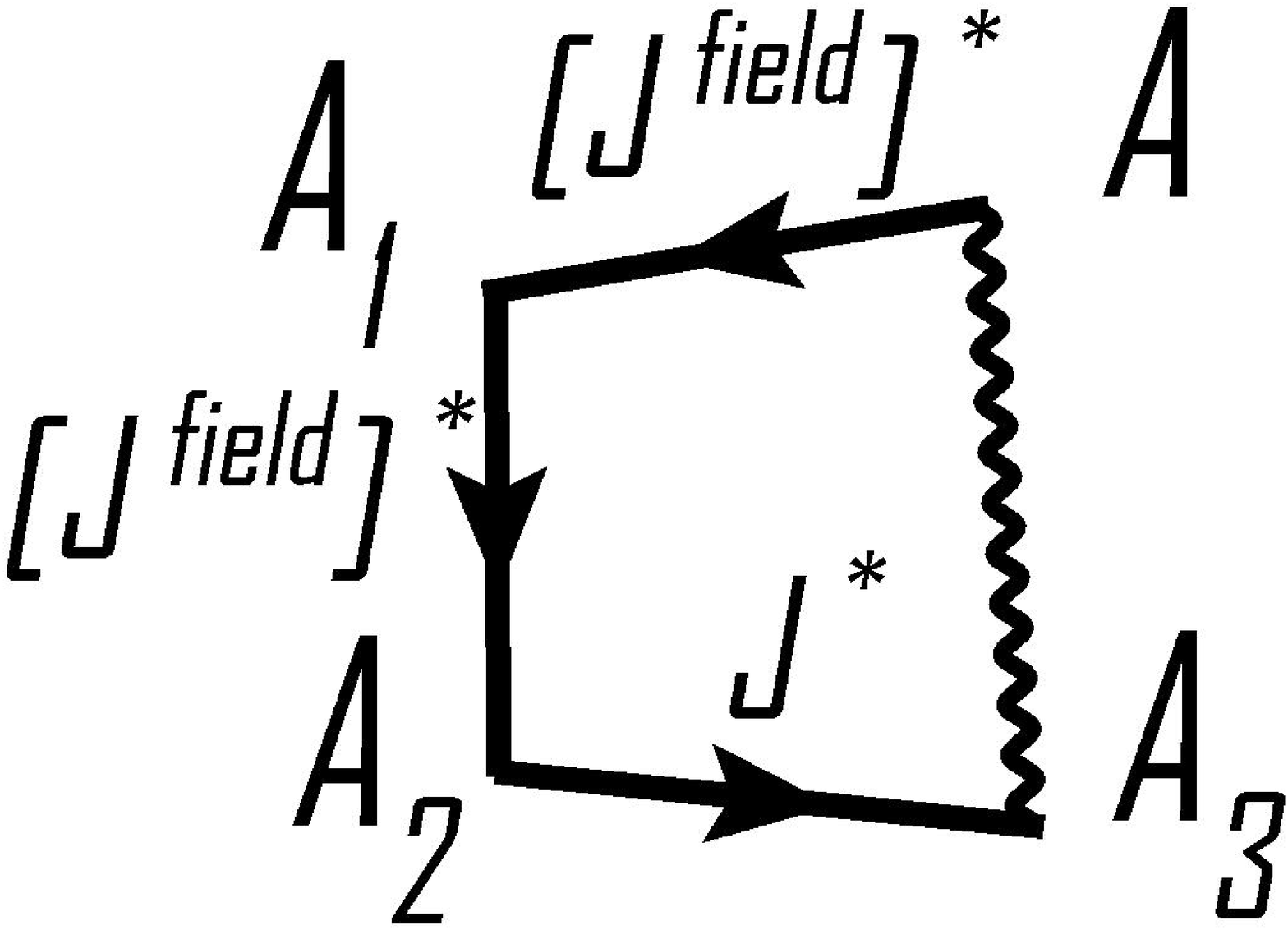}\label{ffJ}}\hfill
\subfloat[]{\includegraphics[width=.2\linewidth,height=2.5cm]{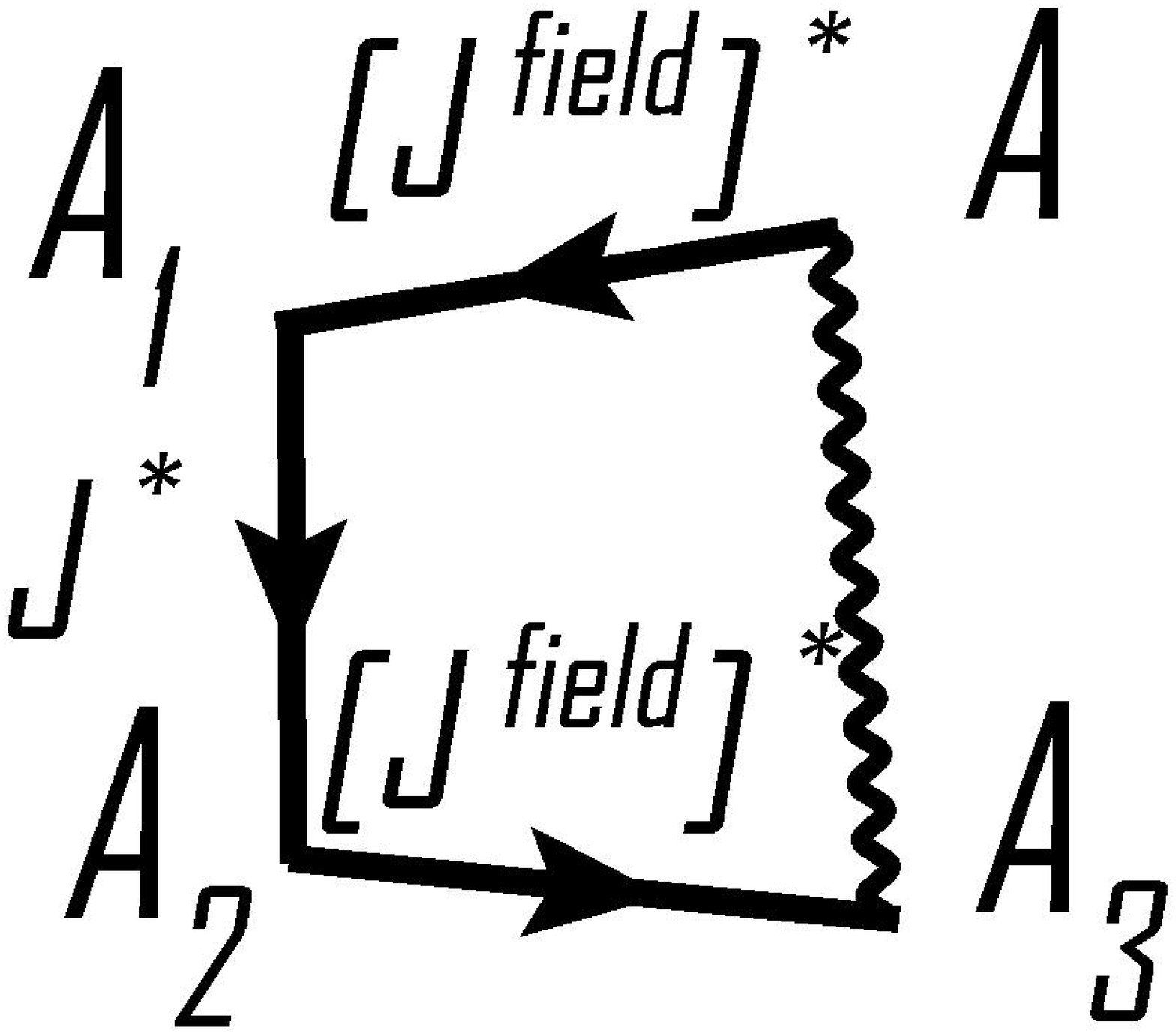}\label{fJf}}\hfill %
\subfloat[]{\includegraphics[width=.2\linewidth,height=2.5cm]{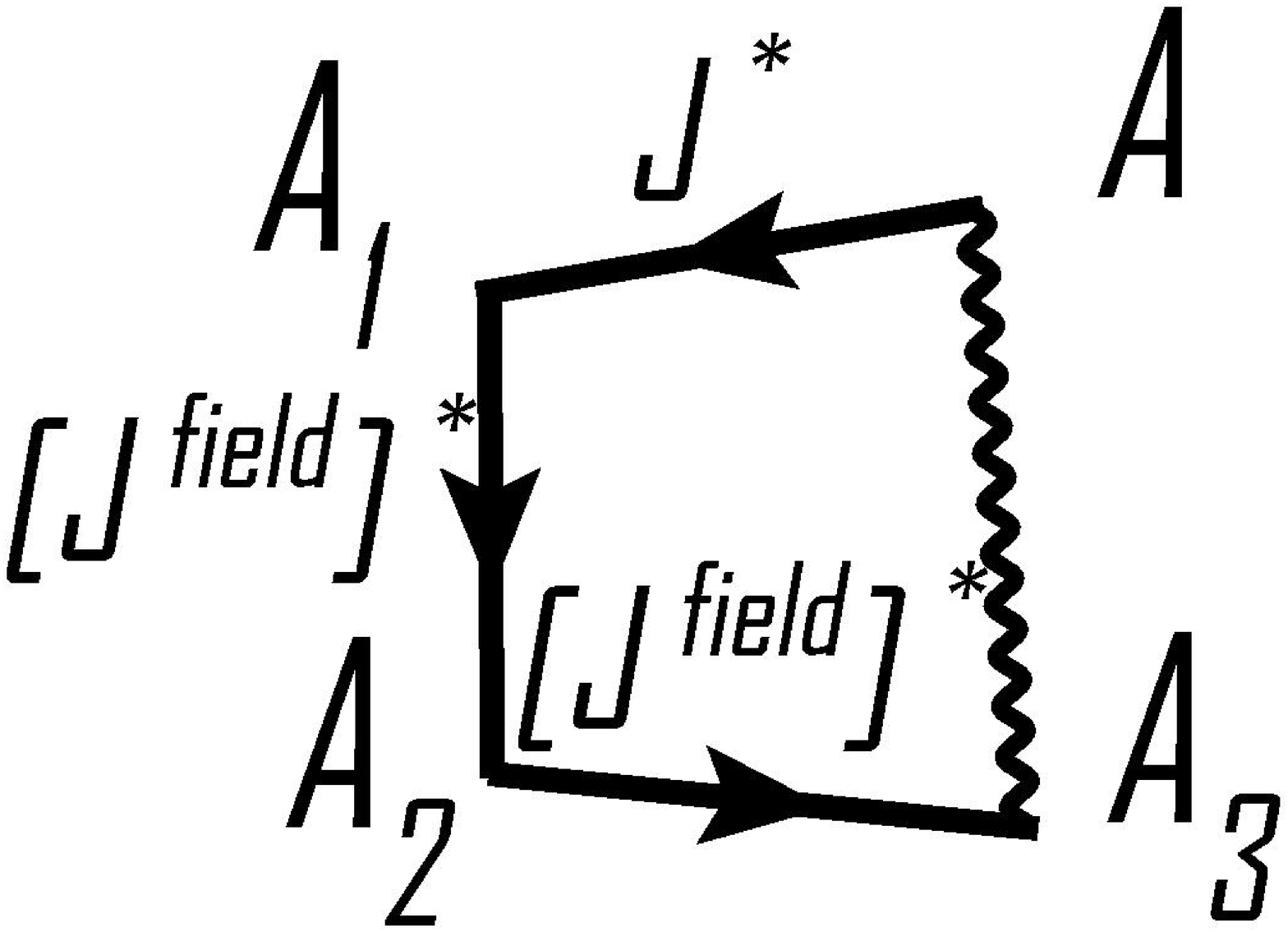}\label{Jff}}\hfill
\caption{Order J contributions to $\protect\sigma_{yx}$: \protect\ref{Z2}
results from $(\protect\psi_{J^{1}}^{(0)}\protect\nabla\protect\psi%
_{J^{0}}^{(2)\ast}-\protect\psi_{J^{0}}^{(2)\ast}\protect\nabla\protect\psi%
_{J^{1}}^{(0)})$. \protect\ref{ffJ}, \protect\ref{fJf} and \protect\ref{Jff}
result from $(\protect\psi_{J^{0}}^{(0)}\protect\nabla\protect\psi%
_{J^{1}}^{(2)\ast}-\protect\psi_{J^{1}}^{(2)\ast}\protect\nabla\protect\psi%
_{J^{0}}^{(0)})$.}
\label{2psi}
\end{figure}

By substituting the corresponding $\psi^{(0)}$, $\psi^{(1)}$
and $\psi^{(2)}$ into the second term of Eq.(\ref{1e1}),
one finds the current density in a magnetic field and an electric field.
Various contributions may be visualized by diagrams as for the conductivity:
the transition amplitudes in Table \ref{FHD} are still applicable. The
expression for the two components connected by the current operator $%
(\psi^{(0)}\nabla \psi^{(2)\ast}-\psi^{(2)\ast}\nabla\psi^{(0)})$ is the
same as that for two components connected by $(\psi^{(0)}\nabla\psi^{(1)%
\ast}-\psi^{(1)\ast}\nabla\psi^{(0)})$ in the ordinary conductivity. We still
use a wavy line to depict such a expression: it points from one component of
$\psi^{(0)}$ to one component of $\psi^{(2)\ast}$.\ There is one order J$^{0}
$ process and four order J$^{1}$ processes. In Table \ref{02H} we list the
contributions from $(\psi^{(0)}\nabla\psi^{(2)\ast}-\psi^{(2)\ast}\nabla%
\psi^{(0)})$. If we exchange $\psi^{(0)}$ and $\psi^{(2)\ast}$, the current
density changes sign. The new contribution to current density from $%
\psi^{(1)}\nabla\psi^{(1)\ast}$ does not include this exchange antisymmetry.
We represent it by a dashed arrow line, cf. the last line of table \ref{current}. The dashed arrow line points from
one component in $\psi^{(1)}$ to one component in $\psi^{(1)\ast}$. There is
one order J$^{0}$ process and four order J$^{1}$ processes. We list the
contributions from $\psi^{(1)}\nabla \psi^{(1)\ast}$ in Table \ref{11H}.
Order $J^{0}$ contributions are illustrated in Fig.\ref{UUp}. Order $J$
contributions are illustrated in Fig.\ref{Zp1Zp2} and Fig.\ref{2psi}.
In the symmetric gauge $\mathbf{A}=\frac{1}{2}\mathbf{B}\times\mathbf{r}$,
the coupling between two localized states by the external magnetic field becomes
\begin{equation}
J_{A_{2}A_{1}}^{field}=-\frac{e}{2m}%
B_{z}L_{A_{2}A_{1}}^{z}-eE_{x}x_{A_{2}A_{1}},   \label{ef}
\end{equation}
where $x_{A_{2}A_{1}}=\int d\mathbf{r}\phi_{A_{2}}^{\ast}x\phi_{A_{1}}$, and
$L_{z}^{A_{2}A_{1}}=\int d\mathbf{r}\phi_{A_{2}}^{\ast}L_{z}\phi_{A_{1}}$ is
the matrix element of the $z$ component of electronic orbital angular
momentum. To compute the Hall mobility, we require only the terms which are
proportional to $B_{z}E_{x}$. For example in $(J_{A_{3}A_{2}}^{field})^{\ast
}(J_{A_{2}A_{1}}^{field})^{\ast}$, we only keep%
\begin{equation}
(e^{2}/2m)B_{z}E_{x}(L_{A_{3}A_{2}}^{z}x_{A_{2}A_{1}}+x_{A_{3}A_{2}}L_{A_{2}A_{1}}^{z})^{\ast}.
\label{2f}
\end{equation}


One can extract $\sigma_{yx}$ from $j_{y}$. The corresponding expressions for various processes are given in table \ref{Halt}. The time integrals $U$, $U^{\prime}$, $Z$, $Z^{\prime}$ and $Y$ are given in Sec.5 of Ref.\cite{sup}, they are functions of temperature $T$. The non-diagonal conductivity $\sigma_{yx}$ is a sum of 10 terms listed in table \ref{Halt}.
\begin{table*}
\caption{diagrams and $\sigma_{yx}$}
\resizebox{13cm}{!} {
\begin{tabular}
[c]{lc}%
\hline\hline
diagram & $\sigma_{yx}$\\\hline
\ref{U} & $%
\begin{array}
[c]{c}%
B_{z}\frac{N_{e}e^{3}}{2m\hbar^{2}\Omega_{\mathbf{s}}}\sum_{AA_{2}A_{1}%
}\operatorname{Im}\{U_{AA_{2}A_{1}}f(E_{A})[1-f(E_{A_{2}})]\\
i(w_{y}^{AA_{2}}-v_{y}^{A_{2}A})(L_{z}^{A_{2}A_{1}}x_{A_{1}A}+x_{A_{2}A_{1}%
}L_{z}^{A_{1}A})^{\ast}\}
\end{array}
$\\\hline
\ref{Up} & $%
\begin{array}
[c]{c}%
-B_{z}\frac{N_{e}e^{3}}{2m\hbar^{2}\Omega_{\mathbf{s}}}\sum_{AA_{2}A_{1}%
}\operatorname{Im}\{iw_{y}^{A_{2}A_{1}}(L_{z}^{A_{1}A}x_{A_{2}A}+x_{A_{1}%
A}L_{z}^{A_{2}A})^{\ast}\\
U_{AA_{1}A_{2}}^{\prime}(T)f(E_{A})[1-f(E_{A_{1}})][1-f(E_{A_{2}})]\}
\end{array}
$\\\hline
\ref{Z1} & $%
\begin{array}
[c]{c}%
B_{z}\frac{N_{e}e^{3}}{2m\hbar^{3}\Omega_{\mathbf{s}}}\sum_{AA_{1}A_{2}A_{3}%
}\operatorname{Im}f(E_{A})[1-f(E_{A_{2}})][1-f(E_{A_{3}})]\\
w_{y}^{A_{3}A_{2}}(L_{z}^{A_{3}A}x_{A_{2}A_{1}}^{\ast}J_{A_{1}A}^{\ast
}+x_{A_{3}A}(L_{z}^{A_{2}A_{1}})^{\ast}J_{A_{1}A}^{\ast})Z_{AA_{3}A_{2}A_{1}}%
\end{array}
$\\\hline
\ref{Z3} & $%
\begin{array}
[c]{c}%
B_{z}\frac{N_{e}e^{3}}{2m\hbar^{3}\Omega_{\mathbf{s}}}\sum_{AA_{1}A_{2}A_{3}%
}\operatorname{Im}f(E_{A})[1-f(E_{A_{2}})][1-f(E_{A_{3}})]\\
w_{y}^{A_{3}A_{2}}(x_{A_{3}A}J_{A_{2}A_{1}}^{\ast}(L_{z}^{A_{1}A})^{\ast
}+L_{z}^{A_{3}A}J_{A_{2}A_{1}}^{\ast}x_{A_{1}A}^{\ast})Z_{AA_{3}A_{2}A_{1}}%
\end{array}
$\\\hline
\ref{Zp1} & $%
\begin{array}
[c]{c}%
-B_{z}\frac{N_{e}e^{3}}{2m\hbar^{3}\Omega_{\mathbf{s}}}\sum_{AA_{1}A_{2}A_{3}%
}\operatorname{Im}f(E_{A})[1-f(E_{A_{2}})][1-f(E_{A_{3}})]\\
w_{y}^{A_{2}A_{3}}(x_{A_{3}A}^{\ast}L_{z}^{A_{2}A_{1}}+(L_{z}^{A_{3}A})^{\ast
}x_{A_{2}A_{1}})J_{A_{1}A}Z_{AA_{3}A_{2}A_{1}}^{\prime}%
\end{array}
$\\\hline
\ref{Zp2} & $%
\begin{array}
[c]{c}%
-B_{z}\frac{N_{e}e^{3}}{2m\hbar^{3}\Omega_{\mathbf{s}}}\sum_{AA_{1}A_{2}A_{3}%
}\operatorname{Im}f(E_{A})[1-f(E_{A_{2}})][1-f(E_{A_{3}})]\\
w_{y}^{A_{2}A_{3}}(x_{A_{3}A}^{\ast}J_{A_{2}A_{1}}L_{z}^{A_{1}A}+(L_{z}%
^{A_{3}A})^{\ast}J_{A_{2}A_{1}}x_{A_{1}A})Z_{AA_{3}A_{2}A_{1}}^{\prime}%
\end{array}
$\\\hline
\ref{Z2} & $%
\begin{array}
[c]{c}%
+B_{z}\frac{N_{e}e^{3}}{2m\hbar^{3}\Omega_{\mathbf{s}}}\sum_{AA_{1}A_{2}A_{3}%
}\operatorname{Im}f(E_{A})[1-f(E_{A_{2}})][1-f(E_{A_{3}})]\\
(w_{y}^{A_{3}A_{2}}-v_{y}^{A_{2}A_{3}})J_{A_{3}A}(L_{z}^{A_{2}A_{1}}x_{A_{1}%
A}+x_{A_{2}A_{1}}L_{z}^{A_{1}A})^{\ast}Z_{AA_{3}A_{2}A_{1}}%
\end{array}
$\\\hline
\ref{ffJ} & $%
\begin{array}
[c]{c}%
\frac{N_{e}e^{3}}{\Omega_{\mathbf{s}}2m^{2}\hbar^{2}}B_{z}\sum_{A_{3}%
A_{2}A_{1}A}\operatorname{Im}Y_{A_{3}A_{2}A_{1}A}f(E_{A})[1-f(E_{A_{3}})]\\
iJ_{A_{3}A_{2}}^{\ast}(w_{AA_{3}}^{y}-v_{A_{3}A}^{y})[(L_{A_{2}A_{1}}%
^{z})^{\ast}(x_{A_{1}A})^{\ast}+(x_{A_{2}A_{1}})^{\ast}(L_{A_{1}A}^{z})^{\ast
}]
\end{array}
$\\\hline
\ref{fJf} & $%
\begin{array}
[c]{c}%
\frac{N_{e}e^{3}}{\Omega_{\mathbf{s}}2m^{2}\hbar^{2}}B_{z}E_{x}\sum
_{A_{3}A_{2}A_{1}A}\operatorname{Im}Y_{A_{3}A_{2}A_{1}A}f(E_{A})[1-f(E_{A_{3}%
})]\\
iJ_{A_{2}A_{1}}^{\ast}(w_{AA_{3}}^{y}-v_{A_{3}A}^{y})[(L_{A_{3}A_{2}}%
^{z})^{\ast}(x_{A_{1}A})^{\ast}+(x_{A_{3}A_{2}})^{\ast}(L_{A_{1}A}^{z})^{\ast
}]
\end{array}
$\\\hline
\ref{Jff} & $%
\begin{array}
[c]{c}%
\frac{N_{e}e^{3}}{\Omega_{\mathbf{s}}2m^{2}\hbar^{2}}B_{z}\sum_{A_{3}%
A_{2}A_{1}A}\operatorname{Im}Y_{A_{3}A_{2}A_{1}A}f(E_{A})[1-f(E_{A_{3}})]\\
iJ_{A_{1}A}^{\ast}(w_{AA_{3}}^{y}-v_{A_{3}A}^{y})[(L_{A_{3}A_{2}}^{z})^{\ast
}(x_{A_{2}A_{1}})^{\ast}+(x_{A_{3}A_{2}})^{\ast}(L_{A_{2}A_{1}}^{z})^{\ast}]
\end{array}
$\\%
\hline\hline
\end{tabular}%
}
\label{Halt}
\end{table*}
In the previous theory of the Hall effect\cite{fir,sch65,hol68},
researchers introduced magnetic field dependent Wannier functions as the
basis and expanded the magnetic field dependent phase in the final step.
Such a method excludes the interference between the transition amplitudes
caused by electric field and magnetic field. In the present work, we treat
electric field and magnetic field on the same footing. A mixing similar to
Eq.(\ref{2f}) appears in each term of table \ref{Halt}. In the application
of Kubo formula\cite{fir,sch65,hol68}, Fig.\ref{Up}, \ref{Zp1Zp2}, \ref%
{ffJ}, \ref{fJf}, \ref{Jff} are excluded. These new terms are of the same
order as Fig.\ref{U} and \ref{Z2}, but the temperature dependence are
different. To check the existence of these new term, it requires more experimental data points\cite{short}.

The full Hall effect from the
carriers in localized states and in extended states can be found by eventually changing
the states from A to B, and correspondingly changing the residual
interaction and the transition amplitudes. For example there are 7 more
diagrams for each diagram in Fig.\ref{2psi}. Using tables \ref{current} and \ref{FHD}, one can easily write out the expressions for each diagram\cite{short}.

\section{Conclusion}\label{sum}
The microscopic response method\cite{short,pre,jpcm} is applied to systematically estimate the conductivity (tables 4 and 5) and Hall
coefficient (table 8 and Eq.(\ref{hm})) in amorphous semiconductors. A diagrammatic
representation of transport coefficients is introduced by inspecting the
structure of the observed current density and the perturbation expansion of
wave function about small parameters and about external field. From the
topology of the diagrams, one can easily write down various contributions
for a transport coefficient to a given order of small parameters. This is
helpful for higher order processes. Comparing to the Kubo formula, all important conduction processes can be easily taken into account by the diagrammatic rules in tables \ref{current} amd \ref{FHD}. Impressive advances have been made for computing transport properties from first principles using the Keldysh formalism\cite{transies1,keld,transies2}. The present method is complementary to that work as a means of computing temperature dependence of the transport coefficients rather than I-V curves.

The present work has two obvious advantages over the thermal average method of classical MD: (i) the vibrations of atoms are described by quantum mechanics, the results obtained are valid for arbitrary temperature (at least for which the harmonic approximation is valid), whereas the results of MD are correct only above the Debye temperature of a material;
(ii) In the expressions for conductivity and Hall mobility, the e-ph interaction is already integrated out. To calculate the transport coefficients at a given temperature, one only needs the eigenvectors and eigenvalues of the Kohn-Sham Hamiltonian and the dynamical matrix for one configuration. One does not need to compute the time-consuming MD trajectory or worry about proper equilibration. The method is ideally suited for \textit{ab initio} simulation and is in the process of numerical implementation.

\begin{acknowledgement}
We thank the Army Research Office for support under MURI W91NF-06-2-0026 and the NSF under DMR 09-03225. The diagrams are drawn with JaxoDraw\cite{jax}.
\end{acknowledgement}

\end{document}


\title{Supporting Information for ``The Microscopic Response Method: theory of transport for systems with both topological and thermal disorder''
}


\author{%
MingLiang Zhang,
  D. A. Drabold \textsuperscript{\Ast}}

\authorrunning{MingLiang Zhang and D. A. Drabold}

\mail{e-mail
  \textsf{zhangm@ohio.edu,drabold@ohio.edu}, Phone:
  +01-740-5931715, Fax: +01-740-5930433}

\institute{%
  Department of Physics and Astronomy, Ohio University, Athens, Ohio 45701
}

\received{XXXX, revised XXXX, accepted XXXX} 
\published{XXXX} 


\abstract{%
%
%
%
\abstcol{%
  The external electromagnetic field changes the states of system. The zeroth, first and second order corrections in wave function required for conductivity and Hall mobility are listed. In general, there are several types of processes contributing to a given transport coefficient. In each type of transport process, for a group of given electronic states, the electronic and vibrational contributions are factorized.}{ The vibrational degrees of freedom can be integrated out, and result in a time integral. The time integrals depend on the electronic state, temperature and the frequency of external field. The time integrals for conductivity and Hall mobility are given. The time integrals can be estimated by asymptotic analysis. We sketch how to implement the present results in \textit{ab initio} code.}}
\maketitle   

\section{State and the changes in state caused by external field}\label{wave}
To compute the conductivity, one needs $\psi^{(0)}$ and $\psi^{(1)}$, while to calculate Hall mobility one needs $\psi^{(0)}$, $\psi^{(1)}$ and $\psi^{(2)}$, cf.[text, Eq.(17)]. As discussed in Sec. 4, the lowest order self-consistent approximation requires computing all these quantities to order $S^{0}$ and $S^{1}$.
\subsection{Initial state is a localized state}\label{Lw}
By the procedure sketched in the end of Sec. 4, to first order in small
parameters $J$ and $J^{\prime}$, the state of system is%
\[
\psi^{(0)}(t)=\phi_{A}\Psi_{A}^{\{N_{\alpha}\}}e^{-it\mathcal{E}%
_{A}^{\{N_{\alpha}\}}/\hbar}%
\]%
\[
+\sum_{A_{2}\cdots N_{\alpha}^{\prime\prime}\cdots}\phi_{A_{2}}\Psi_{A_{2}%
}^{\{N_{\alpha}^{\prime\prime}\}}e^{-it\mathcal{E}_{A_{2}}^{\{N_{\alpha
}^{\prime\prime}\}}/\hbar}%
\]%
\[
\times(\frac{-i}{\hbar})J_{A_{2}A}\int_{-\infty}^{t}dt^{\prime}\langle
\Psi_{A_{2}}^{\{N_{\alpha}^{\prime\prime}\}}|\Psi_{A}^{\{N_{\alpha}\}}\rangle
\]%
\[
e^{it^{\prime}(\mathcal{E}_{A_{2}}^{\{N_{\alpha}^{\prime\prime}\}}%
-\mathcal{E}_{A}^{\{N_{\alpha}\}})/\hbar}%
\]%
\begin{equation}
+\sum_{B_{1}\cdots N_{\alpha}^{\prime}\cdots}\xi_{B_{1}}\Xi_{B_{1}%
}^{\{N_{\alpha}^{\prime}\}}e^{-it\mathcal{E}_{B_{1}}^{\{N_{\alpha}^{\prime}%
\}}/\hbar} \label{0a}%
\end{equation}%
\[
\times(\frac{-i}{\hbar})J_{B_{1}A}^{\prime}\int_{-\infty}^{t}dt^{\prime
}\langle\Xi_{B_{1}}^{\{N_{\alpha}^{\prime}\}}|\Psi_{A}^{\{N_{\alpha}\}}\rangle
\]%
\[
e^{it^{\prime}(\mathcal{E}_{B_{1}}^{\{N_{\alpha}^{\prime}\}}-\mathcal{E}%
_{A}^{\{N_{\alpha}\}})/\hbar}.
\]

To first order in external field, the change in state $\psi^{(1)}$ to S$^{0}$
and S is%
\begin{equation}
\psi^{(1)}=\sum_{A_{2}\cdots N_{\alpha}^{\prime\prime}\cdots}\phi_{A_{2}}%
\Psi_{A_{2}}^{\{N_{\alpha}^{\prime\prime}\}}e^{-it\mathcal{E}_{A_{2}%
}^{\{N_{\alpha}^{\prime\prime}\}}/\hbar} \label{fa}%
\end{equation}%
\[
\times(\frac{-i}{\hbar})\int_{-\infty}^{t}dt^{\prime}J_{A_{2}A}^{field}%
(t^{\prime})\langle\Psi_{A_{2}}^{\{N_{\alpha}^{\prime\prime}\}}|\Psi
_{A}^{\{N_{\alpha}\}}\rangle
\]%
\[
e^{it^{\prime}(\mathcal{E}_{A_{2}}^{\{N_{\alpha}^{\prime\prime}\}}%
-\mathcal{E}_{A}^{\{N_{\alpha}\}})/\hbar}%
\]%
\[
+\sum_{B_{1}\cdots N_{\alpha}^{\prime}\cdots}\xi_{B_{1}}\Xi_{B_{1}%
}^{\{N_{\alpha}^{\prime}\}}e^{-it\mathcal{E}_{B_{1}}^{\{N_{\alpha}^{\prime}%
\}}/\hbar}%
\]%
\[
\times(\frac{-i}{\hbar})\int_{-\infty}^{t}dt^{\prime}J_{B_{1}A}^{\prime
field}(t^{\prime})\langle\Xi_{B_{1}}^{\{N_{\alpha}^{\prime}\}}|\Psi
_{A}^{\{N_{\alpha}\}}\rangle
\]%
\[
e^{it^{\prime}(\mathcal{E}_{B_{1}}^{\{N_{\alpha}^{\prime}\}}-\mathcal{E}%
_{A}^{\{N_{\alpha}\}})/\hbar}%
\]%
\[
+\sum_{A_{2}\cdots N_{\alpha}^{\prime\prime}\cdots}\phi_{A_{2}}\Psi_{A_{2}%
}^{\{N_{\alpha}^{\prime\prime}\}}e^{-it\mathcal{E}_{A_{2}}^{\{N_{\alpha
}^{\prime\prime}\}}/\hbar}\sum_{A_{1}\cdots N_{\alpha}^{\prime}\cdots}%
\]%
\[
\times(\frac{-i}{\hbar})\int_{-\infty}^{t}dt^{\prime}J_{A_{2}A_{1}}\langle
\Psi_{A_{2}}^{\{N_{\alpha}^{\prime\prime}\}}|\Psi_{A_{1}}^{\{N_{\alpha
}^{\prime}\}}\rangle
\]%
\[
e^{it^{\prime}(\mathcal{E}_{A_{2}}^{\{N_{\alpha}^{\prime\prime}\}}%
-\mathcal{E}_{A_{1}}^{\{N_{\alpha}^{\prime}\}})/\hbar}%
\]%
\[
\times(\frac{-i}{\hbar})\int_{-\infty}^{t^{\prime}}dt^{\prime\prime}J_{A_{1}%
A}^{field}(t^{\prime\prime})\langle\Psi_{A_{1}}^{\{N_{\alpha}^{\prime}\}}%
|\Psi_{A}^{\{N_{\alpha}\}}\rangle
\]%
\[
e^{it^{\prime\prime}(\mathcal{E}_{A_{1}}^{\{N_{\alpha}^{\prime}\}}%
-\mathcal{E}_{A}^{\{N_{\alpha}\}})/\hbar}%
\]%
\[
+\sum_{A_{2}\cdots N_{\alpha}^{\prime\prime}\cdots}\phi_{A_{2}}\Psi_{A_{2}%
}^{\{N_{\alpha}^{\prime\prime}\}}e^{-it\mathcal{E}_{A_{2}}^{\{N_{\alpha
}^{\prime\prime}\}}/\hbar}\sum_{A_{1}\cdots N_{\alpha}^{\prime}\cdots}%
\]%
\[
\times(\frac{-i}{\hbar})\int_{-\infty}^{t}dt^{\prime}J_{A_{2}A_{1}}%
^{field}(t^{\prime})\langle\Psi_{A_{2}}^{\{N_{\alpha}^{\prime\prime}\}}%
|\Psi_{A_{1}}^{\{N_{\alpha}^{\prime}\}}\rangle
\]%
\[
e^{it^{\prime}(\mathcal{E}_{A_{2}}^{\{N_{\alpha}^{\prime\prime}\}}%
-\mathcal{E}_{A_{1}}^{\{N_{\alpha}^{\prime}\}})/\hbar}%
\]%
\[
\times(\frac{-i}{\hbar})J_{A_{1}A}\int_{-\infty}^{t^{\prime}}dt^{\prime\prime
}\langle\Psi_{A_{1}}^{\{N_{\alpha}^{\prime}\}}|\Psi_{A}^{\{N_{\alpha}%
\}}\rangle
\]%
\[
e^{it^{\prime\prime}(\mathcal{E}_{A_{1}}^{\{N_{\alpha}^{\prime}\}}%
-\mathcal{E}_{A}^{\{N_{\alpha}\}})/\hbar}%
\]%
\[
+\sum_{A_{2}\cdots N_{\alpha}^{\prime\prime}\cdots}\phi_{A_{2}}\Psi_{A_{2}%
}^{\{N_{\alpha}^{\prime\prime}\}}e^{-it\mathcal{E}_{A_{2}}^{\{N_{\alpha
}^{\prime\prime}\}}/\hbar}\sum_{B_{1}\cdots N_{\alpha}^{\prime}\cdots}%
\]%
\[
(\frac{-i}{\hbar})\int_{-\infty}^{t}dt^{\prime}\langle\Psi_{A_{2}%
}^{\{N_{\alpha}^{\prime\prime}\}}|K_{A_{2}B_{1}}^{\prime}|\Xi_{B_{1}%
}^{\{N_{\alpha}^{\prime}\}}\rangle
\]%
\[
e^{it^{\prime}(\mathcal{E}_{A_{2}}^{\{N_{\alpha}^{\prime\prime}\}}%
-\mathcal{E}_{B_{1}}^{\{N_{\alpha}^{\prime}\}})/\hbar}%
\]%
\[
\times(\frac{-i}{\hbar})\int_{-\infty}^{t^{\prime}}dt^{\prime\prime}J_{B_{1}%
A}^{\prime field}(t^{\prime\prime})\langle\Xi_{B_{1}}^{\{N_{\alpha}^{\prime
}\}}|\Psi_{A}^{\{N_{\alpha}\}}\rangle
\]%
\[
e^{it^{\prime\prime}(\mathcal{E}_{B_{1}}^{\{N_{\alpha}^{\prime}\}}%
-\mathcal{E}_{A}^{\{N_{\alpha}\}})/\hbar}%
\]%
\[
+\sum_{A_{2}\cdots N_{\alpha}^{\prime\prime}\cdots}\phi_{A_{2}}\Psi_{A_{2}%
}^{\{N_{\alpha}^{\prime\prime}\}}e^{-it\mathcal{E}_{A_{2}}^{\{N_{\alpha
}^{\prime\prime}\}}/\hbar}\sum_{B_{1}\cdots N_{\alpha}^{\prime}\cdots}%
\]%
\[
(\frac{-i}{\hbar})\int_{-\infty}^{t}dt^{\prime}K_{A_{2}B_{1}}^{\prime
field}(t^{\prime})\langle\Psi_{A_{2}}^{\{N_{\alpha}^{\prime\prime}\}}%
|\Xi_{B_{1}}^{\{N_{\alpha}^{\prime}\}}\rangle
\]%
\[
e^{it^{\prime}(\mathcal{E}_{A_{2}}^{\{N_{\alpha}^{\prime\prime}\}}%
-\mathcal{E}_{B_{1}}^{\{N_{\alpha}^{\prime}\}})/\hbar}%
\]%
\[
\times(\frac{-i}{\hbar})J_{B_{1}A}^{\prime}\int_{-\infty}^{t^{\prime}%
}dt^{\prime\prime}\langle\Xi_{B_{1}}^{\{N_{\alpha}^{\prime}\}}|\Psi
_{A}^{\{N_{\alpha}\}}\rangle
\]%
\[
e^{it^{\prime\prime}(\mathcal{E}_{B_{1}}^{\{N_{\alpha}^{\prime}\}}%
-\mathcal{E}_{A}^{\{N_{\alpha}\}})/\hbar}%
\]%
\[
+\sum_{B_{2}\cdots N_{\alpha}^{\prime\prime}\cdots}\xi_{B_{2}}\Xi_{B_{2}%
}^{\{N_{\alpha}^{\prime\prime}\}}e^{-it\mathcal{E}_{B_{2}}^{\{N_{\alpha
}^{\prime\prime}\}}/\hbar}\sum_{A_{1}\cdots N_{\alpha}^{\prime}\cdots}%
\]%
\[
(\frac{-i}{\hbar})\int_{-\infty}^{t}dt^{\prime}J_{B_{2}A_{1}}^{\prime}%
\langle\Xi_{B_{2}}^{\{N_{\alpha}^{\prime\prime}\}}|\Psi_{A_{1}}^{\{N_{\alpha
}^{\prime}\}}\rangle
\]%
\[
e^{it^{\prime}(\mathcal{E}_{B_{2}}^{\{N_{\alpha}^{\prime\prime}\}}%
-\mathcal{E}_{A_{1}}^{\{N_{\alpha}^{\prime}\}})/\hbar}%
\]%
\[
\times(\frac{-i}{\hbar})\int_{-\infty}^{t^{\prime}}dt^{\prime\prime}J_{A_{1}%
A}^{field}(t^{\prime\prime})\langle\Psi_{A_{1}}^{\{N_{\alpha}^{\prime}\}}%
|\Psi_{A}^{\{N_{\alpha}\}}\rangle
\]%
\[
e^{it^{\prime\prime}(\mathcal{E}_{A_{1}}^{\{N_{\alpha}^{\prime}\}}%
-\mathcal{E}_{A}^{\{N_{\alpha}\}})/\hbar}%
\]%
\[
+\sum_{B_{2}\cdots N_{\alpha}^{\prime\prime}\cdots}\xi_{B_{2}}\Xi_{B_{2}%
}^{\{N_{\alpha}^{\prime\prime}\}}e^{-it\mathcal{E}_{B_{2}}^{\{N_{\alpha
}^{\prime\prime}\}}/\hbar}\sum_{A_{1}\cdots N_{\alpha}^{\prime}\cdots}%
\]%
\[
(\frac{-i}{\hbar})\int_{-\infty}^{t}dt^{\prime}J_{B_{2}A_{1}}^{\prime
field}(t^{\prime})\langle\Xi_{B_{2}}^{\{N_{\alpha}^{\prime\prime}\}}%
|\Psi_{A_{1}}^{\{N_{\alpha}^{\prime}\}}\rangle
\]%
\[
e^{it^{\prime}(\mathcal{E}_{B_{2}}^{\{N_{\alpha}^{\prime\prime}\}}%
-\mathcal{E}_{A_{1}}^{\{N_{\alpha}^{\prime}\}})/\hbar}%
\]%
\[
\times(\frac{-i}{\hbar})J_{A_{1}A}\int_{-\infty}^{t^{\prime}}dt^{\prime\prime
}\langle\Psi_{A_{1}}^{\{N_{\alpha}^{\prime}\}}|\Psi_{A}^{\{N_{\alpha}%
\}}\rangle
\]%
\[
e^{it^{\prime}(\mathcal{E}_{A_{1}}^{\{N_{\alpha}^{\prime}\}}-\mathcal{E}%
_{A}^{\{N_{\alpha}\}})/\hbar}%
\]%
\[
+\sum_{B_{2}\cdots N_{\alpha}^{\prime\prime}\cdots}\xi_{B_{2}}\Xi_{B_{2}%
}^{\{N_{\alpha}^{\prime\prime}\}}e^{-it\mathcal{E}_{B_{2}}^{\{N_{\alpha
}^{\prime\prime}\}}/\hbar}\sum_{B_{1}\cdots N_{\alpha}^{\prime}\cdots}%
\]%
\[
(\frac{-i}{\hbar})\int_{-\infty}^{t}dt^{\prime}\langle\Xi_{B_{2}}%
^{\{N_{\alpha}^{\prime\prime}\}}|K_{B_{2}B_{1}}|\Xi_{B_{1}}^{\{N_{\alpha
}^{\prime}\}}\rangle
\]%
\[
e^{it^{\prime}(\mathcal{E}_{B_{2}}^{\{N_{\alpha}^{\prime\prime}\}}%
-\mathcal{E}_{B_{1}}^{\{N_{\alpha}^{\prime}\}})/\hbar}%
\]%
\[
\times(\frac{-i}{\hbar})\int_{-\infty}^{t^{\prime}}dt^{\prime\prime}J_{B_{1}%
A}^{\prime field}(t^{\prime\prime})\langle\Xi_{B_{1}}^{\{N_{\alpha}^{\prime
}\}}|\Psi_{A}^{\{N_{\alpha}\}}\rangle
\]%
\[
e^{it^{\prime\prime}(\mathcal{E}_{B_{1}}^{\{N_{\alpha}^{\prime}\}}%
-\mathcal{E}_{A}^{\{N_{\alpha}\}})/\hbar}%
\]%
\[
+\sum_{B_{2}\cdots N_{\alpha}^{\prime\prime}\cdots}\xi_{B_{2}}\Xi_{B_{2}%
}^{\{N_{\alpha}^{\prime\prime}\}}e^{-it\mathcal{E}_{B_{2}}^{\{N_{\alpha
}^{\prime\prime}\}}/\hbar}\sum_{B_{1}\cdots N_{\alpha}^{\prime}\cdots}%
\]%
\[
(\frac{-i}{\hbar})\int_{-\infty}^{t}dt^{\prime}K_{B_{2}B_{1}}^{field}%
(t^{\prime})\langle\Xi_{B_{2}}^{\{N_{\alpha}^{\prime\prime}\}}|\Xi_{B_{1}%
}^{\{N_{\alpha}^{\prime}\}}\rangle
\]%
\[
e^{it^{\prime}(\mathcal{E}_{B_{2}}^{\{N_{\alpha}^{\prime\prime}\}}%
-\mathcal{E}_{B_{1}}^{\{N_{\alpha}^{\prime}\}})/\hbar}%
\]%
\[
\times(\frac{-i}{\hbar})J_{B_{1}A}^{\prime}\int_{-\infty}^{t^{\prime}%
}dt^{\prime\prime}\langle\Xi_{B_{1}}^{\{N_{\alpha}^{\prime}\}}|\Psi
_{A}^{\{N_{\alpha}\}}\rangle
\]%
\[
e^{it^{\prime}(\mathcal{E}_{B_{1}}^{\{N_{\alpha}^{\prime}\}}-\mathcal{E}%
_{A}^{\{N_{\alpha}\}})/\hbar}.
\]
Because the carriers obey Fermi statistics, in Eqs.(\ref{0a},\ref{fa}), a
factor $1-f(E_{A_{2}})$ must be included before component $\phi_{A_{2}}$ of
the final state, factor $1-f(E_{B_{1}})$ must be inserted before component
$\xi_{B_{1}}$ of the final state.

The full expression for $\psi^{(2)}$ is too
long to reproduce. Supposing there exist only localized states, to order
$J^{0}$ and $J^{1}$, $\psi^{(2)}$ is:%

\[
\psi^{(2)}(t)=\sum_{A_{2}\cdots N_{\alpha}^{\prime\prime}\cdots}\phi_{A_{2}%
}(\mathbf{r}-\mathbf{R}_{A_{2}})\Psi_{A_{2}}^{\{N_{\alpha}^{\prime\prime}%
\}}e^{-it\mathcal{E}_{A_{2}}^{\{N_{\alpha}^{\prime\prime}\}}/\hbar}%
\]%
\[
\sum_{A_{1}\cdots N_{\alpha}^{\prime}\cdots}(-\frac{i}{\hbar})\int_{-\infty
}^{t}dt^{\prime}J_{A_{2}A_{1}}^{\text{field}}(t^{\prime})\langle\Psi_{A_{2}%
}^{\{N_{\alpha}^{\prime\prime}\}}|\Psi_{A_{1}}^{\{N_{\alpha}^{\prime}%
\}}\rangle
\]%
\[
e^{it^{\prime}(\mathcal{E}_{A_{2}}^{\{N_{\alpha}^{\prime\prime}\}}%
-\mathcal{E}_{A_{1}}^{\{N_{\alpha}^{\prime}\}})/\hbar}%
\]%
\[
(-\frac{i}{\hbar})\int_{-\infty}^{t^{\prime}}dt^{\prime\prime}J_{A_{1}%
A}^{\text{field}}(t^{\prime\prime})\langle\Psi_{A_{1}}^{\{N_{\alpha}^{\prime
}\}}|\Psi_{A}^{\{N_{\alpha}\}}\rangle
\]%
\[
e^{it^{\prime\prime}(\mathcal{E}_{A_{1}}^{\{N_{\alpha}^{\prime}\}}%
-\mathcal{E}_{A}^{\{N_{\alpha}\}})/\hbar}%
\]%
\[
+\sum_{A_{3}\cdots N_{\alpha}^{\prime\prime\prime}\cdots}\phi_{A_{3}%
}(\mathbf{r}-\mathbf{R}_{A_{3}})\Psi_{A_{3}}^{\{N_{\alpha}^{\prime\prime
\prime}\}}e^{-it\mathcal{E}_{A_{3}}^{\{N_{\alpha}^{\prime\prime\prime}%
\}}/\hbar}%
\]%
\[
\sum_{A_{2}\cdots N_{\alpha}^{\prime\prime}\cdots}(-\frac{i}{\hbar}%
)\int_{-\infty}^{t}dt^{\prime}J_{A_{3}A_{2}}\langle\Psi_{A_{3}}^{\{N_{\alpha
}^{\prime\prime\prime}\}}|\Psi_{A_{2}}^{\{N_{\alpha}^{\prime\prime}\}}\rangle
\]%
\[
e^{it^{\prime}(\mathcal{E}_{A_{3}}^{\{N_{\alpha}^{\prime\prime\prime}%
\}}-\mathcal{E}_{A_{2}}^{\{N_{\alpha}^{\prime\prime}\}})/\hbar}%
\]%
\[
\sum_{A_{1}\cdots N_{\alpha}^{\prime}\cdots}(-\frac{i}{\hbar})\int_{-\infty
}^{t^{\prime}}dt^{\prime\prime}J_{A_{2}A_{1}}^{\text{field}}(t^{\prime\prime
})\langle\Psi_{A_{2}}^{\{N_{\alpha}^{\prime\prime}\}}|\Psi_{A_{1}%
}^{\{N_{\alpha}^{\prime}\}}\rangle
\]%
\[
e^{it^{\prime\prime}(\mathcal{E}_{A_{2}}^{\{N_{\alpha}^{\prime\prime}%
\}}-\mathcal{E}_{A_{1}}^{\{N_{\alpha}^{\prime}\}})/\hbar}%
\]%
\[
(-\frac{i}{\hbar})\int_{-\infty}^{t^{\prime\prime}}dt^{\prime\prime\prime
}J_{A_{1}A}^{\text{field}}(t^{\prime\prime\prime})\langle\Psi_{A_{1}%
}^{\{N_{\alpha}^{\prime}\}}|\Psi_{A}^{\{N_{\alpha}\}}\rangle
\]%
\[
e^{it^{\prime\prime\prime}(\mathcal{E}_{A_{1}}^{\{N_{\alpha}^{\prime}%
\}}-\mathcal{E}_{A}^{\{N_{\alpha}\}})/\hbar}%
\]%
\[
+\sum_{A_{3}\cdots N_{\alpha}^{\prime\prime\prime}\cdots}\phi_{A_{3}%
}(\mathbf{r}-\mathbf{R}_{A_{3}})\Psi_{A_{3}}^{\{N_{\alpha}^{\prime\prime
\prime}\}}e^{-it\mathcal{E}_{A_{3}}^{\{N_{\alpha}^{\prime\prime\prime}%
\}}/\hbar}%
\]%
\[
\sum_{A_{2}\cdots N_{\alpha}^{\prime\prime}\cdots}(-\frac{i}{\hbar}%
)\int_{-\infty}^{t}dt^{\prime}J_{A_{3}A_{2}}^{\text{field}}\langle\Psi_{A_{3}%
}^{\{N_{\alpha}^{\prime\prime\prime}\}}|\Psi_{A_{2}}^{\{N_{\alpha}%
^{\prime\prime}\}}\rangle
\]%
\[
e^{it^{\prime}(\mathcal{E}_{A_{3}}^{\{N_{\alpha}^{\prime\prime\prime}%
\}}-\mathcal{E}_{A_{2}}^{\{N_{\alpha}^{\prime\prime}\}})/\hbar}%
\]%
\[
\sum_{A_{1}\cdots N_{\alpha}^{\prime}\cdots}(-\frac{i}{\hbar})\int_{-\infty
}^{t^{\prime}}dt^{\prime\prime}J_{A_{2}A_{1}}\langle\Psi_{A_{2}}^{\{N_{\alpha
}^{\prime\prime}\}}|\Psi_{A_{1}}^{\{N_{\alpha}^{\prime}\}}\rangle
\]%
\[
e^{it^{\prime\prime}(\mathcal{E}_{A_{2}}^{\{N_{\alpha}^{\prime\prime}%
\}}-\mathcal{E}_{A_{1}}^{\{N_{\alpha}^{\prime}\}})/\hbar}%
\]%
\[
(-\frac{i}{\hbar})\int_{-\infty}^{t^{\prime\prime}}dt^{\prime\prime\prime
}J_{A_{1}A}^{\text{field}}(t^{\prime\prime\prime})\langle\Psi_{A_{1}%
}^{\{N_{\alpha}^{\prime}\}}|\Psi_{A}^{\{N_{\alpha}\}}\rangle
\]%
\[
e^{it^{\prime\prime\prime}(\mathcal{E}_{A_{1}}^{\{N_{\alpha}^{\prime}%
\}}-\mathcal{E}_{A}^{\{N_{\alpha}\}})/\hbar}%
\]%
\[
+\sum_{A_{3}\cdots N_{\alpha}^{\prime\prime\prime}\cdots}\phi_{A_{3}%
}(\mathbf{r}-\mathbf{R}_{A_{3}})\Psi_{A_{3}}^{\{N_{\alpha}^{\prime\prime
\prime}\}}e^{-it\mathcal{E}_{A_{3}}^{\{N_{\alpha}^{\prime\prime\prime}%
\}}/\hbar}%
\]%
\[
\sum_{A_{2}\cdots N_{\alpha}^{\prime\prime}\cdots}(-\frac{i}{\hbar}%
)\int_{-\infty}^{t}dt^{\prime}J_{A_{3}A_{2}}^{\text{field}}(t^{\prime}%
)\langle\Psi_{A_{3}}^{\{N_{\alpha}^{\prime\prime\prime}\}}|\Psi_{A_{2}%
}^{\{N_{\alpha}^{\prime\prime}\}}\rangle
\]%
\[
e^{it^{\prime}(\mathcal{E}_{A_{3}}^{\{N_{\alpha}^{\prime\prime\prime}%
\}}-\mathcal{E}_{A_{2}}^{\{N_{\alpha}^{\prime\prime}\}})/\hbar}%
\]%
\[
\sum_{A_{1}\cdots N_{\alpha}^{\prime}\cdots}(-\frac{i}{\hbar})\int_{-\infty
}^{t^{\prime}}dt^{\prime\prime}J_{A_{2}A_{1}}^{\text{field}}(t^{\prime\prime
})\langle\Psi_{A_{2}}^{\{N_{\alpha}^{\prime\prime}\}}|\Psi_{A_{1}%
}^{\{N_{\alpha}^{\prime}\}}\rangle
\]%
\[
e^{it^{\prime\prime}(\mathcal{E}_{A_{2}}^{\{N_{\alpha}^{\prime\prime}%
\}}-\mathcal{E}_{A_{1}}^{\{N_{\alpha}^{\prime}\}})/\hbar}%
\]%
\begin{equation}
(-\frac{i}{\hbar})\int_{-\infty}^{t^{\prime\prime}}dt^{\prime\prime\prime
}J_{A_{1}A}\langle\Psi_{A_{1}}^{\{N_{\alpha}^{\prime}\}}|\Psi_{A}%
^{\{N_{\alpha}\}}\rangle\label{2fL}%
\end{equation}%
\[
e^{it^{\prime\prime\prime}(\mathcal{E}_{A_{1}}^{\{N_{\alpha}^{\prime}%
\}}-\mathcal{E}_{A}^{\{N_{\alpha}\}})/\hbar}%
\]

\subsection{Initial state is extended state}

\label{Ew} If the initial state is $|B\cdots N_{\alpha}\cdots\rangle$, the
state of system at time $t$ without the external field is%
\[
\psi^{(0)}(t)=\xi_{B}\Xi_{B}^{\{N_{\alpha}\}}e^{-it\mathcal{E}_{B}%
^{\{N_{\alpha}\}}/\hbar}%
\]%
\[
+\sum_{A_{2}\cdots N_{\alpha}^{\prime\prime}\cdots}\phi_{A_{2}}\Psi_{A_{2}%
}^{\{N_{\alpha}^{\prime\prime}\}}e^{-it\mathcal{E}_{A_{2}}^{\{N_{\alpha
}^{\prime\prime}\}}/\hbar}%
\]%
\[
\times(\frac{-i}{\hbar})\langle\Psi_{A_{2}}^{\{N_{\alpha}^{\prime\prime}%
\}}|K_{A_{2}B}^{\prime}|\Xi_{B}^{\{N_{\alpha}\}}\rangle\int_{-\infty}%
^{t}dt^{\prime}e^{it^{\prime}(\mathcal{E}_{A_{2}}^{\{N_{\alpha}^{\prime\prime
}\}}-\mathcal{E}_{B}^{\{N_{\alpha}\}})/\hbar}%
\]%
\begin{equation}
+\sum_{B_{1}\cdots N_{\alpha}^{\prime}\cdots}\xi_{B_{1}}\Xi_{B_{1}%
}^{\{N_{\alpha}^{\prime}\}}e^{-it\mathcal{E}_{B_{1}}^{\{N_{\alpha}^{\prime}%
\}}/\hbar}(\frac{-i}{\hbar}) \label{pb}%
\end{equation}%
\[
\times\langle\Xi_{B_{1}}^{\{N_{\alpha}^{\prime}\}}|K_{B_{1}B}|\Xi
_{B}^{\{N_{\alpha}\}}\rangle\int_{-\infty}^{t}dt^{\prime}e^{it^{\prime
}(\mathcal{E}_{B_{1}}^{\{N_{\alpha}^{\prime}\}}-\mathcal{E}_{B}^{\{N_{\alpha
}\}})/\hbar}%
\]%
\[
+\sum_{B_{1}\cdots N_{\alpha}^{\prime}\cdots}\xi_{B_{1}}\Xi_{B_{1}%
}^{\{N_{\alpha}^{\prime}\}}e^{-it\mathcal{E}_{B_{1}}^{\{N_{\alpha}^{\prime}%
\}}/\hbar}\sum_{B_{3}\cdots N_{\alpha}^{\prime\prime\prime}\cdots}%
\]%
\[
(\frac{-i}{\hbar})\int_{-\infty}^{t}dt^{\prime}\langle\Xi_{B_{1}}%
^{\{N_{\alpha}^{\prime}\}}|K_{B_{1}B_{3}}|\Xi_{B_{3}}^{\{N_{\alpha}%
^{\prime\prime\prime}\}}\rangle
\]%
\[
e^{it^{\prime}(\mathcal{E}_{B_{1}}^{\{N_{\alpha}^{\prime}\}}-\mathcal{E}%
_{B_{3}}^{\{N_{\alpha}^{\prime\prime\prime}\}})/\hbar}%
\]%
\[
\times(\frac{-i}{\hbar})\int_{-\infty}^{t^{\prime}}dt^{\prime\prime}\langle
\Xi_{B_{3}}^{\{N_{\alpha}^{\prime\prime\prime}\}}|K_{B_{3}B}|\Xi
_{B}^{\{N_{\alpha}\}}\rangle
\]%
\[
e^{it^{\prime\prime}(\mathcal{E}_{B_{3}}^{\{N_{\alpha}^{\prime\prime\prime}%
\}}-\mathcal{E}_{B}^{\{N_{\alpha}\}})/\hbar}.
\]
The last term is an order $K^{2}$ contribution, and is useful only to
calculate conductivity from EE transition. To first order in the external
field, the change in state caused by the external field is%
\begin{equation}
\psi^{(1)}=\sum_{A_{2}\cdots N_{\alpha}^{\prime\prime}\cdots}\phi_{A_{2}}%
\Psi_{A_{2}}^{\{N_{\alpha}^{\prime\prime}\}}e^{-it\mathcal{E}_{A_{2}%
}^{\{N_{\alpha}^{\prime\prime}\}}/\hbar}(\frac{-i}{\hbar}) \label{fb}%
\end{equation}%
\[
\int_{-\infty}^{t}dt^{\prime}K_{A_{2}B}^{\prime field}(t^{\prime})|\langle
\Psi_{A_{2}}^{\{N_{\alpha}^{\prime\prime}\}}|\Xi_{B}^{\{N_{\alpha}\}}\rangle
\]%
\[
e^{it^{\prime}(\mathcal{E}_{A_{2}}^{\{N_{\alpha}^{\prime\prime}\}}%
-\mathcal{E}_{B}^{\{N_{\alpha}\}})/\hbar}%
\]%
\[
+\sum_{B_{1}\cdots N_{\alpha}^{\prime}\cdots}\xi_{B_{1}}\Xi_{B_{1}%
}^{\{N_{\alpha}^{\prime}\}}e^{-it\mathcal{E}_{B_{1}}^{\{N_{\alpha}^{\prime}%
\}}/\hbar}(\frac{-i}{\hbar})
\]%
\[
\int_{-\infty}^{t}dt^{\prime}K_{B_{1}B}^{field}(t^{\prime})\langle\Xi_{B_{1}%
}^{\{N_{\alpha}^{\prime}\}}|\Xi_{B}^{\{N_{\alpha}\}}\rangle
\]%
\[
e^{it^{\prime}(\mathcal{E}_{B_{1}}^{\{N_{\alpha}^{\prime}\}}-\mathcal{E}%
_{B}^{\{N_{\alpha}\}})/\hbar}%
\]%
\[
+\sum_{A_{2}\cdots N_{\alpha}^{\prime\prime}\cdots}\phi_{A_{2}}\Psi_{A_{2}%
}^{\{N_{\alpha}^{\prime\prime}\}}e^{-it\mathcal{E}_{A_{2}}^{\{N_{\alpha
}^{\prime\prime}\}}/\hbar}\sum_{A_{1}\cdots N_{\alpha}^{\prime}\cdots}%
\]%
\[
(\frac{-i}{\hbar})\int_{-\infty}^{t}dt^{\prime}J_{A_{2}A_{1}}\langle
\Psi_{A_{2}}^{\{N_{\alpha}^{\prime\prime}\}}|\Psi_{A_{1}}^{\{N_{\alpha
}^{\prime}\}}\rangle
\]%
\[
e^{it^{\prime}(\mathcal{E}_{A_{2}}^{\{N_{\alpha}^{\prime\prime}\}}%
-\mathcal{E}_{A_{1}}^{\{N_{\alpha}^{\prime}\}})/\hbar}%
\]%
\[
(\frac{-i}{\hbar})\int_{-\infty}^{t^{\prime}}dt^{\prime\prime}K_{A_{1}%
B}^{\prime field}(t^{\prime\prime})\langle\Psi_{A_{1}}^{\{N_{\alpha}^{\prime
}\}}|\Xi_{B}^{\{N_{\alpha}\}}\rangle
\]%
\[
e^{it^{\prime\prime}(\mathcal{E}_{A_{1}}^{\{N_{\alpha}^{\prime}\}}%
-\mathcal{E}_{B}^{\{N_{\alpha}\}})/\hbar}%
\]%
\[
+\sum_{A_{2}\cdots N_{\alpha}^{\prime\prime}\cdots}\phi_{A_{2}}\Psi_{A_{2}%
}^{\{N_{\alpha}^{\prime\prime}\}}e^{-it\mathcal{E}_{A_{2}}^{\{N_{\alpha
}^{\prime\prime}\}}/\hbar}\sum_{A_{1}\cdots N_{\alpha}^{\prime}\cdots}%
\]%
\[
(\frac{-i}{\hbar})\int_{-\infty}^{t}dt^{\prime}J_{A_{2}A_{1}}^{field}%
(t^{\prime})\langle\Psi_{A_{2}}^{\{N_{\alpha}^{\prime\prime}\}}|\Psi_{A_{1}%
}^{\{N_{\alpha}^{\prime}\}}\rangle
\]%
\[
e^{it^{\prime}(\mathcal{E}_{A_{2}}^{\{N_{\alpha}^{\prime\prime}\}}%
-\mathcal{E}_{A_{1}}^{\{N_{\alpha}^{\prime}\}})/\hbar}%
\]%
\[
(\frac{-i}{\hbar})\int_{-\infty}^{t^{\prime}}dt^{\prime\prime}\langle
\Psi_{A_{1}}^{\{N_{\alpha}^{\prime}\}}|K_{A_{1}B}^{\prime}|\Xi_{B}%
^{\{N_{\alpha}\}}\rangle
\]%
\[
e^{it^{\prime\prime}(\mathcal{E}_{A_{1}}^{\{N_{\alpha}^{\prime}\}}%
-\mathcal{E}_{B}^{\{N_{\alpha}\}})/\hbar}%
\]%
\[
+\sum_{A_{2}\cdots N_{\alpha}^{\prime\prime}\cdots}\phi_{A_{2}}\Psi_{A_{2}%
}^{\{N_{\alpha}^{\prime\prime}\}}e^{-it\mathcal{E}_{A_{2}}^{\{N_{\alpha
}^{\prime\prime}\}}/\hbar}\sum_{B_{1}\cdots N_{\alpha}^{\prime}\cdots}%
\]%
\[
(\frac{-i}{\hbar})\int_{-\infty}^{t}dt^{\prime}\langle\Psi_{A_{2}%
}^{\{N_{\alpha}^{\prime\prime}\}}|K_{A_{2}B_{1}}^{\prime}|\Xi_{B_{1}%
}^{\{N_{\alpha}^{\prime}\}}\rangle
\]%
\[
e^{it^{\prime}(\mathcal{E}_{A_{2}}^{\{N_{\alpha}^{\prime\prime}\}}%
-\mathcal{E}_{B_{1}}^{\{N_{\alpha}^{\prime}\}})/\hbar}%
\]%
\[
(\frac{-i}{\hbar})\int_{-\infty}^{t^{\prime}}dt^{\prime\prime}K_{B_{1}%
B}^{field}(t^{\prime\prime})\langle\Xi_{B_{1}}^{\{N_{\alpha}^{\prime}\}}%
|\Xi_{B}^{\{N_{\alpha}\}}\rangle
\]%
\[
e^{it^{\prime\prime}(\mathcal{E}_{B_{1}}^{\{N_{\alpha}^{\prime}\}}%
-\mathcal{E}_{B}^{\{N_{\alpha}\}})/\hbar}%
\]%
\[
+\sum_{A_{2}\cdots N_{\alpha}^{\prime\prime}\cdots}\phi_{A_{2}}\Psi_{A_{2}%
}^{\{N_{\alpha}^{\prime\prime}\}}e^{-it\mathcal{E}_{A_{2}}^{\{N_{\alpha
}^{\prime\prime}\}}/\hbar}\sum_{B_{1}\cdots N_{\alpha}^{\prime}\cdots}%
\]%
\[
(\frac{-i}{\hbar})\int_{-\infty}^{t}dt^{\prime}K_{A_{2}B_{1}}^{\prime
field}(t^{\prime})\langle\Psi_{A_{2}}^{\{N_{\alpha}^{\prime\prime}\}}%
|\Xi_{B_{1}}^{\{N_{\alpha}^{\prime}\}}\rangle
\]%
\[
e^{it^{\prime}(\mathcal{E}_{A_{2}}^{\{N_{\alpha}^{\prime\prime}\}}%
-\mathcal{E}_{B_{1}}^{\{N_{\alpha}^{\prime}\}})/\hbar}%
\]%
\[
(\frac{-i}{\hbar})\int_{-\infty}^{t^{\prime}}dt^{\prime\prime}\langle
\Xi_{B_{1}}^{\{N_{\alpha}^{\prime}\}}|K_{B_{1}B}|\Xi_{B}^{\{N_{\alpha}%
\}}\rangle
\]%
\[
e^{it^{\prime\prime}(\mathcal{E}_{B_{1}}^{\{N_{\alpha}^{\prime}\}}%
-\mathcal{E}_{B}^{\{N_{\alpha}\}})/\hbar}%
\]%
\[
+\sum_{B_{2}\cdots N_{\alpha}^{\prime\prime}\cdots}\xi_{B_{2}}\Xi_{B_{2}%
}^{\{N_{\alpha}^{\prime\prime}\}}e^{-it\mathcal{E}_{B_{2}}^{\{N_{\alpha
}^{\prime\prime}\}}/\hbar}\sum_{A_{1}\cdots N_{\alpha}^{\prime}\cdots}%
\]%
\[
(\frac{-i}{\hbar})\int_{-\infty}^{t}dt^{\prime}J_{B_{2}A_{1}}^{\prime}%
\langle\Xi_{B_{2}}^{\{N_{\alpha}^{\prime\prime}\}}|\Psi_{A_{1}}^{\{N_{\alpha
}^{\prime}\}}\rangle
\]%
\[
e^{it^{\prime}(\mathcal{E}_{B_{2}}^{\{N_{\alpha}^{\prime\prime}\}}%
-\mathcal{E}_{A_{1}}^{\{N_{\alpha}^{\prime}\}})/\hbar}%
\]%
\[
(\frac{-i}{\hbar})\int_{-\infty}^{t^{\prime}}dt^{\prime\prime}K_{A_{1}%
B}^{\prime field}(t^{\prime\prime})|\langle\Psi_{A_{1}}^{\{N_{\alpha}^{\prime
}\}}|\Xi_{B}^{\{N_{\alpha}\}}\rangle
\]%
\[
e^{it^{\prime\prime}(\mathcal{E}_{A_{1}}^{\{N_{\alpha}^{\prime}\}}%
-\mathcal{E}_{B}^{\{N_{\alpha}\}})/\hbar}%
\]%
\[
+\sum_{B_{2}\cdots N_{\alpha}^{\prime\prime}\cdots}\xi_{B_{2}}\Xi_{B_{2}%
}^{\{N_{\alpha}^{\prime\prime}\}}e^{-it\mathcal{E}_{B_{2}}^{\{N_{\alpha
}^{\prime\prime}\}}/\hbar}\sum_{A_{1}\cdots N_{\alpha}^{\prime}\cdots}%
\]%
\[
(\frac{-i}{\hbar})\int_{-\infty}^{t}dt^{\prime}J_{B_{2}A_{1}}^{\prime
field}(t^{\prime})\langle\Xi_{B_{2}}^{\{N_{\alpha}^{\prime\prime}\}}%
|\Psi_{A_{1}}^{\{N_{\alpha}^{\prime}\}}\rangle
\]%
\[
e^{it^{\prime}(\mathcal{E}_{B_{2}}^{\{N_{\alpha}^{\prime\prime}\}}%
-\mathcal{E}_{A_{1}}^{\{N_{\alpha}^{\prime}\}})/\hbar}%
\]%
\[
\times(\frac{-i}{\hbar})\int_{-\infty}^{t^{\prime}}dt^{\prime\prime}%
\langle\Psi_{A_{1}}^{\{N_{\alpha}^{\prime}\}}|K_{A_{1}B}^{\prime}|\Xi
_{B}^{\{N_{\alpha}\}}\rangle
\]%
\[
e^{it^{\prime\prime}(\mathcal{E}_{A_{1}}^{\{N_{\alpha}^{\prime}\}}%
-\mathcal{E}_{B}^{\{N_{\alpha}\}})/\hbar}%
\]%
\[
+\sum_{B_{2}\cdots N_{\alpha}^{\prime\prime}\cdots}\xi_{B_{2}}\Xi_{B_{2}%
}^{\{N_{\alpha}^{\prime\prime}\}}e^{-it\mathcal{E}_{B_{2}}^{\{N_{\alpha
}^{\prime\prime}\}}/\hbar}\sum_{B_{1}\cdots N_{\alpha}^{\prime}\cdots}%
\]%
\[
(\frac{-i}{\hbar})\int_{-\infty}^{t}dt^{\prime}\langle\Xi_{B_{2}}%
^{\{N_{\alpha}^{\prime\prime}\}}|K_{B_{2}B_{1}}|\Xi_{B_{1}}^{\{N_{\alpha
}^{\prime}\}}\rangle
\]%
\[
e^{it^{\prime}(\mathcal{E}_{B_{2}}^{\{N_{\alpha}^{\prime\prime}\}}%
-\mathcal{E}_{B_{1}}^{\{N_{\alpha}^{\prime}\}})/\hbar}%
\]%
\[
(\frac{-i}{\hbar})\int_{-\infty}^{t^{\prime}}dt^{\prime\prime}K_{B_{1}%
B}^{field}(t^{\prime\prime})\langle\Xi_{B_{1}}^{\{N_{\alpha}^{\prime}\}}%
|\Xi_{B}^{\{N_{\alpha}\}}\rangle
\]%
\[
e^{it^{\prime\prime}(\mathcal{E}_{B_{1}}^{\{N_{\alpha}^{\prime}\}}%
-\mathcal{E}_{B}^{\{N_{\alpha}\}})/\hbar}%
\]%
\[
+\sum_{B_{2}\cdots N_{\alpha}^{\prime\prime}\cdots}\xi_{B_{2}}\Xi_{B_{2}%
}^{\{N_{\alpha}^{\prime\prime}\}}e^{-it\mathcal{E}_{B_{2}}^{\{N_{\alpha
}^{\prime\prime}\}}/\hbar}\sum_{B_{1}\cdots N_{\alpha}^{\prime}\cdots}%
\]%
\[
(\frac{-i}{\hbar})\int_{-\infty}^{t}dt^{\prime}K_{B_{2}B_{1}}^{field}%
(t^{\prime})\langle\Xi_{B_{2}}^{\{N_{\alpha}^{\prime\prime}\}}|\Xi_{B_{1}%
}^{\{N_{\alpha}^{\prime}\}}\rangle
\]%
\[
e^{it^{\prime}(\mathcal{E}_{B_{2}}^{\{N_{\alpha}^{\prime\prime}\}}%
-\mathcal{E}_{B_{1}}^{\{N_{\alpha}^{\prime}\}})/\hbar}%
\]%
\[
(\frac{-i}{\hbar})\int_{-\infty}^{t^{\prime}}dt^{\prime\prime}\langle
\Xi_{B_{1}}^{\{N_{\alpha}^{\prime}\}}|K_{B_{1}B}|\Xi_{B}^{\{N_{\alpha}%
\}}\rangle
\]%
\[
e^{it^{\prime\prime}(\mathcal{E}_{B_{1}}^{\{N_{\alpha}^{\prime}\}}%
-\mathcal{E}_{B}^{\{N_{\alpha}\}})/\hbar}.
\]
In Eqs.(\ref{pb},\ref{fb}), a factor $[1-f(E_{A_{2}})]$ must inserted before
component $\phi_{A_{2}}$ of the final state, a factor $[1-f(E_{B_{1}})]$ must
be included before component $\xi_{B_{1}}$ of the final state.

\section{Sum over final phonon states and average over initial phonon state}\label{suav}

For processes involving only $J$ and $J^{\prime}$, one first calculates inner
products to second order in the static displacements ($\theta^{A}$,
$\theta^{A_{1}}$, $\theta^{A_{2}}$, $\theta^{A_{3}}$). Then one multiplies the
inner products, neglects the terms of third and higher order in ($\theta^{A}%
$, $\theta^{A_{1}}$, $\theta^{A_{2}}$, $\theta^{A_{3}}$) and the terms which
involve more than one phonon changed in one mode. One can first compute
the sum over phonon final phonon states and average over initial phonon for a
single mode, and then multiply all modes together.

For processes involving e-ph interaction $K^{\prime}$ or $K$, the matrix
element of $K^{\prime}$ or $K$ has shape of $[\sum_{\alpha^{\prime}%
}f_{N_{\alpha^{\prime}}^{\prime\prime},N_{\alpha^{\prime}}^{\prime}}%
{\displaystyle\prod\limits_{\alpha(\neq\alpha^{\prime})}}
p_{N_{\alpha^{\prime}}^{\prime\prime},N_{\alpha^{\prime}}^{\prime}}]$. The sum
over final phonon states and average initial phonon state takes the form:%
\[
\sum_{\cdots N_{\alpha}\cdots}%
{\displaystyle\prod\limits_{\alpha}}
P(N_{\alpha})\sum_{\cdots N_{\alpha}^{\prime\prime}\cdots}[%
{\displaystyle\prod\limits_{\alpha}}
g_{N_{\alpha}^{\prime\prime},N_{\alpha}}]\sum_{\cdots N_{\alpha}^{\prime
}\cdots}%
\]%
\begin{equation}
\lbrack\sum_{\alpha^{\prime}}f_{N_{\alpha^{\prime}}^{\prime\prime}%
,N_{\alpha^{\prime}}^{\prime}}%
{\displaystyle\prod\limits_{\alpha(\neq\alpha^{\prime})}}
p_{N_{\alpha^{\prime}}^{\prime\prime},N_{\alpha^{\prime}}^{\prime}}][%
{\displaystyle\prod\limits_{\alpha}}
h_{N_{\alpha}^{\prime},N_{\alpha}}]. \label{y0}%
\end{equation}
Using the induction method, one can see that Eq.(\ref{y0}) is equivalent to:%
\[
\lbrack\sum_{\alpha^{\prime}}\frac{\sum_{N_{\alpha^{\prime}}^{\prime}%
N_{\alpha^{\prime}}^{\prime\prime}}f_{N_{\alpha^{\prime}}^{\prime\prime
},N_{\alpha^{\prime}}^{\prime}}g_{N_{\alpha^{\prime}}^{\prime\prime}%
,N_{\alpha^{\prime}}}h_{N_{\alpha^{\prime}}^{\prime},N_{\alpha^{\prime}}}%
}{\sum_{N_{\alpha^{\prime}}^{\prime}N_{\alpha^{\prime}}^{\prime\prime}%
}p_{N_{\alpha^{\prime}}^{\prime\prime},N_{\alpha^{\prime}}^{\prime}%
}g_{N_{\alpha^{\prime}}^{\prime\prime},N_{\alpha^{\prime}}}h_{N_{\alpha
^{\prime}}^{\prime},N_{\alpha^{\prime}}}}]
\]%
\begin{equation}
\lbrack%
{\displaystyle\prod\limits_{\alpha}}
(\sum_{N_{\alpha}^{\prime}N_{\alpha}^{\prime\prime}}p_{N_{\alpha}%
^{\prime\prime},N_{\alpha}^{\prime}}g_{N_{\alpha}^{\prime\prime},N_{\alpha}%
}h_{N_{\alpha}^{\prime},N_{\alpha}})]. \label{y1}%
\end{equation}
In the first square bracket in Eq.(\ref{y1}),
each term in the denominator is a pure number (does not include Kronecker delta signs). 
One can expand each fraction into a power series in
the static displacements ($\theta^{A}$, $\theta^{A_{1}}$, $\theta^{A_{2}}$,
$\theta^{A_{3}}$), neglecting terms of third and higher order in ($\theta^{A}%
$, $\theta^{A_{1}}$, $\theta^{A_{2}}$, $\theta^{A_{3}}$) and the terms in
which more than one phonon changed in one mode. The remaining calculation
is straightforward.

Taking Fig.5(a) as example, according to the rule in tables 2 and 3,
the contribution to current density is%
\[
(-N_{e}\hbar e/m)\sum_{A\cdots N_{\alpha}\cdots}f(E_{A})%
{\displaystyle\prod\limits_{\alpha}}
P(N_{\alpha})
\]%
\[
\sum_{A_{2}\cdots N_{\alpha}^{\prime\prime}\cdots}[1-f(E_{A_{2}})]\sum
_{B_{1}\cdots N_{\alpha}^{\prime}\cdots}%
\]%
\[
\Omega_{\mathbf{s}}^{-1}\int_{\Omega_{\mathbf{s}}}d\mathbf{r}(\phi_{A}%
\nabla_{\mathbf{r}}\phi_{A_{2}}^{\ast}-\phi_{A_{2}}^{\ast}\nabla_{\mathbf{r}%
}\phi_{A})\langle\Psi_{A_{2}}^{\{N_{\alpha}^{\prime\prime}\}}|\Psi
_{A}^{\{N_{\alpha}\}}\rangle
\]%
\[
e^{it(\mathcal{E}_{A_{2}}^{\{N_{\alpha}^{\prime\prime}\}}-\mathcal{E}%
_{A}^{\{N_{\alpha}\}})/\hbar}%
\]%
\[
\frac{i}{\hbar}\langle\Psi_{A_{2}}^{\{N_{\alpha}^{\prime\prime}\}}%
|K_{A_{2}B_{1}}^{\prime}|\Xi_{B_{1}}^{\{N_{\alpha}^{\prime}\}}\rangle^{\ast
}\int_{-\infty}^{t}dt^{\prime}e^{-it^{\prime}(\mathcal{E}_{A_{2}}%
^{\{N_{\alpha}^{\prime\prime}\}}-\mathcal{E}_{B_{1}}^{\{N_{\alpha}^{\prime}%
\}})/\hbar}%
\]%
\begin{equation}
\frac{i}{\hbar}\langle\Xi_{B_{1}}^{\{N_{\alpha}^{\prime}\}}|\Psi
_{A}^{\{N_{\alpha}\}}\rangle^{\ast}\int_{-\infty}^{t^{\prime}}dt^{\prime
\prime}[J_{B_{1}A}^{^{\prime}field}(t^{\prime\prime})]^{\ast} \label{7a}%
\end{equation}%
\[
e^{-it^{\prime\prime}(\mathcal{E}_{B_{1}}^{\{N_{\alpha}^{\prime}%
\}}-\mathcal{E}_{A}^{\{N_{\alpha}\}})/\hbar}.
\]
Expanding $\Psi_{A}^{\{N_{\alpha}\}}$ to second order in the origin shift
$\theta_{\alpha}^{A}$, and using the orthogonality of harmonic oscillator
wave functions, one has
\begin{equation}
\langle\Xi_{B_{1}}^{\{N_{\alpha}^{\prime}\}}|\Psi_{A}^{\{N_{\alpha}\}}\rangle=%
{\displaystyle\prod\limits_{\alpha}}
h_{N_{\alpha}^{\prime},N_{\alpha}}, \label{ab1}%
\end{equation}
where%
\[
h_{N_{\alpha}^{\prime},N_{\alpha}}=\delta_{N_{\alpha}N_{\alpha}^{\prime}%
}[1-\frac{1}{2}(N_{\alpha}+\frac{1}{2})(\theta_{\alpha}^{A})^{2}]
\]%
\begin{equation}
+(\theta_{\alpha}^{A})[(\frac{N_{\alpha}}{2})^{1/2}\delta_{N_{\alpha
}-1,N_{\alpha}^{\prime}}-(\frac{N_{\alpha}+1}{2})^{1/2}\delta_{N_{\alpha
}+1,N_{\alpha}^{\prime}}]. \label{b1a}%
\end{equation}
Similarly, one has%
\begin{equation}
\langle\Psi_{A_{2}}^{\{N_{\alpha}^{\prime\prime}\}}|\Psi_{A}^{\{N_{\alpha}%
\}}\rangle=%
{\displaystyle\prod\limits_{\alpha}}
g_{N_{\alpha}^{\prime\prime},N_{\alpha}}, \label{aa2}%
\end{equation}
where%
\begin{equation}
g_{N_{\alpha}^{\prime\prime},N_{\alpha}}=\delta_{N_{\alpha}N_{\alpha}%
^{\prime\prime}}[1-\frac{1}{2}(N_{\alpha}+\frac{1}{2})(\theta_{\alpha}^{A_{2}%
}-\theta_{\alpha}^{A})^{2}] \label{a2a}%
\end{equation}%
\[
+(\theta_{\alpha}^{A_{2}}-\theta_{\alpha}^{A})[(\frac{N_{\alpha}}{2}%
)^{1/2}\delta_{N_{\alpha}-1,N_{\alpha}^{\prime\prime}}-(\frac{N_{\alpha}+1}%
{2})^{1/2}\delta_{N_{\alpha}+1,N_{\alpha}^{\prime\prime}}],
\]
The matrix element of the e-ph interaction is\cite{epjb}%
\begin{equation}
\langle\Psi_{A_{2}}^{\{N_{\alpha}^{\prime\prime}\}}|K_{A_{2}B}^{\prime}%
|\Xi_{B_{1}}^{\{N_{\alpha}^{\prime}\}}\rangle=[\sum_{\alpha^{\prime}%
}f_{N_{\alpha^{\prime}}^{\prime\prime},N_{\alpha^{\prime}}^{\prime}}]%
{\displaystyle\prod\limits_{\alpha(\neq\alpha^{\prime})}}
p_{N_{\alpha}^{\prime\prime},N_{\alpha}^{\prime}}, \label{kp}%
\end{equation}
where%
\begin{equation}
f_{N_{\alpha^{\prime}}^{\prime\prime},N_{\alpha^{\prime}}^{\prime}}=K_{A_{2}%
B}^{\prime\alpha^{\prime}}(\frac{\hbar}{M_{\alpha^{\prime}}\omega
_{\alpha^{\prime}}})^{1/2} \label{f}%
\end{equation}%
\[
\{(\frac{N_{\alpha^{\prime}}^{\prime}}{2})^{1/2}\delta_{N_{\alpha^{\prime}%
}^{\prime\prime},N_{\alpha^{\prime}}^{\prime}-1}+(\frac{N_{\alpha^{\prime}%
}^{\prime}+1}{2})^{1/2}\delta_{N_{\alpha^{\prime}}^{\prime\prime}%
,N_{\alpha^{\prime}}^{\prime}+1}%
\]%
\[
+\frac{\theta_{\alpha^{\prime}}^{A_{2}}}{2}[N_{\alpha^{\prime}}^{\prime
\prime1/2}N_{\alpha^{\prime}}^{\prime1/2}-(N_{\alpha^{\prime}}^{\prime\prime
}+1)^{1/2}(N_{\alpha^{\prime}}^{\prime}+1)^{1/2}]\delta_{N_{\alpha^{\prime}%
}^{\prime\prime},N_{\alpha^{\prime}}^{\prime}}\},
\]
and%
\begin{equation}
p_{N_{\alpha}^{\prime\prime},N_{\alpha}^{\prime}}=\delta_{N_{\alpha}^{\prime
}N_{\alpha}^{\prime\prime}}[1-\frac{1}{2}(N_{\alpha}^{\prime}+\frac{1}%
{2})(\theta_{\alpha}^{A_{2}})^{2}] \label{p}%
\end{equation}%
\[
+\theta_{\alpha}^{A_{2}}[(\frac{N_{\alpha}^{\prime}}{2})^{1/2}\delta
_{N_{\alpha}^{\prime}-1,N_{\alpha}^{\prime\prime}}-(\frac{N_{\alpha}^{\prime
}+1}{2})^{1/2}\delta_{N_{\alpha}^{\prime}+1,N_{\alpha}^{\prime\prime}}].
\]
A simple and lengthy calculation shows that%
\[
\sum_{N_{\alpha^{\prime}}^{\prime\prime}}g_{N_{\alpha^{\prime}}^{\prime\prime
},N_{\alpha^{\prime}}}e^{-isN_{\alpha^{\prime}}^{\prime\prime}\omega
_{\alpha^{\prime}}}\sum_{N_{\alpha^{\prime}}^{\prime}}f_{N_{\alpha^{\prime}%
}^{\prime\prime},N_{\alpha^{\prime}}^{\prime}}h_{N_{\alpha^{\prime}}^{\prime
},N_{\alpha^{\prime}}}e^{-is^{\prime}N_{\alpha^{\prime}}^{\prime}%
\omega_{\alpha^{\prime}}}%
\]%
\[
=e^{-i(s+s^{\prime})N_{\alpha^{\prime}}\omega_{\alpha^{\prime}}}K_{A_{2}%
B}^{\prime\alpha^{\prime}}(\frac{\hbar}{M_{\alpha^{\prime}}\omega
_{\alpha^{\prime}}})^{1/2}%
\]%
\[
\{\theta_{\alpha^{\prime}}^{A}(e^{is^{\prime}\omega_{\alpha^{\prime}}}%
\frac{N_{\alpha^{\prime}}}{2}-e^{-is^{\prime}\omega_{\alpha^{\prime}}}%
\frac{N_{\alpha^{\prime}}+1}{2})-\frac{\theta_{\alpha^{\prime}}^{A_{2}}}{2}%
\]%
\begin{equation}
+(\theta_{\alpha^{\prime}}^{A_{2}}-\theta_{\alpha^{\prime}}^{A})(\frac
{N_{\alpha^{\prime}}}{2}e^{is\omega_{\alpha^{\prime}}}-\frac{N_{\alpha
^{\prime}}+1}{2}e^{-is\omega_{\alpha^{\prime}}})\}, \label{y3}%
\end{equation}
and%
\begin{equation}
\sum_{N_{\alpha}^{\prime\prime}}g_{N_{\alpha}^{\prime\prime},N_{\alpha}%
}e^{-isN_{\alpha}^{\prime\prime}\omega_{\alpha}}\sum_{N_{\alpha}^{\prime}%
}p_{N_{\alpha}^{\prime\prime},N_{\alpha}^{\prime}}h_{N_{\alpha}^{\prime
},N_{\alpha}}e^{-is^{\prime}N_{\alpha}^{\prime}\omega_{\alpha}} \label{y2}%
\end{equation}%
\[
=e^{-i(s+s^{\prime})N_{\alpha}\omega_{\alpha}}\{1-\frac{1}{2}(N_{\alpha}%
+\frac{1}{2})(\theta_{\alpha}^{A_{2}}-\theta_{\alpha}^{A})^{2}%
\]%
\[
-\frac{1}{2}(N_{\alpha}+\frac{1}{2})(\theta_{\alpha}^{A_{2}})^{2}-\frac{1}%
{2}(N_{\alpha}+\frac{1}{2})(\theta_{\alpha}^{A})^{2}%
\]%
\[
-\theta_{\alpha}^{A}\theta_{\alpha}^{A_{2}}[\frac{N_{\alpha}}{2}e^{is^{\prime
}\omega_{\alpha}}+e^{-is^{\prime}\omega_{\alpha}}\frac{N_{\alpha}+1}{2}]
\]%
\[
+(\theta_{\alpha}^{A_{2}}-\theta_{\alpha}^{A})\theta_{\alpha}^{A}%
[\frac{N_{\alpha}}{2}e^{i(s+s^{\prime})\omega_{\alpha}}+\frac{N_{\alpha}+1}%
{2}e^{-i(s+s^{\prime})\omega_{\alpha}}]
\]%
\[
+(\theta_{\alpha}^{A_{2}}-\theta_{\alpha}^{A})\theta_{\alpha}^{A_{2}}%
[\frac{N_{\alpha}}{2}e^{is\omega_{\alpha}}+\frac{N_{\alpha}+1}{2}%
e^{-is\omega_{\alpha}}]\}.
\]
Combine (\ref{y3}) and (\ref{y2}), we have%
\[
\frac{\sum_{N_{\alpha^{\prime}}^{\prime}N_{\alpha^{\prime}}^{\prime\prime}%
}f_{N_{\alpha^{\prime}}^{\prime\prime},N_{\alpha^{\prime}}^{\prime}%
}g_{N_{\alpha^{\prime}}^{\prime\prime},N_{\alpha^{\prime}}}h_{N_{\alpha
^{\prime}}^{\prime},N_{\alpha^{\prime}}}e^{-is^{\prime}N_{\alpha^{\prime}%
}^{\prime}\omega_{\alpha^{\prime}}-isN_{\alpha^{\prime}}^{\prime\prime}%
\omega_{\alpha^{\prime}}}}{\sum_{N_{\alpha^{\prime}}^{\prime}N_{\alpha
^{\prime}}^{\prime\prime}}p_{N_{\alpha^{\prime}}^{\prime\prime},N_{\alpha
^{\prime}}^{\prime}}g_{N_{\alpha^{\prime}}^{\prime\prime},N_{\alpha^{\prime}}%
}h_{N_{\alpha^{\prime}}^{\prime},N_{\alpha^{\prime}}}e^{-is^{\prime}%
N_{\alpha^{\prime}}^{\prime}\omega_{\alpha^{\prime}}-isN_{\alpha^{\prime}%
}^{\prime\prime}\omega_{\alpha^{\prime}}}}%
\]%
\[
=K_{A_{2}B}^{\prime\alpha^{\prime}}(\frac{\hbar}{M_{\alpha^{\prime}}%
\omega_{\alpha^{\prime}}})^{1/2}\{\theta_{\alpha^{\prime}}^{A}(e^{is^{\prime
}\omega_{\alpha^{\prime}}}\frac{N_{\alpha^{\prime}}}{2}-e^{-is^{\prime}%
\omega_{\alpha^{\prime}}}\frac{N_{\alpha^{\prime}}+1}{2})
\]%
\begin{equation}
-\frac{\theta_{\alpha^{\prime}}^{A_{2}}}{2}+(\theta_{\alpha^{\prime}}^{A_{2}%
}-\theta_{\alpha^{\prime}}^{A})(\frac{N_{\alpha^{\prime}}}{2}e^{is\omega
_{\alpha^{\prime}}}-\frac{N_{\alpha^{\prime}}+1}{2}e^{-is\omega_{\alpha
^{\prime}}})\} \label{y5}%
\end{equation}
Making use of Eqs.(\ref{y1}), (\ref{y2}) and (\ref{y5}), one reaches
Eq.(\ref{l5}).

To fulfil integrating out the vibrational degrees of freedom, two approximations are involved\cite{hol59}.
When computing inner product $\langle
\Psi_{A_{2}}^{\{N_{\alpha}^{\prime\prime}\}}|\Psi_{A}^{\{N_{\alpha}\}}\rangle$
and matrix element $\langle\Psi_{A_{2}}^{\{N_{\alpha}^{\prime\prime}%
\}}|K_{A_{2}B}^{\prime}|\Xi_{B_{1}}^{\{N_{\alpha}^{\prime}\}}\rangle$ etc, we
neglected (1) the terms which are $(\theta_{\alpha}^{A})^{3}$ and higher
order; and (2) the terms in which two or more phonons are changed in one mode.
When temperature is close to the melting point, the present results may be
only qualitative.
\section{Time integrals for conductivity}\label{timej}
\subsection{Localized state as initial state}\label{Ltj}
Fig.2(a):%
\begin{equation}
I_{A_{1}A\pm}(\omega)=\int_{-\infty}^{0}dse^{\pm i\omega s}e^{-is(E_{A_{1}%
}^{\prime}-E_{A}^{\prime})/\hbar} \label{r0+}%
\end{equation}%
\[
\exp\{-\sum_{\alpha}\frac{(\theta_{\alpha}^{A_{1}}-\theta_{\alpha}^{A})^{2}%
}{2}[\coth\frac{\beta\hbar\omega_{\alpha}}{2}(1-\cos\omega_{\alpha}%
s)-i\sin\omega_{\alpha}s]\}
\]
Fig.2(b):%
\begin{equation}
I_{B_{1}A\pm}=\int_{-\infty}^{0}dse^{\pm i\omega s}e^{-is(E_{B_{1}}%
-E_{A})/\hbar} \label{A2pn}%
\end{equation}%
\[
\exp(-\sum_{\alpha}\{\frac{1}{2}\coth\frac{\beta\hbar\omega_{\alpha}}%
{2}(\theta_{\alpha}^{A})^{2}%
\]%
\[
-\frac{1}{2}(\theta_{\alpha}^{A})^{2}[\coth\frac{\beta\hbar\omega_{\alpha}}%
{2}\cos s\omega_{\alpha}-i\sin s\omega_{\alpha}]\}),
\]
Fig.3(a):%
\begin{equation}
I_{A_{3}A_{1}A\pm}=\int_{-\infty}^{0}ds\int_{-\infty}^{0}ds^{\prime}e^{\pm
i\omega s^{\prime}} \label{A3pn}%
\end{equation}%
\begin{align*}
&  e^{is(E_{A_{1}}-E_{A})/\hbar-is^{\prime}(E_{A_{3}}-E_{A})/\hbar}\exp
(-\sum_{\alpha}\{\frac{1}{4}\coth\frac{\beta\hbar\omega_{\alpha}}{2}\\
&  [(\theta_{\alpha}^{A_{3}}-\theta_{\alpha}^{A_{1}})^{2}+(\theta_{\alpha
}^{A_{3}}-\theta_{\alpha}^{A})^{2}+(\theta_{\alpha}^{A_{1}}-\theta_{\alpha
}^{A})^{2}]
\end{align*}%
\[
-\frac{1}{2}(\theta_{\alpha}^{A_{3}}-\theta_{\alpha}^{A})(\theta_{\alpha
}^{A_{3}}-\theta_{\alpha}^{A_{1}})(\coth\frac{\beta\hbar\omega_{\alpha}}%
{2}\cos s^{\prime}\omega_{\alpha}-i\sin s^{\prime}\omega_{\alpha})
\]%
\begin{align*}
&  -\frac{1}{2}(\theta_{\alpha}^{A_{1}}-\theta_{\alpha}^{A})(\theta_{\alpha
}^{A_{3}}-\theta_{\alpha}^{A})\\
&  [\coth\frac{\beta\hbar\omega_{\alpha}}{2}\cos(s-s^{\prime})\omega_{\alpha
}+i\sin(s-s^{\prime})\omega_{\alpha}]
\end{align*}%
\[
+\frac{1}{2}(\theta_{\alpha}^{A_{1}}-\theta_{\alpha}^{A})(\theta_{\alpha
}^{A_{3}}-\theta_{\alpha}^{A_{1}})(\coth\frac{\beta\hbar\omega_{\alpha}}%
{2}\cos s\omega_{\alpha}+i\sin s\omega_{\alpha})\})
\]
Fig.3(b):%
\begin{equation}
I_{B_{1}A_{3}A\pm}=\int_{-\infty}^{0}ds\int_{-\infty}^{0}ds^{\prime}e^{\pm
i\omega s^{\prime}} \label{A4pn}%
\end{equation}%
\[
e^{is(E_{A_{3}}-E_{A})/\hbar-is^{\prime}(E_{B_{1}}-E_{A})/\hbar}%
\]%
\[
\exp(-\sum_{\alpha}\{\frac{1}{4}\coth\frac{\beta\hbar\omega_{\alpha}}%
{2}[(\theta_{\alpha}^{A_{3}})^{2}+(\theta_{\alpha}^{A})^{2}+(\theta_{\alpha
}^{A_{3}}-\theta_{\alpha}^{A})^{2}]
\]%
\[
-\frac{1}{2}\theta_{\alpha}^{A}\theta_{\alpha}^{A_{3}}(\coth\frac{\beta
\hbar\omega_{\alpha}}{2}\cos s^{\prime}\omega_{\alpha}-i\sin s^{\prime}%
\omega_{\alpha})
\]%
\[
+\frac{1}{2}\theta_{\alpha}^{A_{3}}(\theta_{\alpha}^{A_{3}}-\theta_{\alpha
}^{A})(\coth\frac{\beta\hbar\omega_{\alpha}}{2}\cos s\omega_{\alpha}+i\sin
s\omega_{\alpha})
\]%
\[
-\frac{1}{2}\theta_{\alpha}^{A}(\theta_{\alpha}^{A_{3}}-\theta_{\alpha}%
^{A})[\coth\frac{\beta\hbar\omega_{\alpha}}{2}\cos(s-s^{\prime})\omega
_{\alpha}+i\sin(s-s^{\prime})\omega_{\alpha}]\})
\]
Fig.3(c):%
\begin{equation}
I_{B_{3}A_{2}A\pm}=\int_{-\infty}^{0}ds\int_{-\infty}^{0}ds^{\prime}e^{\pm
i\omega s^{\prime}} \label{A5pn}%
\end{equation}%
\[
e^{is(E_{B_{3}}-E_{A})/\hbar-is^{\prime}(E_{A_{2}}-E_{A})/\hbar}%
\]%
\[
\exp(-\sum_{\alpha}\{\frac{1}{4}\coth\frac{\beta\hbar\omega_{\alpha}}%
{2}[(\theta_{\alpha}^{A_{2}}-\theta_{\alpha}^{A})^{2}+(\theta_{\alpha}^{A_{2}%
})^{2}+(\theta_{\alpha}^{A})^{2}]
\]%
\[
+\frac{1}{2}\theta_{\alpha}^{A_{2}}(\theta_{\alpha}^{A_{2}}-\theta_{\alpha
}^{A})(\coth\frac{\beta\hbar\omega_{\alpha}}{2}\cos s^{\prime}\omega_{\alpha
}-i\sin s^{\prime}\omega_{\alpha})
\]%
\[
-\frac{1}{2}\theta_{\alpha}^{A}(\theta_{\alpha}^{A_{2}}-\theta_{\alpha}%
^{A})\]
\[(\coth\frac{\beta\hbar\omega_{\alpha}}{2}\cos(s-s^{\prime})\omega
_{\alpha}+i\sin(s-s^{\prime})\omega_{\alpha})
\]%
\[
-\frac{1}{2}\theta_{\alpha}^{A}\theta_{\alpha}^{A_{2}}(\coth\frac{\beta
\hbar\omega_{\alpha}}{2}\cos s\omega_{\alpha}+i\sin s\omega_{\alpha})\})
\]
Fig.3(d):%
\begin{equation}
I_{B_{3}B_{1}A\pm}=\int_{-\infty}^{0}ds\int_{-\infty}^{0}ds^{\prime}e^{\pm
i\omega s^{\prime}} \label{A6pn}%
\end{equation}%
\[
e^{is(E_{B_{3}}-E_{A})/\hbar-is^{\prime}(E_{B_{1}}-E_{A})/\hbar}\exp
\{-\frac{1}{2}\sum_{\alpha}(\theta_{\alpha}^{A})^{2}%
\]%
\[
\lbrack\coth\frac{\beta\hbar\omega_{\alpha}}{2}-\coth\frac{\beta\hbar
\omega_{\alpha}}{2}\cos(s-s^{\prime})\omega_{\alpha}-i\sin(s-s^{\prime}%
)\omega_{\alpha}]\}.
\]
Fig.4(a):%
\begin{equation}
Q_{1A_{2}A_{1}A\pm}(\omega,T)=\int_{-\infty}^{0}ds\int_{-\infty}^{0}%
ds^{\prime}e^{\pm is\omega}e^{-is(E_{A_{2}}-E_{A_{1}})/\hbar} \label{l1}%
\end{equation}%
\[
e^{-i(s^{\prime}+s)(E_{A_{1}}-E_{A})/\hbar}W_{A_{2}A_{1}A}(s,s^{\prime},T),
\]
where%
\begin{equation}
W_{A_{2}A_{1}A}(s,s^{\prime},T)=\exp\{-\frac{1}{4}\sum_{\alpha}\coth
\frac{\beta\hbar\omega_{\alpha}}{2} \label{u6}%
\end{equation}%
\[
\lbrack(\theta_{\alpha}^{A_{2}}-\theta_{\alpha}^{A})^{2}+(\theta_{\alpha
}^{A_{2}}-\theta_{\alpha}^{A_{1}})^{2}+(\theta_{\alpha}^{A_{1}}-\theta
_{\alpha}^{A})^{2}]
\]%
\[
-\frac{1}{2}\sum_{\alpha}(\theta_{\alpha}^{A_{2}}-\theta_{\alpha}^{A_{1}%
})(\theta_{\alpha}^{A_{1}}-\theta_{\alpha}^{A})\]
\[[\coth\frac{\beta\hbar
\omega_{\alpha}}{2}\cos s^{\prime}\omega_{\alpha}-i\sin s^{\prime}%
\omega_{\alpha}]
\]%
\[
+\frac{1}{2}\sum_{\alpha}(\theta_{\alpha}^{A_{1}}-\theta_{\alpha}^{A}%
)(\theta_{\alpha}^{A_{2}}-\theta_{\alpha}^{A})
\]%
\[
\lbrack\coth\frac{\beta\hbar\omega_{\alpha}}{2}\cos(s+s^{\prime}%
)\omega_{\alpha}-i\sin(s+s^{\prime})\omega_{\alpha}]
\]%
\[
+\frac{1}{2}\sum_{\alpha}(\theta_{\alpha}^{A_{2}}-\theta_{\alpha}^{A}%
)(\theta_{\alpha}^{A_{2}}-\theta_{\alpha}^{A_{1}})\]
\[[\coth\frac{\beta\hbar
\omega_{\alpha}}{2}\cos s\omega_{\alpha}-i\sin s\omega_{\alpha}]\}.
\]
Fig.4(b):%
\begin{equation}
Q_{2A_{2}A_{1}A\pm}(\omega,T)=\int_{-\infty}^{0}dse^{-is(E_{A_{2}}-E_{A_{1}%
})/\hbar} \label{l2}%
\end{equation}%
\[
\int_{-\infty}^{0}ds^{\prime}e^{\pm i\omega(s^{\prime}+s)}e^{-i(s^{\prime
}+s)(E_{A_{1}}-E_{A})/\hbar}W_{A_{2}A_{1}A}(s,s^{\prime},T).
\]
Fig.4(c):%
\begin{equation}
Q_{1B_{2}A_{1}A\pm}(\omega,T)=\int_{-\infty}^{0}dse^{\pm is\omega
}e^{-is(E_{B_{2}}-E_{A_{1}})/\hbar} \label{l3}%
\end{equation}%
\[
\int_{-\infty}^{0}ds^{\prime}e^{-i(s+s^{\prime})(E_{A_{1}}-E_{A})/\hbar
}W_{B_{2}A_{1}A}(s,s^{\prime},T),
\]
where%
\begin{equation}
W_{B_{2}A_{1}A}(s,s^{\prime},T)= \label{u7}%
\end{equation}%
\[
\exp\{-\frac{1}{4}\sum_{\alpha}\coth\frac{\beta\hbar\omega_{\alpha}}%
{2}[(\theta_{\alpha}^{A})^{2}+(\theta_{\alpha}^{A_{1}})^{2}+(\theta_{\alpha
}^{A_{1}}-\theta_{\alpha}^{A})^{2}]
\]%
\[
+\frac{1}{2}\sum_{\alpha}\theta_{\alpha}^{A_{1}}(\theta_{\alpha}^{A_{1}%
}-\theta_{\alpha}^{A})[\coth\frac{\beta\hbar\omega_{\alpha}}{2}\cos s^{\prime
}\omega_{\alpha}-i\sin s^{\prime}\omega_{\alpha}]
\]%
\[
-\frac{1}{2}\sum_{\alpha}(\theta_{\alpha}^{A_{1}}-\theta_{\alpha}^{A}%
)\theta_{\alpha}^{A}[\coth\frac{\beta\hbar\omega_{\alpha}}{2}\cos(s+s^{\prime
})\omega_{\alpha}-i\sin(s+s^{\prime})\omega_{\alpha}]
\]%
\[
+\frac{1}{2}\sum_{\alpha}\theta_{\alpha}^{A}\theta_{\alpha}^{A_{1}}[\coth
\frac{\beta\hbar\omega_{\alpha}}{2}\cos s\omega_{\alpha}-i\sin s\omega
_{\alpha}]\}.
\]
Fig.4(d):%
\begin{equation}
Q_{2B_{2}A_{1}A\pm}(\omega,T)=\int_{-\infty}^{0}dse^{-is(E_{B_{2}}-E_{A_{1}%
})/\hbar} \label{l4}%
\end{equation}%
\[
\int_{-\infty}^{0}ds^{\prime}e^{\pm i(s+s^{\prime})\omega}e^{-i(s+s^{\prime
})(E_{A_{1}}-E_{A})/\hbar}W_{B_{2}A_{1}A}(s,s^{\prime},T).
\]
Fig.5(a):%
\begin{equation}
Q_{1A_{2}B_{1}A\pm}^{K^{\prime}}(\omega,T)=\int_{-\infty}^{0}dse^{-is(E_{A_{2}%
}-E_{B_{1}})/\hbar} \label{l5}%
\end{equation}%
\[
\int_{-\infty}^{0}ds^{\prime}e^{\pm i(s+s^{\prime})\omega}e^{-i(s+s^{\prime
})(E_{B_{1}}-E_{A})/\hbar}%
\]
\[
\sum_{\alpha^{\prime}}[K_{A_{2}B}^{\prime\alpha^{\prime}}(\frac{\hbar
}{M_{\alpha^{\prime}}\omega_{\alpha^{\prime}}})^{1/2}\frac{1}{2}%
\]%
\[
\{\theta_{\alpha^{\prime}}^{A}(i\coth\frac{\beta\hbar\omega_{\alpha^{\prime}}%
}{2}\sin s^{\prime}\omega_{\alpha^{\prime}}-\cos s^{\prime}\omega
_{\alpha^{\prime}})-\theta_{\alpha^{\prime}}^{A_{2}}%
\]%
\[
+(\theta_{\alpha^{\prime}}^{A_{2}}-\theta_{\alpha^{\prime}}^{A})(i\coth
\frac{\beta\hbar\omega_{\alpha^{\prime}}}{2}\sin s\omega_{\alpha^{\prime}%
}-\cos s\omega_{\alpha^{\prime}})\}]
\]%
\[
\exp\{-\frac{1}{4}\sum_{\alpha}\coth\frac{\beta\hbar\omega_{\alpha}}%
{2}[(\theta_{\alpha}^{A_{2}}-\theta_{\alpha}^{A})^{2}+(\theta_{\alpha}^{A_{2}%
})^{2}+(\theta_{\alpha}^{A})^{2}]
\]%
\[
-\sum_{\alpha}\frac{\theta_{\alpha}^{A}\theta_{\alpha}^{A_{2}}}{2}[\coth
\frac{\beta\hbar\omega_{\alpha}}{2}\cos s^{\prime}\omega_{\alpha}-i\sin
s^{\prime}\omega_{\alpha}]
\]%
\[
+\sum_{\alpha}(\theta_{\alpha}^{A_{2}}-\theta_{\alpha}^{A})\frac
{\theta_{\alpha}^{A}}{2}\]
\[[\coth\frac{\beta\hbar\omega_{\alpha^{\prime}}}{2}%
\cos(s+s^{\prime})\omega_{\alpha}-i\sin(s+s^{\prime})\omega_{\alpha}]
\]%
\[
+\sum_{\alpha}(\theta_{\alpha}^{A_{2}}-\theta_{\alpha}^{A})\frac
{\theta_{\alpha}^{A_{2}}}{2}[\coth\frac{\beta\hbar\omega_{\alpha}}{2}\cos
s\omega_{\alpha}-i\sin s\omega_{\alpha}]\}.
\]
Fig.5(b):%
\begin{equation}
Q_{2A_{2}B_{1}A\pm}(\omega,T)=\int_{-\infty}^{0}dse^{\pm is\omega
}e^{-is(E_{A_{2}}-E_{B_{1}})/\hbar} \label{l6}%
\end{equation}%
\[
\int_{-\infty}^{0}ds^{\prime}e^{-i(s+s^{\prime})(E_{B_{1}}-E_{A})/\hbar
}W_{A_{2}B_{1}A}(s,s^{\prime},T),
\]
where%
\begin{equation}
W_{A_{2}B_{1}A}(s,s^{\prime},T)= \label{u8}%
\end{equation}%
\[
\exp\{-\frac{1}{4}\sum_{\alpha}\coth\frac{\beta\hbar\omega_{\alpha}}%
{2}[(\theta_{\alpha}^{A_{2}}-\theta_{\alpha}^{A})^{2}+(\theta_{\alpha}^{A_{2}%
})^{2}+(\theta_{\alpha}^{A})^{2}]
\]%
\[
+\frac{1}{2}\sum_{\alpha}\theta_{\alpha}^{A_{2}}\theta_{\alpha}^{A}[\coth
\frac{\beta\hbar\omega_{\alpha}}{2}\cos s^{\prime}\omega_{\alpha}-i\sin
s^{\prime}\omega_{\alpha}]
\]%
\[
-\frac{1}{2}\sum_{\alpha}\theta_{\alpha}^{A}(\theta_{\alpha}^{A_{2}}%
-\theta_{\alpha}^{A})\]
\[[\coth\frac{\beta\hbar\omega_{\alpha}}{2}\cos
(s+s^{\prime})\omega_{\alpha}-i\sin(s+s^{\prime})\omega_{\alpha}]
\]%
\[
+\frac{1}{2}\sum_{\alpha}(\theta_{\alpha}^{A_{2}}-\theta_{\alpha}^{A}%
)\theta_{\alpha}^{A_{2}}[\coth\frac{\beta\hbar\omega_{\alpha}}{2}\cos
s\omega_{\alpha}-i\sin s\omega_{\alpha}]\}.
\]
Fig.5(c):%
\begin{equation}
Q_{1B_{2}B_{1}A\pm}(\omega,T)=\int_{-\infty}^{0}dse^{-is(E_{B_{2}}-E_{B_{1}%
})/\hbar} \label{l61}%
\end{equation}%
\[
\int_{-\infty}^{0}ds^{\prime}e^{\pm i\omega(s+s^{\prime})}e^{-i(s+s^{\prime
})(E_{B_{1}}-E_{A})/\hbar}%
\]%
\[
\lbrack\sum_{\alpha^{\prime}}\theta_{\alpha^{\prime}}^{A}K_{B_{2}B_{1}%
}^{\alpha^{\prime}}(\frac{\hbar}{M_{\alpha^{\prime}}\omega_{\alpha^{\prime}}%
})^{1/2}\frac{1}{2}%
\]%
\[
\lbrack(\coth\frac{\beta\hbar\omega_{\alpha}}{2}\cos s^{\prime}\omega
_{\alpha^{\prime}}-i\sin s^{\prime}\omega_{\alpha^{\prime}})
\]%
\[
+(i\coth\frac{\beta\hbar\omega_{\alpha}}{2}\sin s\omega_{\alpha^{\prime}}-\cos
s\omega_{\alpha^{\prime}})]
\]%
\[
\exp\{-\frac{1}{2}\sum_{\alpha}\coth\frac{\beta\hbar\omega_{\alpha}}{2}%
(\theta_{\alpha}^{A})^{2}%
\]%
\[
+\frac{1}{2}\sum_{\alpha}(\theta_{\alpha}^{A})^{2}\]
\[[\coth\frac{\beta\hbar
\omega_{\alpha}}{2}\cos(s+s^{\prime})\omega_{\alpha}-i\sin(s+s^{\prime}%
)\omega_{\alpha}]\}.
\]
Fig.5(d):%
\begin{equation}
Q_{2B_{2}B_{1}A\pm}(\omega,T)=\int_{-\infty}^{0}dse^{\pm is\omega
}e^{-is(E_{B_{2}}-E_{B_{1}})/\hbar} \label{l7}%
\end{equation}%
\[
\int_{-\infty}^{0}ds^{\prime}e^{-i(s+s^{\prime})(E_{B_{1}}-E_{A})/\hbar}%
\exp\{-\frac{1}{2}\sum_{\alpha}\coth\frac{\beta\hbar\omega_{\alpha}}{2}%
(\theta_{\alpha}^{A})^{2}%
\]%
\[
+\frac{1}{2}\sum_{\alpha}(\theta_{\alpha}^{A})^{2}[\coth\frac{\beta\hbar
\omega_{\alpha}}{2}\cos(s+s^{\prime})\omega_{\alpha}-i\sin(s+s^{\prime}%
)\omega_{\alpha}]\}.
\]
\subsection{Extended state as initial state}
Fig.6(a):%
\begin{equation}
I_{A_{2}B\pm}=\int_{-\infty}^{0}dse^{\pm i\omega s}e^{-is(E_{A_{2}}%
-E_{B})/\hbar} \label{B1pn}%
\end{equation}%
\begin{align*}
&  \exp\{-\frac{1}{2}\coth\frac{\beta\hbar\omega_{\alpha}}{2}(\theta_{\alpha
}^{A_{2}})^{2}\\
&  +\frac{1}{2}(\theta_{\alpha}^{A_{2}})^{2}[\coth\frac{\beta\hbar
\omega_{\alpha}}{2}\cos s\omega_{\alpha}-i\sin s\omega_{\alpha}]\}.
\end{align*}
Fig.6(b):%
\begin{equation}
I_{B_{1}B\pm}=\int_{-\infty}^{0}dse^{\pm i\omega s}e^{-is(E_{B_{1}}%
-E_{B})/\hbar} \label{B2pn}%
\end{equation}%
\[
=\int_{-\infty}^{0}dse^{is(\pm\hbar\omega-E_{B_{1}}+E_{B}-i0^{+})/\hbar}%
\]
\[
=\frac{\hbar}{i(\pm\hbar\omega-E_{B_{1}}+E_{B}-i0^{+})}e^{is(\pm\hbar
\omega-E_{B_{1}}+E_{B}-i0^{+})/\hbar}%
\]%
\[
\rightarrow\frac{\hbar}{i(\pm\hbar\omega-E_{B_{1}}+E_{B}-i0^{+})}%
\]
Fig.7(a):%
\begin{equation}
I_{A_{3}A_{2}B\pm}=-[\sum_{\alpha^{\prime}}K_{A_{3}B}^{\prime\alpha^{\prime}%
}(\frac{\hbar}{M_{\alpha^{\prime}}\omega_{\alpha^{\prime}}})^{1/2}\frac
{\theta_{\alpha^{\prime}}^{A_{3}}}{2}] \label{B3pn}%
\end{equation}%
\[
\int_{-\infty}^{0}ds\int_{-\infty}^{0}dse^{\pm i\omega s^{\prime}%
}e^{-is^{\prime}(E_{A_{2}}-E_{B})/\hbar+is(E_{A_{3}}-E_{B})/\hbar}%
\]%
\[
\exp(-\sum_{\alpha}\{\frac{1}{4}\coth\frac{\beta\hbar\omega_{\alpha}}%
{2}[(\theta_{\alpha}^{A_{3}})^{2}+(\theta_{\alpha}^{A_{2}})^{2}+(\theta
_{\alpha}^{A_{2}}-\theta_{\alpha}^{A_{3}})^{2}]
\]%
\[
+\frac{1}{2}\theta_{\alpha}^{A_{2}}(\theta_{\alpha}^{A_{2}}-\theta_{\alpha
}^{A_{3}})[\coth\frac{\beta\hbar\omega_{\alpha}}{2}\cos s^{\prime}%
\omega_{\alpha}-i\sin s^{\prime}\omega_{\alpha}]
\]%
\[
+\frac{1}{2}\theta_{\alpha}^{A_{3}}(\theta_{\alpha}^{A_{2}}-\theta_{\alpha
}^{A_{3}})[\coth\frac{\beta\hbar\omega_{\alpha}}{2}\cos s\omega_{\alpha}+i\sin
s\omega_{\alpha}]
\]%
\[
+\frac{1}{2}\theta_{\alpha}^{A_{2}}\theta_{\alpha}^{A_{3}}[\coth\frac
{\beta\hbar\omega_{\alpha}}{2}\cos(s-s^{\prime})\omega_{\alpha}+i\sin
(s-s^{\prime})\omega_{\alpha}]\}).
\]
Fig.7(b):%
\begin{equation}
I_{A_{3}B_{1}B\pm}=-\sum_{\alpha^{\prime}}K_{A_{3}B}^{\prime\alpha^{\prime}%
}(\frac{\hbar}{M_{\alpha^{\prime}}\omega_{\alpha^{\prime}}})^{1/2}\frac
{\theta_{\alpha^{\prime}}^{A_{3}}}{2} \label{B4pn}%
\end{equation}%
\[
\int_{-\infty}^{0}ds\int_{-\infty}^{0}dse^{\pm i\omega s^{\prime}%
}e^{is(E_{A_{3}}-E_{B})/\hbar-is^{\prime}(E_{B_{1}}-E_{B})/\hbar}%
\]%
\[
\exp(-\sum_{\alpha}(\theta_{\alpha}^{A_{3}})^{2}[(N_{\alpha}+\frac{1}%
{2})+\frac{N_{\alpha}}{2}e^{-is\omega_{\alpha}}+\frac{N_{\alpha}+1}%
{2}e^{is\omega_{\alpha}}])
\]
Fig.7(c):%
\begin{equation}
I_{A_{2}B_{3}B\pm}=\int_{-\infty}^{0}ds\int_{-\infty}^{0}dse^{\pm i\omega
s^{\prime}} \label{B5pn}%
\end{equation}%
\[
e^{is(E_{B_{3}}-E_{B})/\hbar-is^{\prime}(E_{A_{2}}-E_{B})/\hbar}\sum
_{\alpha^{\prime}}K_{B_{3}B}^{\alpha^{\prime}}(\frac{\hbar}{M_{\alpha^{\prime
}}\omega_{\alpha^{\prime}}})^{1/2}\frac{(\theta_{\alpha^{\prime}}^{A_{2}})}{2}%
\]%
\[
(i\coth\frac{\beta\hbar\omega_{\alpha}}{2}[\sin(s-s^{\prime})\omega
_{\alpha^{\prime}}-\sin s\omega_{\alpha^{\prime}}]\]
\[+\cos(s-s^{\prime}%
)\omega_{\alpha^{\prime}}-\cos s\omega_{\alpha^{\prime}})
\]%
\[
\exp\{-\sum_{\alpha}\frac{(\theta_{\alpha}^{A_{2}})^{2}}{2}[\coth\frac
{\beta\hbar\omega_{\alpha}}{2}\]
\[-\coth\frac{\beta\hbar\omega_{\alpha}}{2}\cos
s\omega_{\alpha}+i\sin s\omega_{\alpha}]\},
\]
The time integral of Fig.7(d) is $0$.

Fig.8(a):%
\begin{equation}
S_{1A_{2}A_{1}B\pm}(\omega,T)=\int_{-\infty}^{0}dse^{\pm is\omega
}e^{-is(E_{A_{2}}-E_{A_{1}})/\hbar} \label{e1}%
\end{equation}
\[
\int_{-\infty}^{0}ds^{\prime}e^{-i(s+s^{\prime})(E_{A_{1}}-E_{B})/\hbar}%
[\sum_{\alpha^{\prime}}K_{A_{2}B_{1}}^{\prime\alpha^{\prime}}(\frac{\hbar
}{M_{\alpha^{\prime}}\omega_{\alpha^{\prime}}})^{1/2}\frac{1}{2}%
\]%
\[
\{(\theta_{\alpha^{\prime}}^{A_{1}}-\theta_{\alpha^{\prime}}^{A_{2}}%
)[i\coth\frac{\beta\hbar\omega_{\alpha^{\prime}}}{2}\sin s^{\prime}%
\omega_{\alpha^{\prime}}-\cos s^{\prime}\omega_{\alpha^{\prime}}]
\]%
\[
+\theta_{\alpha^{\prime}}^{A_{2}}[i\coth\frac{\beta\hbar\omega_{\alpha
^{\prime}}}{2}\sin(s+s^{\prime})\omega_{\alpha^{\prime}}-\cos(s+s^{\prime
})\omega_{\alpha^{\prime}}]-\theta_{\alpha^{\prime}}^{A_{1}}\}]
\]%
\[
\exp\{-\frac{1}{2}\sum_{\alpha}[(\theta_{\alpha}^{A_{2}}-\theta_{\alpha
}^{A_{1}})^{2}+(\theta_{\alpha}^{A_{2}})^{2}+(\theta_{\alpha}^{A_{1}}%
)^{2}]\coth\frac{\beta\hbar\omega_{\alpha}}{2}%
\]%
\[
+\frac{1}{2}\sum_{\alpha}\theta_{\alpha}^{A_{2}}(\theta_{\alpha}^{A_{2}%
}-\theta_{\alpha}^{A_{1}})(\coth\frac{\beta\hbar\omega_{\alpha}}{2}\cos
s\omega_{\alpha}-i\sin s\omega_{\alpha})
\]%
\[
+\frac{1}{2}\sum_{\alpha}\theta_{\alpha}^{A_{2}}\theta_{\alpha}^{A_{1}}%
(\coth\frac{\beta\hbar\omega_{\alpha}}{2}\cos(s+s^{\prime})\omega_{\alpha
}-i\sin(s+s^{\prime})\omega_{\alpha})
\]%
\[
-\frac{1}{2}\sum_{\alpha}\theta_{\alpha}^{A_{1}}(\theta_{\alpha}^{A_{2}%
}-\theta_{\alpha}^{A_{1}})(\coth\frac{\beta\hbar\omega_{\alpha}}{2}\cos
s^{\prime}\omega_{\alpha}-i\sin s^{\prime}\omega_{\alpha})\}
\]
Fig.8(b):%
\begin{equation}
S_{A_{2}A_{1}B\pm}(\omega,T)=\int_{-\infty}^{0}dse^{-is(E_{A_{2}}-E_{A_{1}%
})/\hbar} \label{e2}%
\end{equation}%
\[
\int_{-\infty}^{0}ds^{\prime}e^{\pm i(s+s^{\prime})\omega}e^{-i(s+s^{\prime
})(E_{A_{1}}-E_{B})/\hbar}W_{A_{2}A_{1}B}(s,s^{\prime},T),
\]
where%
\[
W_{A_{2}A_{1}B}(s,s^{\prime},T)=
\]%
\[
\exp\{-\frac{1}{4}\sum_{\alpha}\coth\frac{\beta\hbar\omega_{\alpha}}%
{2}[(\theta_{\alpha}^{A_{2}})^{2}+(\theta_{\alpha}^{A_{2}}-\theta_{\alpha
}^{A_{1}})^{2}+(\theta_{\alpha}^{A_{1}})^{2}]
\]%
\[
-\frac{1}{2}\sum_{\alpha}(\theta_{\alpha}^{A_{2}}-\theta_{\alpha}^{A_{1}%
})\theta_{\alpha}^{A_{1}}[\coth\frac{\beta\hbar\omega_{\alpha}}{2}\cos
s^{\prime}\omega_{\alpha}-i\sin s^{\prime}\omega_{\alpha}]
\]%
\[
+\frac{1}{2}\sum_{\alpha}\theta_{\alpha}^{A_{1}}\theta_{\alpha}^{A_{2}}%
[\coth\frac{\beta\hbar\omega_{\alpha}}{2}\cos(s+s^{\prime})\omega_{\alpha
}-i\sin(s+s^{\prime})\omega_{\alpha}]
\]%
\begin{equation}
+\frac{1}{2}\sum_{\alpha}\theta_{\alpha}^{A_{2}}(\theta_{\alpha}^{A_{2}%
}-\theta_{\alpha}^{A_{1}})[\coth\frac{\beta\hbar\omega_{\alpha}}{2}\cos
s\omega_{\alpha}-i\sin s\omega_{\alpha}]\} \label{u9}%
\end{equation}
Fig.8(c):%
\begin{equation}
S_{1B_{2}A_{1}B\pm}(\omega,T)=\int_{-\infty}^{0}dse^{-is(E_{B_{2}}-E_{A_{1}%
})/\hbar} \label{e3}%
\end{equation}%
\[
\int_{-\infty}^{0}ds^{\prime}e^{\pm i(s+s^{\prime})\omega}e^{-i(s+s^{\prime
})(E_{A_{1}}-E_{B})/\hbar}%
\]%
\begin{align*}
&  \exp\{-\frac{1}{2}\sum_{\alpha}\coth\frac{\beta\hbar\omega_{\alpha}}%
{2}(\theta_{\alpha}^{A_{1}})^{2}\\
&  +\frac{1}{2}\sum_{\alpha}(\theta_{\alpha}^{A_{1}})^{2}[\coth\frac
{\beta\hbar\omega_{\alpha}}{2}\cos s^{\prime}\omega_{\alpha}-i\sin s^{\prime
}\omega_{\alpha}]\}.
\end{align*}
Fig.8(d):%
\begin{equation}
S_{2B_{2}A_{1}B\pm}(\omega,T)=\int_{-\infty}^{0}dse^{\pm is\omega
}e^{-is(E_{B_{2}}-E_{A_{1}})/\hbar} \label{e4}%
\end{equation}%
\begin{align*}
&  \int_{-\infty}^{0}ds^{\prime}e^{-i(s+s^{\prime})(E_{A_{1}}-E_{B})/\hbar
}[\sum_{\alpha^{\prime}}K_{A_{2}B_{1}}^{\prime\alpha^{\prime}}(\frac{\hbar
}{M_{\alpha^{\prime}}\omega_{\alpha^{\prime}}})^{1/2}\frac{\theta_{\alpha
}^{A_{1}}}{2}\\
&  (\cos s^{\prime}\omega_{\alpha}-i\coth\frac{\beta\hbar\omega_{\alpha}}%
{2}\sin s^{\prime}\omega_{\alpha}-1)]
\end{align*}%
\begin{align*}
&  \exp\{-\frac{1}{2}\sum_{\alpha}(\theta_{\alpha}^{A_{1}})^{2}\coth
\frac{\beta\hbar\omega_{\alpha}}{2}\\
&  -\frac{1}{2}\sum_{\alpha}(\theta_{\alpha}^{A_{1}})^{2}(\coth\frac
{\beta\hbar\omega_{\alpha}}{2}\cos s^{\prime}\omega_{\alpha}-i\sin s^{\prime
}\omega_{\alpha})\}.
\end{align*}
Fig.9(a):%
\[
S_{1A_{2}B_{1}B\pm}(\omega,T)=\int_{-\infty}^{0}dse^{\pm is\omega
}e^{-is(E_{A_{2}}-E_{B_{1}})/\hbar}%
\]%
\[
\int_{-\infty}^{0}ds^{\prime}e^{-i(s+s^{\prime})(E_{B_{1}}-E_{B})/\hbar}%
\{\sum_{\alpha^{\prime}}K_{B_{2}B}^{\alpha^{\prime}}(\frac{\hbar}%
{M_{\alpha^{\prime}}\omega_{\alpha^{\prime}}})^{1/2}\theta_{\alpha^{\prime}%
}^{A_{2}}\frac{1}{2}%
\]%
\begin{align*}
&  [(i\coth\frac{\beta\hbar\omega_{\alpha}}{2}\sin s^{\prime}\omega
_{\alpha^{\prime}}-\cos s^{\prime}\omega_{\alpha^{\prime}})\\
&  -(i\coth\frac{\beta\hbar\omega_{\alpha}}{2}\sin(s+s^{\prime})\omega
_{\alpha^{\prime}}-\cos(s+s^{\prime})\omega_{\alpha^{\prime}})]\}
\end{align*}%
\[
\exp\{-\frac{1}{2}\sum_{\alpha}\coth\frac{\beta\hbar\omega_{\alpha}}{2}%
(\theta_{\alpha}^{A_{2}})^{2}%
\]%
\begin{equation}
+\frac{1}{2}\sum_{\alpha}(\theta_{\alpha}^{A_{2}})^{2}(\coth\frac{\beta
\hbar\omega_{\alpha}}{2}\cos s\omega_{\alpha}-i\sin s\omega_{\alpha})\}
\label{x11}%
\end{equation}
Fig.9(b):%
\begin{equation}
S_{2A_{2}B_{1}B\pm}(\omega,T)=\int_{-\infty}^{0}dse^{-is(E_{A_{2}}-E_{B_{1}%
})/\hbar} \label{e5}%
\end{equation}%
\[
\int_{-\infty}^{0}ds^{\prime}e^{\pm i(s+s^{\prime})\omega}e^{-i(s+s^{\prime
})(E_{B_{1}}-E_{B})/\hbar}%
\]%
\begin{align*}
&  [\sum_{\alpha^{\prime}}K_{A_{2}B_{1}}^{\prime\alpha^{\prime}}(\frac{\hbar
}{M_{\alpha^{\prime}}\omega_{\alpha^{\prime}}})^{1/2}\frac{\theta
_{\alpha^{\prime}}^{A_{2}}}{2}\\
&  (\cos s\omega_{\alpha^{\prime}}-i\coth\frac{\beta\hbar\omega_{\alpha
^{\prime}}}{2}\sin s\omega_{\alpha^{\prime}}-1)]
\end{align*}%
\begin{align*}
&  \exp\{-\frac{1}{2}\sum_{\alpha}(\theta_{\alpha}^{A_{2}})^{2}\coth
\frac{\beta\hbar\omega_{\alpha}}{2}\\
&  -\frac{1}{2}\sum_{\alpha}(\theta_{\alpha}^{A_{2}})^{2}[\coth\frac
{\beta\hbar\omega_{\alpha}}{2}\cos s\omega_{\alpha}-i\sin s\omega_{\alpha}]\}.
\end{align*}
The time integral of Fig.9(c) is$\ 0$.
The time integral of Fig.9(d) is $0$.
\subsection{Order K$^{2}$ contributions}
Fig.10(a):%
\begin{equation}
I_{B_{3}B_{2}B_{1}B\pm}=\int_{-\infty}^{0}ds\int_{-\infty}^{0}ds^{\prime}%
\int_{-\infty}^{0}ds^{\prime\prime}e^{\pm i\omega s^{\prime\prime}}
\label{B7pn}%
\end{equation}%
\[
\exp\{-is^{\prime\prime}(E_{B_{1}}-E_{B})/\hbar+is(E_{B_{3}}-E_{B}%
)/\hbar+is^{\prime}(E_{B_{2}}-E_{B})/\hbar\}
\]%
\begin{align*}
&  \sum_{\alpha^{\prime}}K_{B_{3}B_{2}}^{\alpha^{\prime}}K_{B_{2}B}%
^{\alpha^{\prime}}\frac{\hbar}{M_{\alpha^{\prime}}\omega_{\alpha^{\prime}}%
}\frac{1}{2}\\
&  (\coth\frac{\beta\hbar\omega_{\alpha^{\prime}}}{2}\cos s^{\prime}%
\omega_{\alpha^{\prime}}+i\sin s^{\prime}\omega_{\alpha^{\prime}}).
\end{align*}
adiabatic introduce interaction%
\[
E_{B_{3}}-E_{B}\rightarrow E_{B_{3}}-E_{B}-i0^{+}%
\]%
\[
\int_{-\infty}^{0}dse^{is(E_{B_{3}}-E_{B})/\hbar}=\frac{\hbar}{i(E_{B_{3}%
}-E_{B})}%
\]%
\[
\pm\hbar\omega-E_{B_{1}}+E_{B}\rightarrow\pm\hbar\omega-E_{B_{1}}+E_{B}-i0^{+}%
\]%
\[
\int_{-\infty}^{0}ds^{\prime\prime}e^{\pm i\omega s^{\prime\prime}%
-is^{\prime\prime}(E_{B_{1}}-E_{B})/\hbar}=\frac{\hbar}{i(\pm\hbar
\omega-E_{B_{1}}+E_{B})}%
\]
consider%
\[
\int_{-\infty}^{0}ds^{\prime}e^{is^{\prime}(E_{B_{2}}-E_{B})/\hbar}\cos
s^{\prime}\omega_{\alpha^{\prime}}%
\]%
\[
\int_{-\infty}^{0}ds^{\prime}e^{is^{\prime}\omega_{1}}\cos s^{\prime}%
\omega,\text{ \ \ }\omega=\omega_{\alpha^{\prime}}\text{, \ \ \ }\omega
_{1}=(E_{B_{2}}-E_{B})/\hbar
\]%
\[
\int_{-\infty}^{0}ds^{\prime}e^{is^{\prime}\omega_{1}}\cos s^{\prime}%
\omega\rightarrow\int_{-\infty}^{0}ds^{\prime}e^{is^{\prime}(\omega_{1}%
-i0^{+})}\cos s^{\prime}\omega
\]%
\[
=\int_{-\infty}^{0}ds^{\prime}\cos s^{\prime}(\omega_{1}-i0^{+})\cos
s^{\prime}\omega\]
\[+i\int_{-\infty}^{0}ds^{\prime}\sin s^{\prime}(\omega
_{1}-i0^{+})\cos s^{\prime}\omega
\]%
\[
=\frac{1}{2}\int_{-\infty}^{0}ds^{\prime}[\cos s^{\prime}(\omega_{1}%
-i0^{+}+\omega)+\cos s^{\prime}(\omega_{1}-i0^{+}-\omega)]
\]%
\[
+\frac{i}{2}\int_{-\infty}^{0}ds^{\prime}[\sin s^{\prime}(\omega_{1}%
-i0^{+}+\omega)+\sin s^{\prime}(\omega_{1}-i0^{+}-\omega)]
\]%
\[
=\frac{1}{2}\frac{1}{\omega_{1}-i0^{+}+\omega}[\sin s^{\prime}(\omega
_{1}-i0^{+}+\omega)\]
\[+i(-)\cos s^{\prime}(\omega_{1}-i0^{+}+\omega)]
\]%
\[
+\frac{1}{2}\frac{1}{\omega_{1}-i0^{+}-\omega}[\sin s^{\prime}(\omega
_{1}-i0^{+}-\omega)\]
\[+i(-)\cos s^{\prime}(\omega_{1}-i0^{+}-\omega)]
\]%
\[
=\frac{-i}{2}\frac{1}{\omega_{1}-i0^{+}+\omega}[\cos s^{\prime}(\omega
_{1}-i0^{+}+\omega)+i\sin s^{\prime}(\omega_{1}-i0^{+}+\omega)]
\]%
\[
-\frac{i}{2}\frac{1}{\omega_{1}-i0^{+}-\omega}[\cos s^{\prime}(\omega
_{1}-i0^{+}-\omega)+i\sin s^{\prime}(\omega_{1}-i0^{+}-\omega)]
\]%
\[
=\frac{-i}{2}\frac{e^{is^{\prime}(\omega_{1}-i0^{+}+\omega)}}{\omega
_{1}-i0^{+}+\omega}-\frac{i}{2}\frac{e^{is^{\prime}(\omega_{1}-i0^{+}-\omega
)}}{\omega_{1}-i0^{+}-\omega}%
\]%
\[
=\frac{-i}{2}[\frac{1}{\omega_{1}-i0^{+}+\omega}+\frac{1}{\omega_{1}%
-i0^{+}-\omega}]
\]
consider%
\[
\int_{-\infty}^{0}ds^{\prime}e^{is^{\prime}(E_{B_{2}}-E_{B})/\hbar}\sin
s^{\prime}\omega_{\alpha^{\prime}}%
\]%
\[
\int_{-\infty}^{0}ds^{\prime}e^{is^{\prime}\omega_{1}}\sin s^{\prime}%
\omega=\int_{-\infty}^{0}ds^{\prime}\cos s^{\prime}\omega_{1}\sin s^{\prime
}\omega\]
\[+i\int_{-\infty}^{0}ds^{\prime}\sin s^{\prime}\omega_{1}\sin s^{\prime
}\omega
\]%
\[
=\frac{1}{2}\int_{-\infty}^{0}ds^{\prime}[\sin s^{\prime}(\omega_{1}%
-i0^{+}+\omega)-\sin s^{\prime}(\omega_{1}-i0^{+}-\omega)]
\]%
\[
+\frac{i}{2}\int_{-\infty}^{0}ds^{\prime}[\cos s^{\prime}(\omega_{1}%
-i0^{+}-\omega)-\cos s^{\prime}(\omega_{1}-i0^{+}+\omega)]
\]%
\[
=-\frac{\cos s^{\prime}(\omega_{1}-i0^{+}+\omega)}{2(\omega_{1}-i0^{+}%
+\omega)}-\frac{i\sin s^{\prime}(\omega_{1}-i0^{+}+\omega)}{2(\omega
_{1}-i0^{+}+\omega)}%
\]%
\[
+\frac{\cos s^{\prime}(\omega_{1}-i0^{+}-\omega)}{2(\omega_{1}-i0^{+}-\omega
)}+\frac{i\sin s^{\prime}(\omega_{1}-i0^{+}-\omega)}{2(\omega_{1}%
-i0^{+}-\omega)}%
\]%
\[
=-\frac{e^{is^{\prime}(\omega_{1}-i0^{+}+\omega)}}{2(\omega_{1}-i0^{+}%
+\omega)}+\frac{e^{is^{\prime}(\omega_{1}-i0^{+}-\omega)}}{2(\omega_{1}%
-i0^{+}-\omega)}%
\]%
\[
=\frac{1}{2}[\frac{1}{\omega_{1}-i0^{+}-\omega}-\frac{1}{\omega_{1}%
-i0^{+}+\omega}]
\]
in summary%
\[
\int_{-\infty}^{0}ds^{\prime}e^{is^{\prime}(E_{B_{2}}-E_{B})/\hbar}(\coth
\frac{\beta\hbar\omega_{\alpha^{\prime}}}{2}\cos s^{\prime}\omega
_{\alpha^{\prime}}+i\sin s^{\prime}\omega_{\alpha^{\prime}})
\]%
\[
=\frac{-i}{2}\coth\frac{\beta\hbar\omega_{\alpha^{\prime}}}{2}[\frac{1}%
{\omega_{1}-i0^{+}+\omega}+\frac{1}{\omega_{1}-i0^{+}-\omega}]
\]%
\[
+\frac{i}{2}[\frac{1}{\omega_{1}-i0^{+}-\omega}-\frac{1}{\omega_{1}%
-i0^{+}+\omega}]
\]
Fig.10(b):%
\[
K_{1B_{3}B_{2}B_{1}B\pm}(\omega,T)=\int_{-\infty}^{0}dse^{-is(E_{B_{3}%
}-E_{B_{2}})/\hbar}%
\]%
\begin{equation}
\int_{-\infty}^{0}ds^{\prime}e^{-i(s+s^{\prime})(E_{B_{2}}-E_{B_{1}})/\hbar}
\label{fKK2}%
\end{equation}%
\[
\int_{-\infty}^{0}ds^{\prime\prime}e^{\pm i(s+s^{\prime}+s^{\prime\prime
})\omega}e^{-i(s+s^{\prime}+s^{\prime\prime})(E_{B_{1}}-E_{B})/\hbar}%
\]%
\[
\frac{1}{2}\sum_{\alpha^{\prime}}K_{B_{3}B_{2}}^{\alpha^{\prime}}K_{B_{2}%
B_{1}}^{\alpha^{\prime}}\frac{\hbar}{M_{\alpha^{\prime}}\omega_{\alpha
^{\prime}}}\]
\[[\coth\frac{\beta\hbar\omega_{\alpha}}{2}\cos s^{\prime}%
\omega_{\alpha}-i\sin s^{\prime}\omega_{\alpha}].
\]
Fig.10(c):%
\[
K_{2B_{3}B_{2}B_{1}B\pm}(\omega,T)=\int_{-\infty}^{0}dse^{-is(E_{B_{3}%
}-E_{B_{2}})/\hbar}%
\]%
\[
\int_{-\infty}^{0}ds^{\prime}e^{\pm i(s+s^{\prime})\omega}e^{-i(s+s^{\prime
})(E_{B_{2}}-E_{B_{1}})/\hbar}%
\]%
\begin{equation}
\int_{-\infty}^{0}ds^{\prime\prime}e^{-i(s+s^{\prime}+s^{\prime\prime
})(E_{B_{1}}-E_{B})/\hbar} \label{KfK2}%
\end{equation}%
\begin{align*}
&  \frac{1}{2}\sum_{\alpha^{\prime}}K_{B_{3}B_{2}}^{\alpha^{\prime}}K_{B_{1}%
B}^{\alpha^{\prime}}\frac{\hbar}{M_{\alpha^{\prime}}\omega_{\alpha^{\prime}}%
}\\
&  \lbrack\coth\frac{\beta\hbar\omega_{\alpha}}{2}\cos(s^{\prime}%
+s^{\prime\prime})\omega_{\alpha}-i\sin(s^{\prime}+s^{\prime\prime}%
)\omega_{\alpha}].
\end{align*}
Fig.10(d):%
\[
K_{3B_{3}B_{2}B_{1}B\pm}(\omega,T)=\int_{-\infty}^{0}dse^{\pm is\omega
}e^{-is(E_{B_{3}}-E_{B_{2}})/\hbar}%
\]%
\begin{equation}
\int_{-\infty}^{0}ds^{\prime}e^{-i(s+s^{\prime})(E_{B_{2}}-E_{B_{1}})/\hbar}
\label{KKf2}%
\end{equation}%
\[
\int_{-\infty}^{0}ds^{\prime\prime}e^{-i(s+s^{\prime}+s^{\prime\prime
})(E_{B_{1}}-E_{B})/\hbar}%
\]%
\[
\frac{1}{2}\sum_{\alpha^{\prime}}K_{B_{2}B_{1}}^{\alpha^{\prime}}K_{B_{1}%
B}^{\alpha^{\prime}}\frac{\hbar}{M_{\alpha^{\prime}}\omega_{\alpha^{\prime}}%
}\]
\[[\coth\frac{\beta\hbar\omega_{\alpha}}{2}\cos s^{\prime\prime}\omega_{\alpha
}-i\sin s^{\prime\prime}\omega_{\alpha}].
\]
\section{Asymptotic calculation of the time integrals}\label{sad}

To illustrate how to carry out time integrals, we take Eq.(\ref{r0+}) as an
example. Extend the integrand to the whole complex$-s$ plane and write
$s=u+iv$. In $(u,v)$ language,
\[
\coth\frac{\beta\hbar\omega}{2}\cos\omega s+i\sin\omega s
\]%
\[
=\cos\omega u(\coth\frac{\beta\hbar\omega}{2}\cosh\omega v-\sinh\omega v)
\]%
\begin{equation}
+i\sin\omega u(\cosh\omega v-\coth\frac{\beta\hbar\omega}{2}\sinh\omega v).
\label{rho}%
\end{equation}
At one end point ($0,0$) of the integral, the imaginary part of Eq.(\ref{rho})
is zero. There are two paths on which the imaginary part of Eq.(\ref{rho}) is
also zero: (1) $u=0$ and (2) $v=\beta\hbar/2$. They are the steepest descent paths.

Because the integrand is analytic in the whole complex$-s$ plane, according to the
Cauchy theorem, we can deform the original contour C: ($-\infty,0$%
)$\rightarrow(0,0)$ to steepest descent path C$_{1}$+C$_{2}$+C$_{3}$, where
C$_{1}$: $(-X,0)\rightarrow(-X,\beta\hbar/2)$, C$_{2}$: $(-X,\beta
\hbar/2)\rightarrow(0,\beta\hbar/2)$, C$_{3}$: $(0,\beta\hbar/2)\rightarrow
(0,0)$, In the end, let $X\rightarrow\infty$. Obviously, $\int_{C}=\int
_{C_{1}}+\int_{C_{2}}+\int_{C_{3}}$.

Along path C$_{1}$: $s=-X+iv$, where $v$: $0\rightarrow\beta\hbar/2$. The
integral becomes%
\[
e^{-iX(\omega-\omega_{A_{1}A}^{\prime})}\int_{0}^{\beta\hbar/2}dve^{-v(\omega
-\omega_{A_{1}A}^{\prime})}\exp\{\frac{1}{2}\sum_{\alpha}(\theta_{\alpha
}^{A_{1}}-\theta_{\alpha}^{A})^{2}%
\]%
\[
\lbrack\cos\omega_{\alpha}X(\coth\frac{\beta\hbar\omega_{\alpha}}{2}%
\cosh\omega_{\alpha}v-\sinh\omega_{\alpha}v)
\]%
\begin{equation}
+i\sin\omega_{\alpha}X(\coth\frac{\beta\hbar\omega_{\alpha}}{2}\sinh
\omega_{\alpha}v-\cosh\omega_{\alpha}v)]\}. \label{c1}%
\end{equation}
Because $X$\ is a large positive number, the phase of the integrand changes
sign wildly due to $\sin\omega_{\alpha}X$, and Eq.(\ref{c1}) is negligible.

Along C$_{2}$, $s=u+i\frac{\beta\hbar}{2}$, where $u$: $-X\rightarrow0$.
Excepting the factor $\exp\{-\sum_{\alpha}\frac{(\theta_{\alpha}^{A_{1}%
}-\theta_{\alpha}^{A})^{2}}{2}\coth\frac{\beta\hbar\omega_{\alpha}}{2}\}$, the
integral\ becomes%
\[
e^{-\beta\hbar\omega/2}e^{\beta\hbar\omega_{A_{1}A}^{\prime}/2}\int_{-\infty
}^{0}due^{i\omega u-iu\omega_{A_{1}A}^{\prime}}%
\]%
\begin{equation}
\exp\{\frac{1}{2}\sum_{\alpha}(\theta_{\alpha}^{A_{1}}-\theta_{\alpha}%
^{A})^{2}\csc h\frac{\beta\hbar\omega_{\alpha}}{2}\cos\omega_{\alpha}u\}.
\label{c2}%
\end{equation}
The main contribution comes from the neighborhood of $u=0$. Expand $\cos
\omega_{\alpha}u\approx1-u^{2}\frac{\omega_{\alpha}^{2}}{2}$, then
Eq.(\ref{c2}) is proportional to%
\begin{equation}
\int_{-\infty}^{0}due^{i\omega u-iu\omega_{A_{1}A}^{\prime}}e^{-Cu^{2}},\text{
\ } \label{c0}%
\end{equation}
where%
\[
C=\frac{1}{4}\sum_{\alpha}\omega_{\alpha}^{2}(\theta_{\alpha}^{A_{1}}%
-\theta_{\alpha}^{A})^{2}\csc h\frac{\beta\hbar\omega_{\alpha}}{2}.
\]
Around $u=0$, the Taylor series of $e^{i\omega u-iu\omega_{A_{1}A}^{\prime}}$
converges rapidly. One can then easily carry out (\ref{c0}).

Along $C_{3}^{-}$: $(0,0)\rightarrow(0,\frac{\beta\hbar}{2})$, the integral
becomes%
\[
i\int_{0}^{\beta\hbar/2}dve^{-v(\omega-\omega_{A_{1}A}^{\prime})}\exp
\{\frac{1}{2}\sum_{\alpha}(\theta_{\alpha}^{A_{1}}-\theta_{\alpha}^{A})^{2}%
\]%
\begin{equation}
(\coth\frac{\beta\hbar\omega_{\alpha}}{2}\cosh\omega_{\alpha}v-\sinh
\omega_{\alpha}v)\}. \label{c3}%
\end{equation}
According to the Laplace method, the main contribution comes from the
neighborhood of $v=\beta\hbar/2$. The leading behavior of (\ref{c3}) is then%
\[
i\{\sum_{\alpha}\frac{1}{2}(\theta_{\alpha}^{A_{1}}-\theta_{\alpha}^{A}%
)^{2}\omega_{\alpha}\cosh\frac{\beta\hbar\omega_{\alpha}}{2}\}^{-1}%
e^{-\beta\hbar(\omega-\omega_{A_{1}A}^{\prime})/2}%
\]%
\begin{equation}
\exp\{\sum_{\alpha}\frac{1}{2}(\theta_{\alpha}^{A_{1}}-\theta_{\alpha}%
^{A})^{2}\csc h\frac{\beta\hbar\omega_{\alpha}}{2}\}. \label{c3a}%
\end{equation}

\section{Time integrals for Hall mobility}\label{tjH}

The time integral for Fig.11(a) is%
\begin{equation}
U_{AA_{2}A_{1}}=\int_{-\infty}^{0}dse^{is(E_{A}-E_{A_{2}})} \label{u}%
\end{equation}%
\begin{align*}
&  \int_{-\infty}^{0}ds^{\prime}e^{is^{\prime}(E_{A}-E_{A_{1}})}\exp
(\sum_{\alpha}\{-\frac{1}{4}\coth\frac{\beta\hbar\omega_{\alpha}}{2}\\
&  \lbrack(\theta_{\alpha}^{A_{2}}-\theta_{\alpha}^{A_{1}})^{2}+(\theta
_{\alpha}^{A_{2}}-\theta_{\alpha}^{A})^{2}+(\theta_{\alpha}^{A_{1}}%
-\theta_{\alpha}^{A})^{2}]
\end{align*}%
\[
+\frac{1}{2}(\theta_{\alpha}^{A_{2}}-\theta_{\alpha}^{A_{1}})(\theta_{\alpha
}^{A_{2}}-\theta_{\alpha}^{A})\]
\[[\coth\frac{\beta\hbar\omega_{\alpha}}{2}\cos
s\omega_{\alpha}-i\sin s\omega_{\alpha}]
\]%
\[
+\frac{1}{2}(\theta_{\alpha}^{A_{1}}-\theta_{\alpha}^{A})(\theta_{\alpha
}^{A_{2}}-\theta_{\alpha}^{A_{1}})\]
\[[\cos s\omega_{\alpha}-\coth\frac{\beta
\hbar\omega_{\alpha}}{2}i\sin s\omega_{\alpha}]
\]%
\begin{align*}
&  -\frac{1}{2}(\theta_{\alpha}^{A_{1}}-\theta_{\alpha}^{A})(\theta_{\alpha
}^{A_{2}}-\theta_{\alpha}^{A})\\
&  \lbrack\cos(s+s^{\prime})\omega_{\alpha}-\coth\frac{\beta\hbar
\omega_{\alpha}}{2}i\sin(s+s^{\prime})\omega_{\alpha}]\}).
\end{align*}
The time integral for Fig.11(b) is%
\begin{equation}
U_{AA_{1}A_{2}}^{\prime}=\int_{-\infty}^{0}dse^{is(E_{A_{2}}-E_{A_{1}})/\hbar}
\label{up}%
\end{equation}%
\begin{align*}
&  \int_{-\infty}^{0}ds^{\prime}e^{-is^{\prime}(E_{A_{1}}-E_{A})/\hbar}%
\exp(\sum_{\alpha}\{-\frac{1}{4}\coth\frac{\beta\hbar\omega_{\alpha}}{2}\\
&  \lbrack(\theta_{\alpha}^{A_{2}}-\theta_{\alpha}^{A})^{2}+(\theta_{\alpha
}^{A_{2}}-\theta_{\alpha}^{A_{1}})^{2}+(\theta_{\alpha}^{A_{1}}-\theta
_{\alpha}^{A})^{2}]
\end{align*}%
\[
+\frac{1}{2}(\theta_{\alpha}^{A_{2}}-\theta_{\alpha}^{A})(\theta_{\alpha
}^{A_{2}}-\theta_{\alpha}^{A_{1}})\]
\[[\coth\frac{\beta\hbar\omega_{\alpha}}%
{2}\cos s\omega_{\alpha}+i\sin s\omega_{\alpha}]
\]%
\begin{align*}
&  -\frac{1}{2}(\theta_{\alpha}^{A_{1}}-\theta_{\alpha}^{A})(\theta_{\alpha
}^{A_{2}}-\theta_{\alpha}^{A_{1}})\\
&  \lbrack\coth\frac{\beta\hbar\omega_{\alpha}}{2}\cos(s+s^{\prime}%
)\omega_{\alpha}-i\sin(s+s^{\prime})\omega_{\alpha}]
\end{align*}%
\[
+\frac{1}{2}(\theta_{\alpha}^{A_{1}}-\theta_{\alpha}^{A})(\theta_{\alpha
}^{A_{2}}-\theta_{\alpha}^{A})\]
\[[\coth\frac{\beta\hbar\omega_{\alpha}}{2}\cos
s^{\prime}\omega_{\alpha}-i\sin s^{\prime}\omega_{\alpha}]\}).
\]

The topology of Fig.12(a), Fig.13(a) and Fig.12(b) are the same. They
have same time integral:%
\begin{equation}
Z_{AA_{3}A_{2}A_{1}}=\int_{-\infty}^{0}dse^{-is(E_{A_{2}}-E_{A})/\hbar}
\label{z}%
\end{equation}%
\[
\int_{-\infty}^{0}ds^{\prime}e^{-is^{\prime}(E_{A_{1}}-E_{A})/\hbar}%
\]%
\begin{align*}
&  \int_{-\infty}^{0}ds^{\prime\prime}e^{is^{\prime\prime}(E_{A_{3}}%
-E_{A})/\hbar}\exp(\sum_{\alpha}\{-\frac{1}{4}\coth\frac{\beta\hbar
\omega_{\alpha}}{2}\\
&  [(\theta_{\alpha}^{A_{1}}-\theta_{\alpha}^{A})^{2}+(\theta_{\alpha}^{A_{2}%
}-\theta_{\alpha}^{A_{1}})^{2}+(\theta_{\alpha}^{A_{2}}-\theta_{\alpha}%
^{A_{3}})^{2}+(\theta_{\alpha}^{A_{3}}-\theta_{\alpha}^{A})^{2}]
\end{align*}%
\[
-\frac{1}{2}(\theta_{\alpha}^{A_{2}}-\theta_{\alpha}^{A_{1}})(\theta_{\alpha
}^{A_{1}}-\theta_{\alpha}^{A})[\coth\frac{\beta\hbar\omega_{\alpha}}{2}%
\cos\omega_{\alpha}s^{\prime}-i\sin\omega_{\alpha}s^{\prime}]
\]%
\[
-\frac{1}{2}(\theta_{\alpha}^{A_{3}}-\theta_{\alpha}^{A})(\theta_{\alpha
}^{A_{2}}-\theta_{\alpha}^{A_{3}})[\coth\frac{\beta\hbar\omega_{\alpha}}%
{2}\cos\omega_{\alpha}s^{\prime\prime}+i\sin\omega_{\alpha}s^{\prime\prime}]
\]%
\begin{align*}
&  +\frac{1}{2}(\theta_{\alpha}^{A_{2}}-\theta_{\alpha}^{A_{3}})(\theta
_{\alpha}^{A_{1}}-\theta_{\alpha}^{A})\\
&  [\coth\frac{\beta\hbar\omega_{\alpha}}{2}\cos\omega_{\alpha}(s+s^{\prime
})-i\sin\omega_{\alpha}(s+s^{\prime})]
\end{align*}%
\[
+\frac{1}{2}(\theta_{\alpha}^{A_{2}}-\theta_{\alpha}^{A_{3}})(\theta_{\alpha
}^{A_{2}}-\theta_{\alpha}^{A_{1}})[\coth\frac{\beta\hbar\omega_{\alpha}}%
{2}\cos\omega_{\alpha}s-i\sin\omega_{\alpha}s]
\]%
\begin{align*}
&  +\frac{1}{2}(\theta_{\alpha}^{A_{3}}-\theta_{\alpha}^{A})(\theta_{\alpha
}^{A_{1}}-\theta_{\alpha}^{A})\\
&  [\coth\frac{\beta\hbar\omega_{\alpha}}{2}\cos\omega_{\alpha}(s+s^{\prime
}-s^{\prime\prime})-i\sin\omega_{\alpha}(s+s^{\prime}-s^{\prime\prime})]
\end{align*}%
\begin{align*}
&  +\frac{1}{2}(\theta_{\alpha}^{A_{3}}-\theta_{\alpha}^{A})(\theta_{\alpha
}^{A_{2}}-\theta_{\alpha}^{A_{1}})\\
&  [\coth\frac{\beta\hbar\omega_{\alpha}}{2}\cos\omega_{\alpha}(s-s^{\prime
\prime})-i\sin\omega_{\alpha}(s-s^{\prime\prime})]\}).
\end{align*}

The topology of Fig.12(c) and Fig.12(d) are the same. They have same
time integral:%
\begin{equation}
Z_{AA_{3}A_{2}A_{1}}^{\prime}=\int_{-\infty}^{0}dse^{is(E_{A_{2}}-E_{A}%
)/\hbar} \label{zp}%
\end{equation}%
\[
\int_{-\infty}^{0}ds^{\prime}e^{is^{\prime}(E_{A_{1}}-E_{A})/\hbar}%
\]%
\[
 \int_{-\infty}^{0}ds^{\prime\prime}e^{-is^{\prime\prime}(E_{A_{3}}%
-E_{A})/\hbar}\]
\[\exp(\sum_{\alpha}\{-\frac{1}{4}\coth\frac{\beta\hbar
\omega_{\alpha}}{2}
  [(\theta_{\alpha}^{A_{1}}-\theta_{\alpha}^{A})^{2}\]
  \[+(\theta_{\alpha}^{A_{2}%
}-\theta_{\alpha}^{A_{1}})^{2}+(\theta_{\alpha}^{A_{2}}-\theta_{\alpha}%
^{A_{3}})^{2}+(\theta_{\alpha}^{A_{3}}-\theta_{\alpha}^{A})^{2}]
\]
\[
-\frac{1}{2}(\theta_{\alpha}^{A_{2}}-\theta_{\alpha}^{A_{1}})(\theta_{\alpha
}^{A_{1}}-\theta_{\alpha}^{A})\]
\[[\coth\frac{\beta\hbar\omega_{\alpha}}{2}%
\cos\omega_{\alpha}s^{\prime}+i\sin\omega_{\alpha}s^{\prime}]
\]%
\[
-\frac{1}{2}(\theta_{\alpha}^{A_{3}}-\theta_{\alpha}^{A})(\theta_{\alpha
}^{A_{2}}-\theta_{\alpha}^{A_{3}})\]
\[[\coth\frac{\beta\hbar\omega_{\alpha}}%
{2}\cos\omega_{\alpha}s^{\prime\prime}-i\sin\omega_{\alpha}s^{\prime\prime}]
\]%
\[
 +\frac{1}{2}(\theta_{\alpha}^{A_{2}}-\theta_{\alpha}^{A_{3}})(\theta
_{\alpha}^{A_{1}}-\theta_{\alpha}^{A})\]
\[
  [\coth\frac{\beta\hbar\omega_{\alpha}}{2}\cos\omega_{\alpha}(s+s^{\prime
})+i\sin\omega_{\alpha}(s+s^{\prime})]
\]%
\[
+\frac{1}{2}(\theta_{\alpha}^{A_{2}}-\theta_{\alpha}^{A_{3}})(\theta_{\alpha
}^{A_{2}}-\theta_{\alpha}^{A_{1}})\]
\[[\coth\frac{\beta\hbar\omega_{\alpha}}%
{2}\cos\omega_{\alpha}s+i\sin\omega_{\alpha}s]
\]%
\[
  +\frac{1}{2}(\theta_{\alpha}^{A_{3}}-\theta_{\alpha}^{A})(\theta_{\alpha
}^{A_{1}}-\theta_{\alpha}^{A})\]
\[
  [\coth\frac{\beta\hbar\omega_{\alpha}}{2}\cos\omega_{\alpha}(s+s^{\prime
}-s^{\prime\prime})+i\sin\omega_{\alpha}(s+s^{\prime}-s^{\prime\prime})]
\]%
\[
  +\frac{1}{2}(\theta_{\alpha}^{A_{3}}-\theta_{\alpha}^{A})(\theta_{\alpha
}^{A_{2}}-\theta_{\alpha}^{A_{1}})\]
\[
  [\coth\frac{\beta\hbar\omega_{\alpha}}{2}\cos\omega_{\alpha}(s-s^{\prime
\prime})+i\sin\omega_{\alpha}(s-s^{\prime\prime})]\}).
\]

The topology of Fig.13(b), Fig.13(c) and Fig.13(d) are the same.
They have same time integral:%
\begin{equation}
Y_{A_{3}A_{2}A_{1}A}=\int_{-\infty}^{0}dse^{-is(E_{A_{3}}-E_{A_{2}})/\hbar}
\label{y}%
\end{equation}%
\[
\int_{-\infty}^{0}ds^{\prime}e^{-i(s+s^{\prime})(E_{A_{2}}-E_{A_{1}})/\hbar}%
\]%
\[
\int_{-\infty}^{0}ds^{\prime\prime}e^{-i(s+s^{\prime}+s^{\prime\prime
})(E_{A_{1}}-E_{A})/\hbar}\exp\{-\frac{1}{4}\sum_{\alpha}\coth\frac{\beta
\hbar\omega_{\alpha}}{2}%
\]%
\[
\lbrack(\theta_{\alpha}^{A_{3}}-\theta_{\alpha}^{A})^{2}+(\theta_{\alpha
}^{A_{2}}-\theta_{\alpha}^{A_{3}})^{2}+(\theta_{\alpha}^{A_{2}}-\theta
_{\alpha}^{A_{1}})^{2}+(\theta_{\alpha}^{A_{1}}-\theta_{\alpha}^{A})^{2}]
\]

\begin{align*}
&  +\frac{1}{2}\sum_{\alpha}(\theta_{\alpha}^{A_{1}}-\theta_{\alpha}%
^{A})(\theta_{\alpha}^{A_{2}}-\theta_{\alpha}^{A_{3}})\\
&  [\coth\frac{\beta\hbar\omega_{\alpha}}{2}\cos(s^{\prime}+s^{\prime\prime
})\omega_{\alpha}-i\sin(s^{\prime}+s^{\prime\prime})\omega_{\alpha}]
\end{align*}%
\[
+\frac{1}{2}\sum_{\alpha}(\theta_{\alpha}^{A_{2}}-\theta_{\alpha}^{A_{3}%
})(\theta_{\alpha}^{A_{2}}-\theta_{\alpha}^{A_{1}})[\coth\frac{\beta
\hbar\omega_{\alpha}}{2}\cos s^{\prime}\omega_{\alpha}-i\sin s^{\prime}%
\omega_{\alpha}]
\]%
\[
-\frac{1}{2}\sum_{\alpha}(\theta_{\alpha}^{A_{2}}-\theta_{\alpha}^{A_{1}%
})(\theta_{\alpha}^{A_{1}}-\theta_{\alpha}^{A})[\coth\frac{\beta\hbar
\omega_{\alpha}}{2}\cos s^{\prime\prime}\omega_{\alpha}-i\sin s^{\prime\prime
}\omega_{\alpha}]
\]%
\begin{align*}
&  +\frac{1}{2}\sum_{\alpha}(\theta_{\alpha}^{A_{3}}-\theta_{\alpha}%
^{A})(\theta_{\alpha}^{A_{1}}-\theta_{\alpha}^{A})\\
&  [\coth\frac{\beta\hbar\omega_{\alpha}}{2}\cos(s+s^{\prime}+s^{\prime\prime
})\omega_{\alpha}-\sin(s+s^{\prime}+s^{\prime\prime})\omega_{\alpha}]
\end{align*}%
\begin{align*}
&  +\frac{1}{2}\sum_{\alpha}(\theta_{\alpha}^{A_{2}}-\theta_{\alpha}^{A_{1}%
})(\theta_{\alpha}^{A_{3}}-\theta_{\alpha}^{A})\\
&  [\coth\frac{\beta\hbar\omega_{\alpha}}{2}\cos(s+s^{\prime})\omega_{\alpha
}-i\sin(s+s^{\prime})\omega_{\alpha}]
\end{align*}%
\[
-\frac{1}{2}\sum_{\alpha}(\theta_{\alpha}^{A_{2}}-\theta_{\alpha}^{A_{3}%
})(\theta_{\alpha}^{A_{3}}-\theta_{\alpha}^{A})[\coth\frac{\beta\hbar
\omega_{\alpha}}{2}\cos s\omega_{\alpha}-i\sin s\omega_{\alpha}]\}.
\]
\section{Potential energy partition}
For extended states, we separate\cite{epjb} $h_{e}$ into two parts: $h_{e}=h_{e}^{0}+h_{e-ph}$, where
\begin{equation}
h_{e}^{0}=\frac{-\hbar^{2}}{2m}\nabla_{\mathbf{r}}^{2}+\sum_{\mathbf{n}%
=1}^{\mathcal{N}}U(\mathbf{r}-\mathcal{R}_{\mathbf{n}}), \label{fix}%
\end{equation}
and%
\begin{equation}
h_{e-ph}=-\sum_{\mathbf{n}=1}^{\mathcal{N}}\mathbf{u}_{\mathbf{n}}\cdot
\nabla_{\mathbf{r}}U(\mathbf{r}-\mathcal{R}_{\mathbf{n}})=\sum_{j=1}%
^{3\mathcal{N}}x_{j}\frac{\partial U}{\partial X_{j}}. \label{eph}%
\end{equation}
The eigenvectors {$\xi_{B}$} and eigenvalues {$E_{B}$} of $h_{e}^{0}$ are taken as those of the Kohn-Sham Hamiltonian.
To make partition (\ref{fix},\ref{eph}) work, one needs: (1) $h_{e-ph}$ does
not change the gross feature of the phonon spectrum and the normal modes; (2)
the changes in the eigenvalues and eigenvectors of $h_{e}^{0}$ caused by
$h_{e-ph}$ are small.

For a localized state, we partition the potential energy as
\[
\sum_{\mathbf{n}}U(r-\mathcal{R}_{\mathbf{n}},\mathbf{u}_{v}^{\mathbf{n}%
})=\sum_{\mathbf{n}\in D_{A_{1}}}U(r-\mathcal{R}_{\mathbf{n}},\mathbf{u}%
_{v}^{\mathbf{n}})
\]%
\begin{equation}
+\sum_{\mathbf{p}\notin D_{A_{1}}}U(r-\mathcal{R}_{\mathbf{p}},\mathbf{u}%
_{v}^{\mathbf{p}}). \label{sl}%
\end{equation}
Because the wave function of a localized state is confined in a finite region, the e-ph interaction from the atoms outside $D_{A}$ is negligible:
\begin{equation}
\sum_{\mathbf{p}\notin D_{A_{1}}}U(r-\mathcal{R}_{\mathbf{p}},\mathbf{u}%
_{v}^{\mathbf{p}})\thickapprox\sum_{\mathbf{p}\notin D_{A_{1}}}U(r-\mathcal{R}%
_{\mathbf{p}}). \label{outi}%
\end{equation}
In partition (\ref{sl}), the zero-order Hamiltonian for localized state
$A_{1}$ is%
\begin{equation}
h_{A_{1}}^{0}=\frac{-\hbar^{2}}{2m}\nabla_{\mathbf{r}}^{2}+\sum_{\mathbf{n}\in
D_{A_{1}}}U(r-\mathcal{R}_{\mathbf{n}},\mathbf{u}_{v}^{\mathbf{n}}).
\label{a10}%
\end{equation}
Denote the lowest eigenvalue and corresponding eigenvector of $h_{A_{1}%
}^{0}$ as $E_{A_{1}}$ and $\phi_{A_{1}}$, using stationary perturbation theory, we have
$\phi_{A_{1}}\approx\phi_{A_{1}}^{0}$, and%
\begin{equation}
E_{A_{1}}=E_{A_{1}}^{0}-\sum_{p_{A_{1}}\mathbf{\in}D_{A_{1}}}d_{p_{A_{1}}%
}x_{p_{A_{1}}}^{A_{1}}, \label{apL}%
\end{equation}
where $\phi_{A_{1}}^{0}$ and $E_{A_{1}}^{0}$ are the Kohn-Sham eigenvector and eigenvalue.
\begin{equation}
d_{p_{A_{1}}}=\int d\mathbf{r}|\phi_{A_{1}}^{0}(\mathbf{r})|^{2}\partial
U/\partial X_{p_{A_{1}}}, \label{epn}%
\end{equation}
is the e-ph coupling for the $p_{A_{1}}^{th}$ vibrational degree of freedom.
Here $\partial U/\partial X_{p_{A_{1}}}$ is the derivative of the
single-electron potential energy $U$ respect to the $p_{A_{1}}^{\text{th}}$
nuclear degree of freedom.

The partition (\ref{sl}) is justified if%
\[
|\int dr\phi_{A_{2}}^{\ast}(\mathbf{r}-\mathcal{R}_{A_{2}})\sum_{\mathbf{p}%
\notin D_{A_{1}}}U(\mathbf{r}-\mathcal{R}_{\mathbf{p}})\phi_{A_{1}}%
(\mathbf{r}-\mathcal{R}_{A_{1}})|
\]%
\begin{equation}
<|\sum_{p_{A_{1}}\mathbf{\in}D_{A_{1}}}d_{p_{A_{1}}}x_{p_{A_{1}}}|,
\label{condi}%
\end{equation}
and%
\[
|\int dr\xi_{B_{2}}^{\ast}(\mathbf{r})\sum_{\mathbf{p}\notin D_{A_{1}}%
}U(\mathbf{r}-\mathcal{R}_{\mathbf{p}})\phi_{A_{1}}(\mathbf{r}-\mathcal{R}%
_{A_{1}})|
\]%
\begin{equation}
<|\sum_{p_{A_{1}}\mathbf{\in}D_{A_{1}}}d_{p_{A_{1}}}x_{p_{A_{1}}}|.
\label{excon}%
\end{equation}
One can easily estimate the RHS of Eqs.(\ref{condi},\ref{excon}), the total
e-ph interaction energy for a localized state $\phi_{A_{1}}$. The number of
atoms inside $D_{A_{1}}$ is $\left(  \xi/d\right)  ^{3}$, where $\xi$ is the
localization length of $\phi_{A_{1}}$.
The RHS of (\ref{condi}) is $\thicksim\left(  \xi/d\right)
^{3}(4\pi\epsilon_{0})^{-1}(Z^{\ast}e^{2}/\xi^{2})a_{v}\\=(4\pi\epsilon
_{0})^{-1}(Z^{\ast}e^{2}/d)(\xi/d)(a_{v}/d)$. The attractive energy between a
nucleus in $D_{A_{2}}$ and the carrier in $D_{A_{1}}$ is $\thicksim
(4\pi\epsilon_{0})^{-1}Z^{\ast}e^{2}/R_{12}$, where $R_{12}$ is the distance
between the centers of two localized states. The LHS of (\ref{condi}) is\\
$\left(  \xi_{2}/d\right)  ^{3}e^{-R_{12}/d}(4\pi\epsilon_{0})^{-1}Z^{\ast
}e^{2}/R_{12}$, so that (\ref{condi}) is satisfied. The LHS of (\ref{excon})
is \\$(\xi/L)^{3/2}(\xi/d)(4\pi\epsilon_{0})^{-1}Z^{\ast}e^{2}/d$, where
$Z^{\ast}$ is typical effective nuclear charge, $L$ is the linear size of
sample. For a well-localized state $\xi<<L$, and factor $(\xi/L)^{3/2}$
assures that condition (\ref{excon}) is satisfied.
Since partition (\ref{sl}) includes polaron as an example, we can also apply the present \textit{ansatz} to obtain the
polaron states from first principles by computing the static displacements caused by e-ph interaction.

\section{Scenario for simulation}\label{num}
We describe how to implement the present method in \textit{ab inito} code.
For an amorphous semiconductor, using a super-cell structural model, one first
calculates the eigenvectors $\{\phi_{C}^{0}\}$ and eigenvalues $\{E_{C}^{0}\}$
of Kohn-Sham Hamiltonian $h_{e}$ at $\Gamma$ point. From the vibrational
Hamiltonian $h_{v}$ or dynamical matrix, one can obtain vibrational spectrum
$\{\omega_{\alpha}\}$ and eigenvectors $\{\Delta_{m\alpha}\}$.

For a given state $\phi_{C}^{0}$, the electron-vibration coupling constant for
the $p^{th}$ nuclear degree of freedom is given by:%
\begin{equation}
d_{p}^{C}=\int d\mathbf{r}|\phi_{C}^{0}(\mathbf{r})|^{2}\partial U/\partial
X_{p},\text{ \ \ \ }p=1,2,\cdots3\mathcal{N}, \label{cp}%
\end{equation}
where $U$ is the effective single-particle potential energy. The static
displacement $x_{m}^{0}$ of the $m^{th}$ degree of freedom is
\begin{equation}
x_{m}^{0}=\sum_{p}d_{p}(k^{-1})_{mp},~m=1,2,\cdots3\mathcal{N}, \label{stad}%
\end{equation}
where $k^{-1}$ is the inverse matrix of the dynamical matrix $k$. The
vibrational amplitude $x_{m}$ of the $m^{th}$ degree of freedom is determined
by the amplitudes $\{\Theta_{\alpha}\}$ of normal modes:
\begin{equation}
x_{m}=\sum_{\alpha}\Delta_{m\alpha}\Theta_{\alpha},\text{ }\Theta_{\alpha
}=\max\{\sqrt{\frac{kT}{M\omega_{\alpha}^{2}}},\sqrt{\frac{\hbar}%
{M\omega_{\alpha}}}\}, \label{va}%
\end{equation}
where $\Delta_{m\alpha}$ is the minor of the determinant\\ $\left\vert
k_{ik}-\omega^{2}M_{i}\delta_{ik}\right\vert =0$.
If for some indexes $x_{m}^{0}>x_{m}$, we say
state $\phi_{C}^{0}$ is a localized state $\phi_{A}$. If none of the three degrees of freedom of an atom satisfies $x_{m}^{0}>x_{m}$, we say that this atom does not belong to $D_{A}$. Otherwise we say that the atom belongs to $D_{A}$. We will treat the e-ph interaction in zeroth order together with $H_{v}$ and consider the polarization effect expressed by $x_{m}^{0}$. The polarization effect of a localized state is like a small polaron, but the electronic localization in AS is caused by the static disorder in $h^{0}_{e}$ rather than $h_{e-ph}$. The topological disorder contained in $\{{\mathcal{R}}_{n}\}$ is much larger than the static displacement $\{x^{0}_{m}\}$ induced by e-ph interaction, the binding force caused by the static disorder is much larger than the self-trapping force caused by the e-ph interaction\cite{epjb}.
If for a state $\phi_{C}$, none of vibrational degrees of freedom satisfies $x_{m}^{0}>x_{m}$, we say this state $\phi_{C}^{0}$ is an extended state and treat e-ph interaction as perturbation.


From $\{x_{p_{A_{1}}}^{0}\}$, the shift $\theta_{\alpha}^{A_{1}}$ of origin
for the $\alpha^{th}$ mode is then obtained\cite{epjb}. With the shift in
origin $\{\theta_{\alpha}^{A_{1}}\}$ for each localized state, the time
integrals in Sec.\ref{timej} and \ref{tjH} can be found numerically.
The small parameters $J,$ $J^{\prime},$ $K^{\prime}$ and $K$ for the residual
interactions are computed\cite{epjb} by the expressions below [text Eq.(16)].
Finally one obtains conductivity in tables 4 and 5 and Hall mobility in
[text Eq.(23) and table 8]. The present results are applicable to crystalline
materials and amorphous metals if the scattering mechanisms are only the
disorder potential and the electron-phonon interaction. Although we did not
explicitly deal with the scattering effect of the static disorder, the
scattering of Bloch waves by the disorder potential is already taken into
account by [text, Eq.(17)]. The reason is: the basis set $\{\xi_{B}\}$ for
$\psi^{(0)}$, $\psi^{(1)}$ and $\psi^{(2)}$ are solved from the Kohn-Sham
Hamiltonian, so that $\psi^{(0)}$, $\psi^{(1)}$ and $\psi^{(2)}$ are fully
dressed by the disorder potential.